\documentclass[12pt]{article}
\usepackage{graphicx} %
\usepackage{amssymb}
\usepackage{color}
\usepackage[authoryear,round]{natbib}
\usepackage{todonotes}
\usepackage{float}
\usepackage{setspace}
\usepackage{amsmath}
\usepackage{enumitem}
\usepackage{amssymb,amsmath}
\usepackage{wrapfig}
\usepackage{dsfont, mathrsfs}
\usepackage{hyperref}
\usepackage{cleveref}
\usepackage{amsthm}

\newtheorem{definition}{Definition}
\newtheorem{proposition}{Proposition}
\newtheorem{corollary}{Corollary}
\newtheorem{theorem}{Theorem}
\newtheorem{assumption}{Assumption}
\newtheorem{lemma}{Lemma}
\newtheorem{optimisation}{Optimisation}
\newtheorem{remark}{Remark}

\newcommand{\R}{{\mathds{R}}}

\newcommand{\Id}{{\mathds{1}}}

\renewcommand{\L}{{\mathcal{L}}}
\newcommand{\M}{{\mathcal{M}}}

\newcommand{\Minv}{{\breve{\mathcal{M}}}}

\newcommand{\DB}{{\mathscr{B}}}

\newcommand{\ep}{{\varepsilon}}
\renewcommand{\d}{{\mathrm{d}}}

\DeclareMathOperator*{\argmin}{argmin}
\DeclareMathOperator*{\argmax}{argmax}

\renewcommand{\Finv}{{\Breve{F}}}
\newcommand{\Ginv}{{\Breve{G}}}
\newcommand{\Hinv}{{\Breve{H}}}

\usepackage{geometry}
\usepackage{xcolor}
\numberwithin{theorem}{section}
\newtheorem{proposition}[theorem]{Proposition}
\newtheorem{lemma}[theorem]{Lemma}
\newtheorem{corollary}[theorem]{Corollary}
\newtheorem{definition}[theorem]{Definition}
\newtheorem{remark}[theorem]{Remark}
\newtheorem{assumption}[theorem]{Assumption}
\newtheorem{optimisation}[theorem]{Optimisation}
\newtheorem{example}[theorem]{Example} 
\hypersetup{
  colorlinks   = true, 
  urlcolor     = blue, 
  linkcolor    = blue, 
  citecolor   = blue 
}

\title{Robust distortion riskmetrics over Bregman–Wasserstein balls and their intersections
}
\author{Peng Liu, Steven Vanduffel and Carole Bernard }
\date{\today}

\begin{document}

\maketitle

\begin{abstract}
We study the worst-case values of distortion riskmetrics over Bregman--Wasserstein balls around a reference distribution. Unlike the squared $2$-Wasserstein distance, Bregman--Wasserstein divergences can penalize upward and downward deviations differently. Our main result applies to generalized distortion functions $g:[0,1]\to\R$ of bounded variation, with $g(0)=0$ and suitable endpoint regularity, which need not be monotone, normalized, or upper semicontinuous. For strictly convex generators satisfying a two-sided $p$-growth condition, we obtain a one-dimensional dual representation of the worst-case value. When a positive Lagrange multiplier makes the divergence constraint binding, we identify a unique quantile function through a least concave majorant and characterize when it attains the supremum. For absolutely continuous generalized distortion functions, we characterize the solution through an isotonic projection; we also study more general generators and shape constraints, including unimodality. Finally, we extend the analysis to intersections of Bregman--Wasserstein balls, describing models acceptable to several parties involved in decision making, and illustrate the results numerically.
\end{abstract}




\section{Introduction}\label{sec:intro}

Let $F$ be a reference distribution of losses and let $\phi$
be a strictly convex differentiable function. We study the problem
\[
\sup\big\{H_g(G):\ \DB_\phi(G,F)\leq\ep\big\},
\]
and its counterpart for the infimum, where $G$ ranges over
distributions satisfying the relevant integrability conditions and where $H_g$ is a distortion riskmetric, i.e., a signed Choquet integral with respect to a generalized distortion function $g$ of bounded variation, $\DB_\phi$ is the Bregman--Wasserstein divergence generated by $\phi$, and $\ep>0$. Distortion riskmetrics include (Range) Value-at-Risk, spectral and other distortion risk measures, as well as non-monotone functionals such as the Gini deviation and mean-median deviation. The worst-case and best-case values measure the model risk of a risk assessment that is based on the reference distribution $F$, which is never known exactly, in the spirit of distributionally robust optimization.

A standard choice of an ambiguity set is a Wasserstein ball. However, the
$2$-Wasserstein distance is symmetric: lowering a quantile is penalized
as much as raising it by the same amount. In risk management, however, underestimating losses is typically more concerning than overestimating them, and different stakeholders may have different sensitivities to model deviations. For instance, a regulator may be primarily concerned with underestimating tail risk, whereas a company may place greater emphasis on achieving a good overall fit to its data. The Bregman--Wasserstein divergence replaces the squared distance $(z_1-z_2)^2$ in the definition of the $2$-Wasserstein distance by the Bregman divergence $B_\phi(z_1,z_2)$ of a convex function $\phi$; it is asymmetric unless $\phi$ is quadratic, and reduces to the squared $2$-Wasserstein distance for $\phi(x)=x^2$. Moreover, as the comonotonic coupling is optimal \citep{pesenti2024optimal}, the Bregman--Wasserstein divergence is an integral of a function of the quantile functions, which makes worst-case problems over Bregman--Wasserstein balls tractable.

Our work relates to two strands of literature. First, Wasserstein balls are a common choice of ambiguity set in distributionally robust optimization \citep{esfahani2018data,gao2023distributionally}. Worst-case values of distortion risk measures over such balls are obtained by \citet{bernard2022robust} under additional mean and variance constraints, and by \citet{liu2025unimodal} and \citet{liu2026wasserstein} for distortion riskmetrics, with unimodality constraints and over $p$-Wasserstein balls, respectively. Second, Bregman--Wasserstein (BW) divergences have recently been employed in portfolio selection \citep{pesenti2024payoff,pesenti2026outperforming}. By replacing the conventional Wasserstein transport cost with a Bregman divergence generated by a convex function, BW divergences offer greater flexibility in measuring deviations from a benchmark distribution. In particular, their generally asymmetric nature allows upward and downward deviations to be penalized differently, while the choice of the generating function also determines how sensitive the divergence is to deviations at different loss levels. These properties make BW divergences particularly appealing in risk management, where underestimating large losses may be substantially more consequential than overestimating them.

Our first contribution concerns the worst-case problem over a single Bregman--Wasserstein ball. For strictly convex generators satisfying a two-sided $p$-growth condition, $p\in(1,\infty)$, we solve this problem over distributions with finite $p$-th moments for generalized distortion functions of bounded variation satisfying a corresponding endpoint growth condition. The distortion function need not be monotone, concave, continuous or upper semicontinuous, and may have countably many jumps, and the reference distribution may be unbounded (Theorem~\ref{thm:general}). The worst-case value is given by a one-dimensional dual formula, and it equals the maximum of the riskmetric of the upper semicontinuous envelope of the distortion function. Whenever a positive Lagrange multiplier makes the divergence constraint binding, this maximum is attained at a unique and explicit quantile function, obtained from the right derivative of the least concave majorant of the distortion function plus a term that depends only on the reference distribution and the generator. The supremum is attained if and only if this quantile function does not jump at the levels at which the distortion function lies strictly below its envelope; for Value-at-Risk, it jumps, so that the worst case is not attained (Remark~\ref{rem:VaR-envelope}). A binding multiplier exists for every radius, unless one minimizes a non-negative and location-invariant riskmetric such as the Gini deviation; then it exists for radii below a threshold $\bar\ep$, and possibly at it, whereas the bound is $0$ for radii at or above it (Proposition~\ref{prop:general-multiplier}). When such a multiplier exists, maximizing sequences converge to the worst case of the envelope, with an explicit lower bound on the optimality gap in terms of the Bregman--Wasserstein divergence (Proposition~\ref{prop:general-stability}).

For absolutely continuous generalized distortion functions,
we obtain the worst-case quantile function
\[
(\phi')^{-1}\big((\phi'(\Finv)+\gamma/\lambda^*)^{\uparrow}\big),
\]
where $\gamma$ is the weight function, $(\cdot)^\uparrow$
denotes the isotonic projection, and $\lambda^*>0$ is
a multiplier for which the divergence constraint binds
(Corollary~\ref{cor:general-ac}). This result requires
a two-sided quadratic growth condition on $\phi$, as
well as suitable conditions on the distortion function.
For distortion risk measures, whose weight functions
are nonnegative, we show that the same formula holds
for a broader class of strictly convex generators,
under conditions on the candidate quantile functions
(Theorem~\ref{thm:WC-RM-BW}). For signed weight
functions, such as those of the Gini deviation,
the formula remains optimal for these generators
whenever the candidate is well defined and a positive
binding multiplier exists
(Proposition~\ref{prop:signed-gamma}). This formula has appeared before in special cases.
For strictly concave distortion functions, $\gamma$
is non-decreasing, so that no isotonic projection is
needed, and the formula reduces to that of
\citet{pesenti2024optimal}.
\citet{tam2025bounds} extend it to two cases:
signed Choquet integrals with a non-decreasing weight
function, i.e., concave but not necessarily monotone
distortion functions, for which again no projection
is needed, and distortion risk measures with
$\phi'(\Finv)+\gamma\in L^2(0,1)$.
In both cases, $\phi'$ is assumed to map $\R$ onto
$\R$, and the result is stated conditionally:
if a worst-case quantile function exists, then it
is given by the formula.

Our results go beyond these earlier
characterizations, both in scope and in the strength of their conclusions.
First, regarding scope, for generators with growth
of order $2$ and square-integrable weight functions,
Corollary~\ref{cor:general-ac} covers
signed weight functions that need not be
non-decreasing, subject to the endpoint condition,
such as that of the best-case Gini deviation, for which
neither case of \citet{tam2025bounds} applies.
Moreover, Theorem~\ref{thm:general} covers general
distortion functions of bounded variation, including
discontinuous ones such as Value-at-Risk, for which the
isotonic projection is replaced by a least concave majorant.
It provides a dual representation of the worst-case value
and, whenever a binding multiplier exists, necessary and sufficient conditions
for attainment, which fail, for instance, for Value-at-Risk.

Second, regarding the strength of the conclusions, we establish
existence and uniqueness of the worst-case quantile function under
explicit conditions, rather than characterizing
an optimizer conditional on its existence. For absolutely
continuous distortion functions, a positive binding
multiplier yields the unique worst-case quantile function.
Moreover, Theorem~\ref{thm:WC-RM-BW} does not require
$\phi'(\R)=\R$, so that it also covers generators such as $\phi(x)=e^{x}$, provided its stated assumptions hold.
For general bounded-variation distortions, a binding
multiplier identifies the unique optimizer for the upper
semicontinuous envelope, while Theorem~\ref{thm:general}(ii)
determines whether the original supremum is attained.
For generators satisfying \eqref{eq:general-p-growth},
Proposition~\ref{prop:general-multiplier}
characterizes the existence of binding
multipliers, including the exceptional case where the
bound equals zero beyond a threshold radius.

Our second contribution concerns shape constraints. We additionally require the distribution to be unimodal with a given inflection point of its quantile function. For a strongly convex generator with a Lipschitz continuous derivative, we show that the worst case exists, is unique, and is given by a Bregman projection onto the unimodal quantile functions, with a multiplier for which the divergence constraint is binding (Theorem~\ref{thm:WC-RM-BW2-general}). In contrast with the unconstrained case, this projection is in general not explicit. For the quadratic generator, it is the $L^2$-projection onto unimodal quantile functions (Corollary~\ref{thm:WC-RM-BW2}), which is the counterpart, without moment constraints, of the corresponding result in \citet{liu2025unimodal}. The same holds for every closed convex set of quantile functions that is invariant under the addition of constants and contains the reference, such as distributions with an increasing or decreasing failure rate. For other generators, the same Lagrangian argument provides a sufficient condition (Proposition~\ref{prop:general-phi-unimodal}).

Value-at-Risk and interquantile difference provide  further applications. The one-dimensional characterization of its worst- and best-case bounds extends to continuously differentiable, strictly convex generators on $\R$ (Proposition~\ref{prop:VaR}); see also \citet{jiang2026insurance} for Value-at-Risk under bounded losses. Under unimodality, we show that Value-at-Risk bounds are limits of Range Value-at-Risk bounds, with explicit convergence rates for quadratic and strongly convex generators (Proposition~\ref{prop:VaR-unimodal} and Corollary~\ref{cor:VaR-unimodal-general}).

Our third contribution concerns worst-case distortion riskmetrics over intersections of Bregman--Wasserstein balls, allowing for different generators and reference distributions. These intersections represent models acceptable to several parties with potentially different assessments of model uncertainty. For example, an insurer and its supervisor may specify different reference models and tolerances when assessing capital requirements. We extend the isotonic solution of Theorem~\ref{thm:WC-RM-BW} to the intersection of Bregman--Wasserstein balls generated by different functions (Theorem~\ref{thm:WC-RM-BW-two-balls}). The worst-case quantile function has the same form as for a single ball, with $\phi$ replaced by a combination $\lambda_1^*\phi_1+\lambda_2^*\phi_2$ whose weights follow from the Karush-Kuhn-Tucker conditions. As different generators penalize deviations in different parts of the distribution, this allows, for instance, the body and the tail to be controlled separately. For Value-at-Risk, by contrast, the bound over the intersection of balls with a common reference distribution is simply the tightest of the bounds over the individual balls (Corollary~\ref{cor:VaR-two-balls}). Theorem~\ref{thm:WC-RM-BW-two-balls} extends to balls around different reference distributions (Proposition~\ref{prop:two-refs}); for the quadratic generator, the worst-case quantile function is then the isotonic projection of a weighted average of the reference quantile functions, shifted by a multiple of $\gamma$. The Lagrangian approach also applies to intersections of balls combined with unimodality (Proposition~\ref{prop:two-refs-unimodal}).


In a numerical study, we use the insurance portfolio of \citet{bernard2022robust} and eight candidate models with the same mean and variance but very different tails. An insurance company uses a $2$-Wasserstein ball around a model that fits the data, and its regulator uses a gamma-deviance ball around a heavy-tailed model. The intersection yields much narrower intervals for Value-at-Risk and Range Value-at-Risk in the tail than a single ball around a central model, and the distribution in the intersection with the largest Range Value-at-Risk in the tail provides a consensus model. Moreover, restricting a $2$-Wasserstein ball to unimodal distributions significantly tightens the bounds on Value-at-Risk, whereas the precise location of the inflection point hardly matters.

The paper is organized as follows. Section~\ref{sec:prelim} introduces the notation and the isotonic projection. Section~\ref{sec:WC-DRM} solves the worst-case problem over a single Bregman--Wasserstein ball, first for general distortion riskmetrics and then for absolutely continuous distortion functions and general generators, with and without shape constraints, and finally for Value-at-Risk and interquantile difference. Section~\ref{sec:two-BW-balls} treats the intersection of several Bregman--Wasserstein balls. Section~\ref{sec:examples} presents the numerical study. All proofs are relegated to an appendix. 

\section{Preliminaries}\label{sec:prelim}

Denote by $\mathcal{M}$ the set of cdfs of real-valued random variables with finite second moment. For $G\in\mathcal M$, we denote by $\Ginv(u):=\inf\{x\in\mathbb R:G(x)\geq u\}$, $0<u<1$, its left-continuous quantile function, and by $\Minv$ the set of these quantile functions. When needed, we write $\Ginv_+(u):=\inf\{x\in\mathbb R:G(x)>u\}$ for the right-continuous quantile function.

We identify functions that coincide almost everywhere. Every non-decreasing function in $L^2(0,1)$ has a left-continuous version, and this version is the quantile function of a distribution with finite second moment. Thus, $\Minv$ can be seen as the set of all non-decreasing functions in $L^2(0,1)$. This set is a closed convex cone in $L^2(0,1)$. Therefore, every $f\in L^2(0,1)$ has a unique $L^2$-projection onto $\Minv$, which we denote by $f^\uparrow$; we always take its left-continuous version.

\begin{definition}
\label{def_unimodality}
A cdf $G\in \mathcal{M}$ is unimodal if $G$ is convex-concave, i.e., there exists $x_m \in \mathbb{R}$ (called a mode) such that $G$ is convex on $(-\infty,x_m)$ and concave on $(x_m,+\infty)$.
\end{definition}

The following lemma shows how unimodality of a cdf $G$ can be expressed in terms of its quantile function $\Ginv$. Its proof can be found in \citet{bernard2020range}.

\begin{lemma}
\label{lemma-continuous-quantiles}
A cdf $G\in\mathcal M$ is unimodal if and only if $\Ginv$ is continuous on $(0,1)$ and there exists $\eta\in[0,1]$ such that $\Ginv$ is concave on $(0,\eta]$ and convex on $[\eta,1)$, with the convention that $\eta=0$ and $\eta=1$ correspond to convexity and concavity on $(0,1)$, respectively.
\end{lemma}

A quantile function $\Ginv\in\Minv$ is called unimodal if the corresponding distribution function is unimodal. For $0\leq\eta\leq1$, we denote by $\Minv^\eta$ the set of all unimodal quantile functions for which Lemma~\ref{lemma-continuous-quantiles} holds with this $\eta$.

The set $\Minv^\eta$ is convex, since monotonicity, concavity and convexity are preserved under convex combinations. It is also closed in $L^2(0,1)$. Indeed, an $L^2$-convergent sequence $(G_n^{-1})$ in $\Minv^\eta$ has a subsequence that converges almost everywhere. When $0<\eta<1$,  monotonicity of $G_n$ then implies that
\[
|G_n^{-1}(\eta)|
\leq
\frac{\lVert G_n^{-1}\rVert_2}
{\sqrt{\min(\eta,1-\eta)}},
\]
so that a further subsequence converges at $\eta$. Monotonicity, concavity on $(0,\eta]$ and convexity on $[\eta,1)$ are preserved under these limits, and together imply continuity at $\eta$. When $\eta=0$ or $\eta=1$, the limit is convex or concave on $(0,1)$, respectively, and thus also continuous on $(0,1)$. Thus the limit belongs to $\Minv^\eta$ and closedness of $\Minv^\eta$ is obtained. Consequently, every $f\in L^2(0,1)$ has a unique $L^2$-projection onto $\Minv^\eta$, which we denote by $f^{\uparrow,\eta}$.


Let $g\colon[0,1]\to\R$ be of bounded variation with $g(0)=0$; $g$ need not be monotone. The \emph{distortion riskmetric} $H_g$, evaluated at $G\in\mathcal M$, is defined as the signed Choquet integral
\begin{equation}\label{eq:signed-choquet}
H_g(G)=\int_{-\infty}^0\Big(g\big(1-G(x)\big)-g(1)\Big)\,\mathrm{d}x+\int_0^{+\infty}g\big(1-G(x)\big)\,\mathrm{d}x\,,
\end{equation}
whenever both integrals exist and their sum is not of the form $\infty-\infty$.
If $g$ is a \emph{distortion function}, i.e., $g$ is non-decreasing with $g(0)=0$ and $g(1)=1$, then $H_g$ is called a \emph{distortion risk measure}. For discontinuous $g$, the values at discontinuity points may affect $H_g$. For instance, $g(t)=\Id_{\{t>1-\alpha\}}$ yields $H_g(G)=\Ginv(\alpha)$, the Value-at-Risk at level $\alpha\in(0,1)$, whereas $g(t)=\Id_{\{t\geq1-\alpha\}}$ yields $H_g(G)=\Ginv_+(\alpha)$.

If $g$ is absolutely continuous, let $\gamma(u):=g'(1-u)$ for almost every $u\in(0,1)$; we call $\gamma$ the \emph{weight function} of $H_g$. By Fubini's theorem,
\begin{equation}
H_g(G) = \textstyle\int_0^1 \gamma(u)\Ginv(u)\,\mathrm{d}u
\label{eq: distortion Choquet integral}
\end{equation}
for every $G\in\mathcal M$ such that $\int_0^1|\gamma(u)\Ginv(u)|\,\d u<\infty$, in particular whenever $\gamma\in L^2(0,1)$.
For distortion risk measures, $\gamma$ is non-negative, and representation \eqref{eq: distortion Choquet integral} remains valid whenever the integral is well defined, possibly taking an infinite value.
In general, $\gamma$ is signed and satisfies $\int_0^1\gamma(u)\,\d u=g(1)$. An example is the Gini deviation, for which $g(t)=t(1-t)$, so that $\int_0^1\gamma(u)\,\d u=0$.

As $H_g(G)$ depends on $G$ only through $\Ginv$, we also write $H_g(\Ginv)$ instead of $H_g(G)$.
For $0<\alpha<\beta<1$ and $G\in\mathcal M$, we write $\mathrm{VaR}_\alpha(G):=\Ginv(\alpha)$ for the Value-at-Risk, $\mathrm{RVaR}_{\alpha,\beta}(G):=(\beta-\alpha)^{-1}\int_\alpha^\beta\Ginv(u)\,\d u$ for the Range Value-at-Risk and $\mathrm{ES}_\alpha(G):=(1-\alpha)^{-1}\int_\alpha^1\Ginv(u)\,\d u$ for the Expected Shortfall. The latter two are distortion risk measures with weight functions $\gamma=(\beta-\alpha)^{-1}\Id_{(\alpha,\beta)}$ and $\gamma=(1-\alpha)^{-1}\Id_{(\alpha,1)}$, respectively. Moreover, $d_W(G,H):=\lVert\Ginv-\Hinv\rVert_2$ denotes the $2$-Wasserstein distance between $G,H\in\mathcal M$.

Let $\phi:\R\to\R$ be convex and differentiable, with derivative $\phi'$. The Bregman divergence associated with $\phi$ is
\[
B_{\phi}(z_1,z_2):=\phi(z_1)-\phi(z_2)-\phi'(z_2)(z_1-z_2)\geq0,\qquad z_1,z_2\in\R,
\]
and we call $\phi$ its generator. Throughout, generators are strictly convex. For $G,F\in\mathcal M$, the Bregman--Wasserstein divergence is defined by
\[
\DB_\phi(G,F):=\inf_{\pi\in\Pi(G,F)}\int_{\R^2}B_{\phi}(z_1,z_2)\,\pi(\d z_1,\d z_2),
\]
where $\Pi(G,F)$ denotes the set of all bivariate distributions with marginal distributions $G$ and $F$. By Theorem~3.2 of \citet{pesenti2024optimal}, the comonotonic coupling is optimal, so that
\[
\DB_\phi(G,F)=\int_{0}^{1}B_{\phi}\big(\Ginv(t),\Finv(t)\big)\,\d t .
\]

To accommodate strictly positive losses, we also allow continuously differentiable, strictly convex generators defined on an open interval $J\subseteq\R$, such as $\phi(x)=-\log x$ on $J=(0,\infty)$. 
We restrict attention to reference distributions $F$ that assign probability one to $J$ and set $B_\phi(z_1,z_2):=+\infty$ for $z_1\notin J$. The comonotonic representation of $\DB_\phi$ remains valid since $B_\phi$ is non-negative and submodular on $J\times J$, as $\partial_{z_1}B_\phi(z_1,z_2)=\phi'(z_1)-\phi'(z_2)$ is non-increasing in $z_2$. 

\section{Worst-case distortion riskmetrics}\label{sec:WC-DRM}

 In Section~\ref{sec:general-g}, $H_g$ is a general distortion riskmetric as in \eqref{eq:signed-choquet}. In Sections~\ref{sec:ac} and~\ref{sec:shape}, we consider distortion risk measures and riskmetrics for which representation \eqref{eq: distortion Choquet integral} holds, and in Section~\ref{sec:VaR} we deal with Value-at-Risk. We focus on worst-case values; best-case values follow by replacing $g$ with $-g$. Let $F \in \mathcal{M}$ be some given reference cdf satisfying $H_g(\Finv) <\infty$.  \\
 
  We consider the problems \begin{equation}
 \label{opt:WC-RMtoto}
     \argmax_{\Ginv \in\Minv} \; H_g\big(\Ginv\big)
 \,,
     \quad s.t. \quad
     \DB_\phi\big(G, \, F\big) \,\le \,\ep\,,
 \end{equation}

 and 

 \begin{equation}
 \label{opt:WC-RMtoto2}
     \argmax_{\Ginv \in{\Minv^\eta}} \; H_g\big(\Ginv\big)
 \,,
     \quad s.t. \quad
     \DB_\phi\big(G, \, F\big) \,\le \,\ep\,,
 \end{equation}
in which $\ep >0$ and $0\leq \eta \leq 1$. We denote by
\begin{align*}
\mathcal M_\ep(\phi)&:=\big\{G\in\mathcal M:\DB_\phi(G,F)\leq\ep\big\},\\
\mathcal M_\ep^\eta(\phi)&:=\big\{G\in\mathcal M_\ep(\phi):\Ginv\in\Minv^\eta\big\}
\end{align*}
the Bregman-Wasserstein ball of radius $\ep$ around $F$ and its subset of unimodal distributions with inflection point $\eta$. Problems \eqref{opt:WC-RMtoto} and \eqref{opt:WC-RMtoto2} thus maximize $H_g$ over $\mathcal M_\ep(\phi)$ and $\mathcal M_\ep^\eta(\phi)$, respectively. \\

This setting extends the analysis of \citet{bernard2022robust}, who obtain sharp bounds on distortion risk measures when the ambiguity set is a ball defined through the $2$-Wasserstein distance. Their framework corresponds to the special case $\phi(x)=x^2$, for which the Bregman-Wasserstein divergence reduces to the squared $2$-Wasserstein distance. For a general strictly convex $\phi$, the Bregman-Wasserstein divergence is asymmetric and penalizes deviations in the tail differently from deviations in the body of the distribution. \\


\subsection{General distortion riskmetrics}\label{sec:general-g}

We conduct the single-ball analysis for general distortion riskmetrics
as defined in \eqref{eq:signed-choquet} and for quantile functions in $L^p(0,1)$,
$1<p<\infty$. 

Fix $1<p<\infty$ and write $p'=p/(p-1)$. In this subsection, $\mathcal M_p$ denotes the probability distributions on $\R$ with finite $p$-th moment, and $\Minv_p$ their non-decreasing, left-continuous quantile functions on $(0,1)$. Assume that $\phi\in C^1(\R)$ is strictly convex and, for constants $a_\phi,A_\phi>0$ and $b_\phi\geq0$, satisfies
\begin{equation}\label{eq:general-p-growth}
a_\phi|x|^p-b_\phi\leq\phi(x)\leq A_\phi(1+|x|^p),\qquad x\in\R.
\end{equation}
Let $F\in\mathcal M_p$; in this subsection, this replaces the assumption $F\in\mathcal M$ made at the beginning of Section~\ref{sec:WC-DRM}, and $H_g(\Finv)$ is finite by Lemma~\ref{lem:C-basic}(a). Assume that $g:[0,1]\to\R$ has bounded variation, $g(0)=0$, and, for some $K_g>0$, $t_0\in(0,1/2]$ and $\nu>1/p$,
\begin{equation}\label{eq:endpoint}
|g(t)|+|g(1-t)-g(1)|\leq K_g t^\nu,\qquad 0<t\leq t_0.
\end{equation}
Define, locally to this subsection,
\begin{equation}\label{opt:general-Lp}
\mathcal M_{\ep,p}(\phi):=\{G\in\mathcal M_p:\DB_\phi(G,F)\leq\ep\},
\qquad V_g(\ep):=\sup_{G\in\mathcal M_{\ep,p}(\phi)}H_g(G).
\end{equation}
Thus, when $p=2$, \eqref{opt:general-Lp} gives the supremum associated with \eqref{opt:WC-RMtoto}. For $G,H\in\mathcal M_p$, we define
\[
\DB_\phi(G,H):=\int_0^1 B_\phi\big(\Ginv(u),\Hinv(u)\big)\,\d u,
\]
which is finite by Lemma~\ref{lem:C-growth}(d). For $G,H\in\mathcal M$, it coincides with the Bregman-Wasserstein divergence of Section~\ref{sec:prelim}, and, as $B_\phi$ is submodular and of at most $p$th-order growth, the comonotonic coupling remains optimal on $\mathcal M_p$, so that it is the Bregman-Wasserstein divergence there as well. We write $H_g(\Ginv)$ and $H_g(G)$ interchangeably, as in the preceding sections. In particular, $\mathcal M_2=\mathcal M$, $\Minv_2=\Minv$, and $\mathcal M_{\ep,2}(\phi)=\mathcal M_\ep(\phi)$.

Condition \eqref{eq:general-p-growth} allows $\phi(x)=|x|^p/p$ for every $p>1$, including generators whose curvature vanishes or is unbounded. More generally, it allows $\phi(x)=|x|^p/p+h(x)$ for any convex $C^1$ function $h$ of at most $p$th-order growth. For $p=2$, the strictly convex function $\phi(x)=x^2-\log(1+x^2)$ is another example: $\phi''(0)=0$, so uniform strong convexity fails. Condition \eqref{eq:endpoint} holds with $\nu=1$ for distortions that are Lipschitz near the endpoints, including Value-at-Risk, Range Value-at-Risk and the Gini deviation. The present theorem still requires a generator defined on all of $\R$ with two-sided $p$-growth; it does not cover the gamma-deviance or exponential generators.

For clarity, these $L^p$ assumptions and the notation $\mathcal M_p,\Minv_p,\mathcal M_{\ep,p}(\phi)$ are local to this subsection. Section~\ref{sec:shape} uses the stronger condition
\begin{equation}\label{eq:strong-smooth}
c(x-y)^2\leq\big(\phi'(x)-\phi'(y)\big)(x-y)\leq L(x-y)^2,
\qquad x,y\in\R,\quad 0<c\leq L<\infty,
\end{equation}
i.e., $\phi$ is $c$-strongly convex and $\phi'$ is $L$-Lipschitz continuous, which implies \eqref{eq:general-p-growth} with $p=2$ (Lemma~\ref{lem:C-strong-smooth}). The proofs of the results of this subsection are given in Appendix~\ref{app:general}.

Set
\[
\hat g(t):=\max\{g(t-),g(t),g(t+)\},\quad 0<t<1,
\qquad \hat g(0):=0,\quad\hat g(1):=g(1),
\]
and $\mathcal E_g:=\{t\in(0,1):\hat g(t)>g(t)\}$. The envelope $\hat g$ has bounded variation and satisfies \eqref{eq:endpoint}, and $\mathcal E_g$ is at most countable (Lemma~\ref{lem:C-basic}(e)). For $G\in\mathcal M_p$, put
\[
\Delta_u(G):=\Ginv_+(u)-\Ginv(u),\qquad
\mathcal J_g(G):=\sum_{t\in\mathcal E_g}\big(\hat g(t)-g(t)\big)\Delta_{1-t}(G).
\]
These quantities satisfy $H_{\hat g}(G)-H_g(G)=\mathcal J_g(G)<\infty$ for every $G\in\mathcal M_p$ (Lemma~\ref{lem:C-gap}).

For $\lambda>0$, define
\[
T_F(t):=\int_{1-t}^1\phi'(\Finv(v))\,\d v,\qquad
 g_\lambda(t):=g(t)+\lambda T_F(t),\qquad 0\leq t\leq1,
\]
and let $\overline g_\lambda$ be the least concave majorant of $g_\lambda$. Then set
\begin{equation}\label{eq:general-candidate}
\bar k_\lambda(u):=\frac{1}{\lambda}\partial_+\overline g_\lambda(1-u),
\qquad \Ginv_\lambda(u):=(\phi')^{-1}(\bar k_\lambda(u)),\quad 0<u<1,
\end{equation}
where $\partial_+$ is the right derivative, and
\begin{equation}\label{eq:general-W}
W(\lambda):=\lambda\int_0^1\Big\{\phi^*(\bar k_\lambda(u))-\phi^*(\phi'(\Finv(u)))\Big\}\,\d u,
\end{equation}
where $\phi^*(y):=\sup_{x\in\R}\{xy-\phi(x)\}$ is the convex conjugate.

\begin{theorem}[General distortion riskmetrics on a single ball]\label{thm:general} Suppose \eqref{eq:general-p-growth}--\eqref{eq:endpoint} hold  and  $\ep>0$. Then the following statements hold.
\begin{enumerate}[label=(\roman*),leftmargin=*]
\item The worst-case value is
\begin{equation}\label{eq:general-value}
V_g(\ep)=V_{\hat{g}}(\ep)
=\inf_{\lambda>0}\{\lambda\ep+W(\lambda)\}.
\end{equation}
\item If $\DB_\phi(G_{\lambda^*},F)=\ep$ for some $\lambda^*>0$, then $\Ginv_{\lambda^*}$ is the unique maximizer for $V_{\hat{g}}(\epsilon)$ and
\[
V_{\hat{g}}(\ep)=H_{\hat{g}}(G_{\lambda^*})=\lambda^*\ep+W(\lambda^*).
\]
 Such a multiplier exists for every $\ep>0$ if $g(1)\neq0$ or $g(t)>0$ for some $t\in(0,1)$.  
 Moreover, $V_g(\ep)$ is  attained if and only if
\begin{equation}\label{eq:general-attain}
\Delta_{1-t}(G_{\lambda^*})=0\qquad\text{for every }t\in\mathcal E_g,
\end{equation}
in which case $G_{\lambda^*}$ is its unique maximizer.
\end{enumerate}
\end{theorem}

Theorem \ref{thm:general} shows that replacing $g$ by its upper
semicontinuous envelope leaves the worst-case value over the
Bregman--Wasserstein ball unchanged, and provides a dual representation
of this value.
If $g$ is upper semicontinuous and a positive binding multiplier exists,
\eqref{eq:general-candidate} gives the unique worst-case quantile function.
Such a multiplier exists for every $\varepsilon>0$ if $g(1)\neq0$
or $g(t)>0$ for some $t\in(0,1)$.
Without a positive binding multiplier, the maximizer need not be unique,
even when $g$ is upper semicontinuous.
For general $g$, attainment is characterized by Theorem
\ref{thm:general}(ii) whenever a positive binding multiplier exists.
Further details on the existence of binding multipliers are given below.

\begin{proposition}[Multipliers]\label{prop:general-multiplier}
Under the assumptions of Theorem~\ref{thm:general}, 
the function $d(\lambda):=\DB_\phi(G_\lambda,F)$ is continuous and non-increasing, tends to $0$ as $\lambda\to\infty$, and is strictly decreasing if $g(1)\neq0$. If $g(1)\neq0$ or $g(t)>0$ for some $t\in (0,1)$, then $d(\lambda)\to\infty$ as $\lambda\downarrow0$. Otherwise $g(1)=0$, $g\leq0$, and
\[
\lim_{\lambda\downarrow0}d(\lambda)=\bar\ep
:=\min\{\DB_\phi(G,F):\Ginv\in\Minv_p,\ H_{\hat g}(\Ginv)=0\}<\infty,
\]
where the minimum is attained. In this latter case a positive multiplier solving $d(\lambda^*)=\ep$ exists for $0<\ep<\bar\ep$, and $V_g(\ep)=0$ for $\ep\geq\bar\ep$ with $\ep>0$. A positive multiplier may also exist at $\ep=\bar\ep$. Whenever $\DB_\phi(G_{\lambda^*},F)=\ep$  has more than one solution, all solutions give the same $\Ginv_{\lambda^*}$; if $g(1)\neq0$, the multiplier is unique.
\end{proposition}

Proposition \ref{prop:general-multiplier} characterizes the existence
of a positive multiplier $\lambda^*$ satisfying
$\mathcal{B}_{\phi}(G_{\lambda^*},F)=\varepsilon$.
If $g(1)\neq 0$ or $g(t)>0$ for some $t\in(0,1)$, such a multiplier
exists for every $\varepsilon>0$.
If $g(1)=0$ and $g\leq 0$, it exists for
$0<\varepsilon<\bar{\varepsilon}$, may or may not exist at
$\varepsilon=\bar{\varepsilon}$, and does not exist for
$\varepsilon>\bar{\varepsilon}$.
In the latter setting, $V_g(\varepsilon)=0$ whenever
$\varepsilon\geq\bar{\varepsilon}$ and $\varepsilon>0$.
If $g$ is upper semicontinuous, this value is attained,
although the maximizer need not be unique.
Without upper semicontinuity, attainment is not guaranteed.
\begin{proposition}[Stability]\label{prop:general-stability}
Under the assumptions of Theorem~\ref{thm:general}, suppose $d(\lambda^*)=\ep$ with $\lambda^*>0$. Let $\{G_n\}_{n\geq 1}$ be a sequence in $\mathcal M_{\ep,p}(\phi)$ such that $H_g(G_n)\to V_g(\ep)$ as $n\to\infty$.
Then $\Ginv_n\to \Ginv_{\lambda^*}$ in $L^p(0,1)$, $\DB_\phi(G_n,F)\to\ep$ and $\mathcal J_g(G_n)\to0$, even when the original supremum is not attained. 
\end{proposition}

Proposition \ref{prop:general-stability} shows that, if $d(\lambda^*)=\ep$ for some $\lambda^*>0$, then every feasible
sequence $\{G_n\}_{n\geq 1}$ satisfying $H_g(G_n)\to V_g(\ep)$ also satisfies
$\Ginv_n\to\Ginv_{\lambda^*}$ in $L^p(0,1)$.
However, convergence in $L^r(0,1)$ for $r>p$ cannot be expected in general.
For instance, let $p=2$, $\phi(x)=x^2/2$, $F=\delta_0$, $g(t)=\Id_{\{t>1/2\}}$ and $\ep=1/4$.
Then $\Ginv_{\lambda^*}=\Id_{(1/2,1)}$ and $V_g(\ep)=1$. Fix $r>2$ and $\beta\in(1/r,1/2)$, and let
$\delta_n,\rho_n\downarrow0$. For all sufficiently large $n$, define
$
\Ginv_n:=b_n\Id_{(1/2-\rho_n,1)}+\delta_n^{-\beta}\Id_{(1-\delta_n,1)},
$
where $b_n>0$ is chosen so that
$
\left(\frac12+\rho_n\right)b_n^2+2b_n\delta_n^{1-\beta}+\delta_n^{1-2\beta}=\frac12.$
Such a choice exists for all sufficiently large $n$ and satisfies $b_n\to1$.
Consequently, $\lVert\Ginv_n\rVert_2^2=1/2$, so that $G_n$ is feasible, and $H_g(G_n)=b_n\to1=V_g(\ep)$.
Moreover, $\Ginv_n\to\Ginv_{\lambda^*}$ in $L^2(0,1)$. In contrast, for all sufficiently large $n$,
$
\lVert\Ginv_n-\Ginv_{\lambda^*}\rVert_r^r\geq \bigl(\delta_n^{-\beta}-1\bigr)^r\delta_n\longrightarrow\infty.
$

The proofs of Theorem~\ref{thm:general} and Propositions~\ref{prop:general-multiplier} and~\ref{prop:general-stability} are given in Appendix~\ref{app:general}. We first solve the problem for the upper semicontinuous envelope $\hat g$ (Proposition~\ref{prop:C-usc}). For each $\lambda>0$, the penalized problem is solved through the least concave majorant of $\hat g+\lambda T_F$. Its right derivative belongs to $L^{p'}(0,1)$, as established by a convex-order comparison with an explicit concave barrier (Lemmas~\ref{lem:C-majorant} and~\ref{lem:C-penalized}). Moreover, the two-sided $p$-growth condition \eqref{eq:general-p-growth} ensures that divergence balls are bounded in  $L^p(0,1)$ and that  quantile functions converge to a fixed reference quantile in  $L^p(0,1)$ whenever their divergence from that reference tends to zero (Lemma~\ref{lem:C-growth}). As $H_g\leq H_{\hat g}$, this gives an upper bound for $V_g(\ep)$. For the lower bound, a maximizer for $H_{\hat g}$  is first moved slightly towards $\Finv$, which makes its divergence strictly less than $\varepsilon$, and its jumps at finitely many levels $1-t$ with $t\in \mathcal E_g$ are then replaced by short flat pieces at the level at which $g$ approaches $\hat{g}$ (Lemmas~\ref{lem:C-patch} and~\ref{lem:C-recovery}).

\begin{remark}[Value-at-Risk]\label{rem:VaR-envelope}
Let $\alpha\in(0,1)$ and $g(t)=\Id_{\{t>1-\alpha\}}$. Then $H_g(G)=\Ginv(\alpha)$, $\hat g(t)=\Id_{\{t\geq1-\alpha\}}$, $H_{\hat g}(G)=\Ginv_+(\alpha)$ and $\mathcal E_g=\{1-\alpha\}$. Note that $g(1)>0$. Hence, Theorem~\ref{thm:general} implies the existence of $\lambda^*>0$ satisfying $d(\lambda^*)=\epsilon$ for any $\epsilon>0$.  The unique envelope maximizer is the quantile function $\Ginv^*$ of Proposition~\ref{prop:VaR}(i) in Section~\ref{sec:VaR}. As $\Ginv^*(\alpha)=\Finv(\alpha)<\Ginv^*_+(\alpha)$, it jumps at $\alpha$, so that, by Theorem~\ref{thm:general}(ii), the supremum of Value-at-Risk over $\mathcal M_{\ep,p}(\phi)$ is not attained. Nevertheless, every maximizing sequence converges to $\Ginv^*$ in $L^p(0,1)$ by Proposition~\ref{prop:general-stability}.
\end{remark}

\begin{example}[A zero supremum that is not attained]\label{ex:general-zero}
Take $p=2$, $\phi(x)=x^2/2$, $\Finv=\Id_{(1/2,1)}$, and
\[
g(t)=\begin{cases}-t(1-t)|t-1/2|,&t\neq1/2,\\-1,&t=1/2.\end{cases}
\]
Then $g$ satisfies \eqref{eq:endpoint} with $\nu=1$, $g(1)=0$, and $g<0$ on $(0,1)$. Its envelope is $\hat g(t)=-t(1-t)|t-1/2|$, and $H_{\hat g}(\Finv)=\hat g(1/2)=0$. Thus $\bar\ep=0$ and $V_g(\ep)=0$ for every $\ep>0$. The increment representation $H_g(\Ginv)=\int_{(0,1)}g(1-u)\,\d \Ginv(u)$ (Lemma~\ref{lem:C-identity}(b)) shows that $H_g(\Ginv)=0$ only for constant quantiles. Their smallest divergence is
\[
\min_{x\in\R}\frac12\int_0^1(x-\Finv(u))^2\,\d u
=\min_{x\in\R}\frac{x^2+(x-1)^2}{4}=\frac18.
\]
Hence the supremum is not attained for $0<\ep<1/8$, and is attained, e.g., by $\delta_{1/2}$ for $\ep\geq1/8$. Explicit recovery is given by $\Ginv_\varrho=\Id_{(1/2-\varrho,1)}$, $0<\varrho<1/2$, for which $\DB_\phi(G_\varrho,F)=\varrho/2$ and $H_g(\Ginv_\varrho)=-(1/4-\varrho^2)\varrho\to0$.
\end{example}

\begin{remark}\label{rem:general}
\begin{enumerate}[label=(\roman*),leftmargin=*]
\item The endpoint condition for distortion function $g$ is sufficient rather than necessary. It permits countably many discontinuities and singular continuous components. Its exponent is tied to the moment space: the required exponent $\nu>1/p$ depends on $p$.
\item Lower bounds follow by replacing $g$ with $-g$: $\inf_{G\in\mathcal M_{\ep,p}(\phi)}H_g(G)=-V_{-g}(\ep)$. The envelope of $-g$ is the negative of the lower semicontinuous envelope of $g$, with the same endpoint values.
\item The concave-majorant construction applies in $L^p$ without replacing it by a metric projection in $L^p$. The right-derivative of the least concave majorant of $\hat g+\lambda T_F$ belong to the dual space $L^{p'}$, and the inverse derivative of $\phi$ returns a quantile in $L^p$. For $\phi(x)=|x|^p/p$, this inverse is $y\mapsto\operatorname{sgn}(y)|y|^{1/(p-1)}$.
\end{enumerate}
\end{remark}

\subsection{Absolutely continuous distortion functions}\label{sec:ac}

If $g$ is absolutely continuous, then $\hat g=g$, so that the supremum in Theorem~\ref{thm:general} is attained, and the least concave majorant reduces to an isotonic projection. Recall that $k^\uparrow$ denotes the $L^2$-projection of $k$ onto $\Minv$.

\begin{corollary}\label{cor:general-ac}
Let the assumptions of Theorem~\ref{thm:general} hold with $p=2$, let $g$ be absolutely continuous with weight function $\gamma\in L^2(0,1)$, and let $k_\lambda:=\phi'(\Finv)+\gamma/\lambda$. Then $\bar k_\lambda=k_\lambda^\uparrow$ for every $\lambda>0$, so that
\[
\Ginv_\lambda=(\phi')^{-1}\big(k_\lambda^\uparrow\big),\qquad\lambda>0 .
\]
Unless $g(1)=0$ and $g\leq0$, for every $\ep>0$ there exists $\lambda^*>0$ with $\DB_\phi(G_{\lambda^*},F)=\ep$, and $\Ginv_{\lambda^*}$ is the unique solution to problem~\eqref{opt:WC-RMtoto}. If $g(1)=0$ and $g\leq0$, the same holds for $0<\ep<\bar\ep$, where $\bar\ep=\min\{\DB_\phi(G,F):\Ginv\in\Minv,\ H_g(\Ginv)=0\}$, and for $\ep\geq\bar\ep$, the maximum in problem~\eqref{opt:WC-RMtoto} equals $0$ and is attained exactly by the $G\in\mathcal M_\ep(\phi)$ with $H_g(G)=0$.
\end{corollary}

For instance, for the best-case Gini deviation, i.e., the worst case of $g(t)=-t(1-t)$, $H_g(\Ginv)=0$ only if $\Ginv$ is constant, so that $\bar\ep=\min_{x\in\R}\DB_\phi(\delta_x,F)$, and for $\ep\geq\bar\ep$, the best case equals $0$ and is attained by the degenerate distributions in the ball.

Corollary~\ref{cor:general-ac} requires \eqref{eq:general-p-growth} with $p=2$. For distortion risk measures, Theorem~\ref{thm:WC-RM-BW} shows that the isotonic formula remains valid for a larger class of generators, such as $\phi(x)=e^x$, under conditions on the candidate quantile functions. These conditions exclude the gamma deviance on a single ball (Remark~\ref{rem:gamma-single-ball}).

The proof of Theorem~\ref{thm:WC-RM-BW}, as well as those in Sections~\ref{sec:shape} and~\ref{sec:two-BW-balls}, relies on a characterization of isotonic regression under convex losses (Lemma~\ref{lem:iso}), a version of results in \citet{barlow1972}, whose proof is given in Appendix~\ref{app:general}. We write $\langle f,h\rangle:=\int_0^1 f(u)h(u)\,\d u$.

\begin{lemma}[Generalized isotonic minimization]\label{lem:iso}
Let $\psi:\R\to\R$ be continuously differentiable and strictly convex ($\psi$ may also be defined on an open interval $J\subseteq\R$, with $\Ginv$ taking values in $J$), and let $k\in L^2(0,1)$. Assume that $k^\uparrow$ takes values in {$\psi'(J)$, with $J=\R$ if $\psi$ is defined on $\R$,} and that $\Ginv_*:=(\psi')^{-1}(k^\uparrow)\in L^2(0,1)$. Then $\Ginv_*\in\Minv$, and for every $\Ginv\in\Minv$ the function
\[
\ell_{\Ginv}(u):=\psi\big(\Ginv(u)\big)-\psi\big(\Ginv_*(u)\big)-k(u)\big(\Ginv(u)-\Ginv_*(u)\big),\qquad 0<u<1,
\]
has an integrable negative part and satisfies $\int_0^1\ell_{\Ginv}(u)\,\d u\geq0$, with equality if and only if $\Ginv=\Ginv_*$. In particular, $\Ginv_*$ is the unique solution to
\[
\argmin_{\Ginv\in\Minv}\int_0^1\Big(\psi\big(\Ginv(u)\big)-k(u)\Ginv(u)\Big)\,\d u,
\]
where the objective is replaced by $\int_0^1\ell_{\Ginv}(u)\,\d u$ if the integral is not finite at $\Ginv_*$.
\end{lemma}

\begin{theorem}[Solution to problem \eqref{opt:WC-RMtoto} for distortion risk measures and general generators]\label{thm:WC-RM-BW}
Let $g$ be an absolutely continuous distortion function, so that $\gamma\geq0$ and $\int_0^1\gamma(u)\,\d u=1$. Assume that $\phi$ is a continuously differentiable and strictly convex function and that $\gamma,\,\phi'(\Finv) \in L^2(0,1)$. For $\lambda>0$, let
\[
k_\lambda(u):= \phi'\big(\Finv(u)\big)+ \frac{1}{\lambda} \gamma(u)
\qquad\text{and}\qquad
\Ginv_{\lambda}(u):=\left(\phi'\right)^{-1}\big(k^\uparrow_{\lambda}(u)\big),\qquad 0<u<1.
\]
Assume that for all $\lambda>0$, $\Ginv_{\lambda}$ is well-defined and square integrable and that ${\DB_\phi( G_{\lambda},F)} < \infty$. Assume that $\lim_{\lambda \to 0} {\DB_\phi( G_{\lambda},F)}=\infty$. Then the unique solution to optimization problem \eqref{opt:WC-RMtoto} is given as $\Ginv_{\lambda^*}$, where $\lambda^* > 0$ solves $\DB_\phi\big(G_{\lambda^*}, \, F\big) = \ep$.
\end{theorem}

\begin{remark}[Relation to \citet{pesenti2024optimal} and
\citet{tam2025bounds}]\label{rem:literature}
For strictly concave distortion functions, $\gamma$ is strictly
increasing, so that $k_\lambda$ coincides with its isotonic
projection, and Theorem~\ref{thm:WC-RM-BW} recovers the
characterization of \citet{pesenti2024optimal}.
\citet[Theorem~6]{tam2025bounds} extend this result, for
generators with $\phi'(\R)=\R$, to non-decreasing weight
functions and to non-negative weight functions with
$\phi'(\Finv)+\gamma\in L^2(0,1)$.
Their result is stated conditionally: if a worst-case quantile
function exists, then it is unique and given by
$\Ginv_{\lambda^*}$.

Our results strengthen and extend these earlier
characterizations in several respects.
First, Theorem~\ref{thm:WC-RM-BW} and
Corollary~\ref{cor:general-ac} establish existence
and uniqueness under explicit assumptions,
with uniqueness in the latter result when a positive
binding multiplier exists.
The key argument is that $\Ginv_{\lambda^*}$ uniquely
maximizes the Lagrangian and satisfies the constraint
with equality, so that, by weak duality, it is the unique
solution of problem~\eqref{opt:WC-RMtoto}.
Moreover, every $\lambda^*>0$ satisfying
$\DB_\phi(G_{\lambda^*},F)=\ep$ yields the same solution,
and Proposition~\ref{prop:general-multiplier}
establishes the existence of such a multiplier
for generators satisfying \eqref{eq:general-p-growth},
under the conditions stated therein.
In addition, Theorem~\ref{thm:WC-RM-BW}
does not require $\phi'(\R)=\R$, so that it also
covers generators such as $\phi(x)=e^x$ under
the stated conditions.

Second, Corollary~\ref{cor:general-ac} covers weight
functions such as that of the best-case Gini deviation,
$\gamma(u)=1-2u$, which is decreasing and changes sign,
so that neither case of \citet{tam2025bounds} applies.
Proposition~\ref{prop:signed-gamma} further extends
the optimality characterization to such signed weight
functions for more general generators, provided a binding
multiplier exists.

Third, our framework goes beyond the
weight-function setting of these earlier results
to distortion functions of bounded variation,
including discontinuous functions.
We obtain a dual representation of the worst-case
value and, whenever a positive binding multiplier
exists, necessary and sufficient conditions for
attainment (Theorem~\ref{thm:general}(ii)).
Attainment is not automatic: for Value-at-Risk,
the supremum over the ball is not attained
(Remark~\ref{rem:VaR-envelope}).
\end{remark}

\begin{remark}[Sufficient conditions]\label{rem:sufficient}
Assume that $\gamma,\Finv\in L^\infty(0,1)$ and that $\phi$ is continuously differentiable and strictly convex with $\phi'(\R)=\R$. Then the assumptions of Theorem~\ref{thm:WC-RM-BW} are satisfied: for every $\lambda>0$, $\Ginv_\lambda$ is well defined, square integrable and has finite divergence, and $\DB_\phi(G_\lambda,F)\to\infty$ as $\lambda\downarrow0$; see Appendix~\ref{app:general} for the verification.
\end{remark}

\begin{remark}[The gamma deviance on a single ball]
\label{rem:gamma-single-ball}
For a generator defined on an open interval $J\subseteq\R$,
the requirement that $k_\lambda^\uparrow$ take values in
$\phi'(J)$, so that $\Ginv_\lambda$ is well defined, is a
genuine restriction. For the gamma deviance, this condition
fails for sufficiently small $\lambda>0$. Moreover, the
worst-case Expected Shortfall over a single gamma-deviance
ball can be infinite.

To show this, let $\phi(x)=-\log x$ on $(0,\infty)$, the generator of the
gamma deviance, so that $\phi'(x)=-1/x$ has range
$(-\infty,0)$, and let $\Finv>0$ with
$1/\Finv\in L^2(0,1)$. For a weight function
$\gamma\in L^2(0,1)$ with
$\int_0^1\gamma(u)\,\d u=1$, the isotonic projection
preserves integrals, so
\[
\int_0^1 k_\lambda^\uparrow(u)\,\d u
=-\int_0^1\Finv(u)^{-1}\,\d u+\frac{1}{\lambda}>0
\]
for sufficiently small $\lambda>0$. For such $\lambda$,
$k_\lambda^\uparrow$ takes positive values on a set of
positive measure, and $\Ginv_\lambda$ is not well defined.

In fact, the worst case over a single gamma-deviance ball
can be infinite. Let $H_g=\mathrm{ES}_\alpha$, i.e.,
$\gamma=(1-\alpha)^{-1}\Id_{(\alpha,1)}$, and let
$F\in\mathcal M$ have a positive, unbounded quantile
function $\Finv$. For $0<\delta<1-\alpha$, let
\[
\Ginv:=\Finv+M_\delta\Id_{(1-\delta,1)},
\qquad
M_\delta:=\frac{\ep\,\Finv(1-\delta)}{\delta}.
\]
As
$B_\phi(y+M,y)=M/y-\log(1+M/y)\leq M/y$
for $y,M>0$, and $\Finv$ is non-decreasing,
\[
\DB_\phi(G,F)
\leq\int_{1-\delta}^1
\frac{M_\delta}{\Finv(u)}\,\d u
\leq\frac{M_\delta\delta}{\Finv(1-\delta)}
=\ep,
\]
whereas
\[
H_g(\Ginv)-H_g(\Finv)
=\frac{M_\delta\delta}{1-\alpha}
=\frac{\ep\,\Finv(1-\delta)}{1-\alpha}
\longrightarrow\infty
\]
as $\delta\downarrow0$. Hence
\[
\sup_{G\in\mathcal M_\ep(\phi)}
\mathrm{ES}_\alpha(G)=\infty
\]
for every $\ep>0$ and every $F\in\mathcal M$ with a
positive, unbounded quantile function, including those
considered in Section~\ref{sec:examples}.

We note that intersecting the gamma-deviance ball with a $2$-Wasserstein ball, as in Section~\ref{sec:examples}, would remove both problems: the worst-case Expected Shortfall is finite, as the intersection is bounded in $L^2(0,1)$ and $\mathrm{ES}_\alpha(G)\leq\lVert\Ginv\rVert_2/\sqrt{1-\alpha}$, and, when both multipliers are positive, $\psi_{\boldsymbol\lambda}'(x)=-\lambda_1/x+\lambda_2x$ maps $(0,\infty)$ onto $\R$, so that the candidate is well defined (Remark~\ref{rem:two-refs}(iv)).
\end{remark}

The proof of Theorem~\ref{thm:WC-RM-BW} uses that $\gamma$ is non-negative, which is the case for distortion risk measures. Distortion riskmetrics with a non-monotone distortion function, such as the Gini deviation, have a signed weight function. For these, the optimality part of Theorem~\ref{thm:WC-RM-BW} remains valid.

\begin{proposition}\label{prop:signed-gamma}
Let $\phi$ and $\Finv$ be as in Theorem~\ref{thm:WC-RM-BW}, and let $\gamma\in L^2(0,1)$ be not necessarily non-negative, with $H_g(\Ginv)=\int_0^1\gamma(u)\Ginv(u)\,\d u$. Assume that for some $\lambda^*>0$, $\Ginv_{\lambda^*}=(\phi')^{-1}(k_{\lambda^*}^\uparrow)$ is well-defined and square integrable, and that $\DB_\phi(G_{\lambda^*},F)=\ep$. Then $\Ginv_{\lambda^*}$ is the unique solution to problem~\eqref{opt:WC-RMtoto}.
\end{proposition}

The existence of $\lambda^*$, however, no longer follows from the monotonicity argument in the proof of Theorem~\ref{thm:WC-RM-BW}, which uses that $\gamma$ is non-negative. For generators satisfying \eqref{eq:general-p-growth} with $p=2$, e.g., under \eqref{eq:strong-smooth}, and distortion functions satisfying \eqref{eq:endpoint} with $\nu>1/2$, it is established in Corollary~\ref{cor:general-ac}, which also identifies the threshold radius beyond which no multiplier exists. Distortion riskmetrics under $2$-Wasserstein and moment constraints are studied in \citet{liu2025unimodal}.

\subsection{Shape constraints}\label{sec:shape}

We keep the setting of Section~\ref{sec:ac}: $g$ is absolutely continuous with weight function $\gamma\in L^2(0,1)$, so that $H_g(\Ginv)=\int_0^1\gamma(u)\Ginv(u)\,\d u$ by \eqref{eq: distortion Choquet integral}. As in Corollary~\ref{cor:general-ac} and Proposition~\ref{prop:signed-gamma}, $\gamma$ may take negative values; instead of a sign condition, Theorem~\ref{thm:WC-RM-BW2-general} and Corollary~\ref{thm:WC-RM-BW2} require $g(1)=\int_0^1\gamma(u)\,\d u\neq0$.

We next add the shape constraint of problem~\eqref{opt:WC-RMtoto2}. The isotonic projection in Theorem~\ref{thm:WC-RM-BW} is then replaced by a Bregman projection onto $\Minv^\eta$. For a strictly convex and continuously differentiable $\phi$ with $\phi'(\R)=\R$ and a function $k\in L^2(0,1)$, we call
\begin{equation}\label{eq:bregman-proj}
\Pi^\eta_\phi(k):=\argmin_{\Ginv\in\Minv^\eta}\int_0^1 B_\phi\Big(\Ginv(u),(\phi')^{-1}\big(k(u)\big)\Big)\,\d u
\end{equation}
the Bregman projection of $(\phi')^{-1}(k)$ onto $\Minv^\eta$, whenever the minimizer exists and is unique. Under \eqref{eq:strong-smooth}, the minimizer exists and is unique for every $k\in L^2(0,1)$ (Lemma~\ref{lem:bregman-proj} in Appendix~\ref{app:general}). For $\phi(x)=x^2/2$, $\Pi^\eta_\phi(k)=k^{\uparrow,\eta}$.

\begin{theorem}[Solution to problem \eqref{opt:WC-RMtoto2} for strongly convex generators]\label{thm:WC-RM-BW2-general}
Assume that $\phi$ is continuously differentiable and satisfies \eqref{eq:strong-smooth}. Assume that $\Finv\in\Minv^\eta$ and $\gamma\in L^2(0,1)$, with $m_\gamma:=\int_0^1\gamma(u)\,\d u\neq0$. Denote for all $\lambda>0$
$$k_\lambda(u):=\phi'\big(\Finv(u)\big)+\frac1\lambda\gamma(u),\qquad 0<u<1,\qquad\text{and}\qquad \Ginv_\lambda:=\Pi^\eta_\phi(k_\lambda).$$
Then $\Ginv_\lambda$ is well-defined for every $\lambda>0$, there exists a unique $\lambda^*>0$ such that $\DB_\phi(G_{\lambda^*},F)=\ep$, and the unique solution to optimization problem \eqref{opt:WC-RMtoto2} is given as $\Ginv_{\lambda^*}$. Moreover, $\Ginv_\lambda$ is the unique $\Ginv\in\Minv^\eta$ such that
\begin{equation}\label{eq:VI}
\int_0^1\Big(\phi'\big(\Ginv(u)\big)-k_\lambda(u)\Big)\big(\Hinv(u)-\Ginv(u)\big)\,\d u\geq0\qquad\text{for all }\Hinv\in\Minv^\eta .
\end{equation}
\end{theorem}

\begin{remark}\label{rem:WC-RM-BW2-general}
\begin{enumerate}[label=(\roman*),leftmargin=*]
\item For $\phi(x)=x^2/2$, \eqref{eq:strong-smooth} holds with $c=L=1$ and $\Pi^\eta_\phi(k_\lambda)=k_\lambda^{\uparrow,\eta}$; this case is stated in Corollary~\ref{thm:WC-RM-BW2} below. The proof only uses that $\Minv^\eta$ is a closed convex subset of $\Minv$ that is invariant under the addition of constants and contains $\Finv$, and therefore applies to every such set $\mathcal C$, with the Bregman projection onto $\mathcal C$ in place of $\Pi^\eta_\phi$. For $\mathcal C=\Minv$, Lemma~\ref{lem:iso} gives $\Ginv_\lambda=(\phi')^{-1}(k_\lambda^\uparrow)$, so that we recover Theorem~\ref{thm:WC-RM-BW} for generators satisfying \eqref{eq:strong-smooth}, in line with Corollary~\ref{cor:general-ac}. For $\Minv^\eta$, however, $\Pi^\eta_\phi(k_\lambda)$ is in general different from $(\phi')^{-1}(k_\lambda^{\uparrow,\eta})$, which need not even belong to $\Minv^\eta$, because concave-convexity need not be preserved by the nonlinear map $(\phi')^{-1}$.
\item The sign of $\gamma$ is not used, so that the theorem also applies to distortion riskmetrics with a signed weight function and, with $\gamma$ replaced by $-\gamma$, to the best-case problem. The condition $m_\gamma\neq0$, however, cannot be dropped. For example, minimizing the Gini deviation is equivalent to maximizing $H_g$ with $g(t)=-t(1-t)$, for which $m_\gamma=g(1)=0$. If $F=\delta_x$ is degenerate, then $\Finv\in\Minv^\eta$ for every $\eta$, and $H_g(G)\leq0$ for every $G$, with equality if and only if $G$ is degenerate. Hence every degenerate distribution $\delta_y$ with $\DB_\phi(\delta_y,\delta_x)\leq\ep$ solves problem~\eqref{opt:WC-RMtoto2}, so that the solution is not unique and not all solutions lie on the boundary of the ball. Note that in the absence of shape constraints, such riskmetrics with $m_\gamma=0$ are covered by Proposition~\ref{prop:general-multiplier} and Corollary~\ref{cor:general-ac}.

\item The assumption $\Finv\in\Minv^\eta$ can be relaxed. If $\Finv\notin\Minv^\eta$, then $\Ginv_\lambda$ still converges as $\lambda\to\infty$, namely to the Bregman projection of $\Finv$ onto $\Minv^\eta$,
\[
\Ginv_\infty:=\lim_{\lambda\to\infty}\Ginv_\lambda=\Pi^\eta_\phi\big(\phi'(\Finv)\big),
\]
which exists by Lemma~\ref{lem:bregman-proj}. Indeed, by \eqref{eq:VI} and \eqref{eq:strong-smooth}, $\Pi^\eta_\phi$ is $1/c$-Lipschitz continuous in $L^2(0,1)$, so that $\lVert\Ginv_\lambda-\Ginv_\infty\rVert\leq\lVert\gamma\rVert/(c\lambda)$, and $\DB_\phi(G_\lambda,F)\to\DB_\phi(G_\infty,F)$. As the remaining steps of the proof do not use $\Finv\in\Minv^\eta$, the theorem remains valid whenever $\ep>\DB_\phi(G_\infty,F)$. This extension is used in Section~\ref{sec:example-unimodal}.

\item The upper bound $L$ in \eqref{eq:strong-smooth}
ensures that $\DB_\phi(\cdot,F)$ is finite and
continuous on $\Minv$. This condition excludes
generators such as $\phi(x)=e^{x/\theta}$ or the
gamma deviance. For these, Proposition~
\ref{prop:general-phi-unimodal} still provides
the solution whenever a multiplier with a binding
constraint exists and the Lagrangian has a unique
maximizer, which need not be a Bregman projection
as defined in \eqref{eq:bregman-proj}.

\end{enumerate}
\end{remark}
For the quadratic generator, the Bregman projection is the $L^2$-projection. Let $\mathcal C\subseteq\Minv$ be closed and convex in $L^2(0,1)$ and invariant under the addition of constants. Besides $\Minv^\eta$, examples of such sets are $\mathcal C_{\mathrm{IFR}}$ and $\mathcal C_{\mathrm{DFR}}$ defined below, which contain the quantile functions of distributions with an increasing (IFR) resp. decreasing (DFR) failure rate and their shifts.

\begin{corollary}[Solution to problem \eqref{opt:WC-RMtoto2} for the quadratic generator]\label{thm:WC-RM-BW2}\label{prop:general-shape}
Let $\phi(x)=x^2/2$, $\gamma\in L^2(0,1)$ with $\int_0^1\gamma(u)\,\d u\neq0$ and $\ep>0$, and let $\mathcal C$ be as above with $\Finv\in\mathcal C$. For $\lambda>0$, let $\Ginv_\lambda$ be the $L^2$-projection of $k_\lambda:=\Finv+\gamma/\lambda$ onto $\mathcal C$. Then there exists a unique $\lambda^*>0$ such that $\DB_\phi(G_{\lambda^*},F)=\ep$, and $\Ginv_{\lambda^*}$ is the unique solution to
\[
\argmax_{\Ginv\in\mathcal C}\; H_g\big(\Ginv\big),\quad s.t.\quad \DB_\phi\big(G,\,F\big)\le\ep .
\]
In particular, if $\Finv\in\Minv^\eta$, the unique solution to problem~\eqref{opt:WC-RMtoto2} is $\Ginv_{\lambda^*}=k_{\lambda^*}^{\uparrow,\eta}$.
\end{corollary}

A distribution $G$ on $[0,\infty)$ is IFR (DFR) if and only if its cumulative hazard function $-\log(1-G)$ is convex (concave), which holds if and only if its inverse $x\mapsto\Ginv(1-e^{-x})$ is concave (convex) on $(0,\infty)$. The corresponding sets
\[
\mathcal C_{\mathrm{IFR}}:=\big\{\Ginv\in\Minv:\ x\mapsto\Ginv(1-e^{-x})\text{ is concave on }(0,\infty)\big\}
\]
and $\mathcal C_{\mathrm{DFR}}$, defined with ``convex'' instead of ``concave'', are convex and invariant under the addition of constants. They are also closed in $L^2(0,1)$, because a subsequence of an $L^2$-convergent sequence converges a.e., and a.e.\ limits of concave (convex) functions coincide a.e.\ with a concave (convex) function. For $\mathcal C=\Minv^\eta$, Corollary~\ref{thm:WC-RM-BW2} is the counterpart, without moment constraints, of the corresponding result in \citet{liu2025unimodal}.

Lower bounds on $H_g$ follow from the same results, as the infimum of $H_g$ equals minus the supremum of $\Ginv\mapsto\int_0^1(-\gamma(u))\Ginv(u)\,\d u$. The proof of Theorem~\ref{thm:WC-RM-BW2-general}, and hence Corollary~\ref{thm:WC-RM-BW2}, does not use the sign of $\gamma$; it only uses that $\int_0^1\gamma(u)\,\d u\neq0$ (Remark~\ref{rem:WC-RM-BW2-general}(ii)). The proof of Theorem~\ref{thm:WC-RM-BW} uses that $\gamma$ is non-negative, but it carries over to $-\gamma$, under its assumptions with $k_\lambda$ replaced by $\phi'(\Finv)-\gamma/\lambda$, with the pointwise monotonicities reversed: $k_\lambda^\uparrow$ is then non-decreasing in $\lambda$, $\Ginv_\lambda\leq\Finv$, and since $z\mapsto B_\phi(z,y)$ is non-increasing on $(-\infty,y]$, $\DB_\phi(G_\lambda,F)$ is continuous and strictly decreasing in $\lambda$ and tends to $0$ as $\lambda\to\infty$, while its divergence to infinity as $\lambda\downarrow0$ is assumed, as in Theorem~\ref{thm:WC-RM-BW}. Hence, the best-case quantile functions over $\mathcal M_\ep(\phi)$ and, for $\phi(x)=x^2/2$ and $\Finv\in\Minv^\eta$, over $\mathcal M_\ep^\eta(\phi)$ are
\begin{equation}
\label{eq:best-case}
\Ginv_{\lambda^*}=(\phi')^{-1}\Big(\big(\phi'(\Finv)-\tfrac{1}{\lambda^*}\gamma\big)^{\uparrow}\Big)
\qquad\text{and}\qquad
\Ginv_{\lambda^*}=\Big(\Finv-\tfrac{1}{\lambda^*}\gamma\Big)^{\uparrow,\eta},
\end{equation}
respectively, where in each case $\lambda^*>0$ is such that $\DB_\phi(G_{\lambda^*},F)=\ep$.

For generators that do not satisfy \eqref{eq:strong-smooth}, such as the gamma deviance, the Lagrangian approach still yields a sufficient condition. For $\Ginv$ with $\DB_\phi(G,F)<\infty$, define
\[
\mathcal L_\lambda(\Ginv):=H_g(\Ginv)-\lambda\,\DB_\phi(G,F),\qquad\lambda>0 .
\]

\begin{proposition}\label{prop:general-phi-unimodal}
Let $\phi$ be continuously differentiable and strictly convex, $\gamma\in L^2(0,1)$, $\eta\in[0,1]$ and $\ep>0$. Assume that there exist $\lambda^*>0$ and $\Ginv_{\lambda^*}\in\Minv^\eta$ such that $\Ginv_{\lambda^*}$ is the unique maximizer of $\mathcal L_{\lambda^*}$ over $\{\Ginv\in\Minv^\eta:\DB_\phi(G,F)<\infty\}$ and $\DB_\phi(G_{\lambda^*},F)=\ep$. Then $\Ginv_{\lambda^*}$ is the unique solution to problem~\eqref{opt:WC-RMtoto2}.
\end{proposition}

If $\phi'(\R)=\R$, the maximizer of $\mathcal L_\lambda$ is again the Bregman projection $\Pi^\eta_\phi(k_\lambda)$ whenever the latter exists{ and some $\Ginv\in\Minv^\eta$ has both $\DB_\phi(G,F)$ and $\int_0^1B_\phi\big(\Ginv(u),(\phi')^{-1}(k_\lambda(u))\big)\,\d u$ finite, which holds under \eqref{eq:strong-smooth}}; for generators on an open interval, such as the gamma deviance, $(\phi')^{-1}(k_\lambda)$ need not be defined, and the maximizer has to be computed directly. In both cases, it is in general not explicit (Remark~\ref{rem:WC-RM-BW2-general}(i)). In contrast with Theorem~\ref{thm:WC-RM-BW2-general}, the existence of $\lambda^*$ is assumed in Proposition~\ref{prop:general-phi-unimodal}; in applications, $\lambda^*$ can be determined by a one-dimensional search.

\subsection{Value-at-Risk and Interquantile difference}\label{sec:VaR}
Recall that $\mathcal M_\ep(\phi)$ denotes the Bregman-Wasserstein ball of radius $\ep$ around $F$. In contrast with Section~\ref{sec:general-g}, the following result holds for every strictly convex generator defined and continuously differentiable on all of $\R$ and does not require a growth condition.

Let $x\vee y=\max(x,y)$ and $x\wedge y=\min(x,y)$.
 \begin{proposition}\label{prop:VaR} Suppose $\phi$ is continuously differentiable on $\R$ and strictly convex, $\alpha\in (0,1)$ and $\ep>0$. Then we have
 \begin{enumerate}
 \item[(i)] $x^*=\sup_{G\in \mathcal M_\ep(\phi)}\Ginv(\alpha)=\sup_{G\in \mathcal M_\ep(\phi)}\Ginv_+(\alpha)$ is the unique solution of the following equation:
 $$\int_{\alpha}^{1}B_{\phi}(x\vee \Finv(t),\Finv(t))\d t=\ep,$$
 and the worst-case distribution $G^*$ for $\sup_{G\in \mathcal M_\ep(\phi)}\Ginv_+(\alpha)$ is characterized by $\Ginv^*(t)=\Finv(t)$ for $t\leq\alpha$ and $\Ginv^*(t)=x^*\vee \Finv(t)$ for $t>\alpha$. It is the unique maximizer of $\Ginv_+(\alpha)$ over $\mathcal M_\ep(\phi)$, and $\sup_{G\in \mathcal M_\ep(\phi)}\Ginv(\alpha)$ is not attained;
\item[(ii)] $y^*=\inf_{G\in \mathcal M_\ep(\phi)}\Ginv(\alpha)=\inf_{G\in \mathcal M_\ep(\phi)}\Ginv_+(\alpha)$ is the unique  solution of the following equation:
 $$\int_{0}^{\alpha}B_{\phi}(y\wedge \Finv(t),\Finv(t))\d t=\ep,$$
 and the {best-case} distribution $G^*$, which attains $\inf_{G\in \mathcal M_\ep(\phi)}\Ginv(\alpha)$, satisfies  $\Ginv^*(t)=y^*\wedge \Finv(t)$ for $t\leq \alpha$ and $\Ginv^*(t)=\Finv(t)$ for $t>\alpha$.
 \end{enumerate}
 \end{proposition}

\begin{remark}
Proposition~\ref{prop:VaR} is related to Propositions~3.1 and~3.2 of \citet{jiang2026insurance}, who consider non-negative losses with support in $[0,M]$, $M<\infty$, so that their ambiguity set only contains distributions on $[0,M]$. Their bounds equal $\min\{x^*,M\}$ and $\max\{y^*,0\}$: as $B_\phi(x\vee\Finv(t),\Finv(t))=0$ for $t>F(x)$, their equation coincides with the one in (i), and similarly for (ii). Proposition~\ref{prop:VaR} does not require bounded support, so that it also applies to unbounded and heavy-tailed losses such as those in Section~\ref{sec:examples}. Moreover, it shows that the bounds are the same for the left and the right quantile, the worst case being attained for the latter, and its proof extends directly to several balls (Corollary~\ref{cor:VaR-two-balls}). It is complemented by results for unimodal distributions and quadratic or strongly convex generators (Proposition~\ref{prop:VaR-unimodal} and Corollary~\ref{cor:VaR-unimodal-general}).
\end{remark}

Recall that $\mathcal M_\ep^\eta(\phi)=\{G\in\mathcal M_\ep(\phi):\Ginv\in\Minv^\eta\}$. Since $\Ginv$ is continuous on $(0,1)$ for every $G\in\mathcal M_\ep^\eta(\phi)$ by Lemma~\ref{lemma-continuous-quantiles}, there is no need to distinguish between $\Ginv(\alpha)$ and $\Ginv_+(\alpha)$ in the following result.

\begin{proposition}\label{prop:VaR-unimodal}
Let $\phi(x)=x^2/2$, $\eta\in[0,1]$, $\alpha\in(0,1)$ and $\ep>0$, and assume that $\mathcal M_\ep^\eta(\phi)$ is nonempty. Then
\begin{enumerate}
\item[(i)] $\sup_{G\in\mathcal M_\ep^\eta(\phi)}\Ginv(\alpha)=\lim_{h\downarrow0}\sup_{G\in\mathcal M_\ep^\eta(\phi)}\mathrm{RVaR}_{\alpha,\alpha+h}(G)$;
\item[(ii)] $\inf_{G\in\mathcal M_\ep^\eta(\phi)}\Ginv(\alpha)=\lim_{h\downarrow0}\inf_{G\in\mathcal M_\ep^\eta(\phi)}\mathrm{RVaR}_{\alpha-h,\alpha}(G)$.
\end{enumerate}
More precisely, there exist $h_0>0$ and $\Lambda<\infty$, depending only on $\alpha$, $\eta$ and $R:=\lVert\Finv\rVert_2+\sqrt{2\ep}$, such that the differences between the left-hand sides and the corresponding suprema and infima over $\mathcal M_\ep^\eta(\phi)$ are at most $\Lambda h$ for all $0<h\leq h_0$. If, in addition, $\Finv\in\Minv^\eta$, these suprema and infima are given by Corollary~\ref{thm:WC-RM-BW2} and \eqref{eq:best-case} with $\gamma(u)=h^{-1}\Id_{(\alpha,\alpha+h]}(u)$ and $\gamma(u)=h^{-1}\Id_{(\alpha-h,\alpha]}(u)$, respectively.
\end{proposition}

\begin{remark}
Without the unimodality constraint, the uniform Lipschitz bound in the proof fails. Indeed, the worst-case quantile function in Proposition~\ref{prop:VaR} jumps at $\alpha$, and $\sup_{G\in\mathcal M_\ep(\phi)}\Ginv(\alpha)$ is not attained. Under unimodality, the quantile function is continuous, so that the jump has to be replaced by a continuous increase, which requires a larger divergence. This explains why unimodality has a larger effect on the bounds on Value-at-Risk than on the bounds on Range Value-at-Risk.
\end{remark}

Proposition~\ref{prop:VaR-unimodal} extends to strongly convex generators.

\begin{corollary}\label{cor:VaR-unimodal-general}
Let $\phi$ be continuously differentiable and such that $x\mapsto\phi(x)-cx^2/2$ is convex for some $c>0$. Let $\eta\in[0,1]$, $\alpha\in(0,1)$ and $\ep>0$, and assume that $\mathcal M_\ep^\eta(\phi)$ is nonempty. Then the statements (i) and (ii) of Proposition~\ref{prop:VaR-unimodal}, as well as the bound $\Lambda h$ on the approximation error, hold with $R:=\lVert\Finv\rVert_2+\sqrt{2\ep/c}$.
\end{corollary}

The approximating RVaR problems are then of the type considered in Proposition~\ref{prop:general-phi-unimodal}.

The interquantile difference is defined as $$
\mathrm{IQD}_{\alpha}^+(G)=\Ginv_+(1-\alpha)-\Ginv(\alpha):=H_{g_{\mathrm{IQD}^+}}(G),~\alpha\in (0,1/2],
$$
and
$$
\mathrm{IQD}_{\alpha}^-(G)=\Ginv(1-\alpha)-\Ginv_+(\alpha):=H_{g_{\mathrm{IQD}^-}}(G),~\alpha\in (0,1/2),
$$
where $g_{\mathrm{IQD}^+}(t)=\Id_{[\alpha,1-\alpha]}(t),~t\in [0,1]$ and $g_{\mathrm{IQD}^-}(t)=\Id_{(\alpha,1-\alpha)}(t),~t\in [0,1]$.
For $\alpha\in(0,1/2)$, define \[
\begin{aligned}
L(x)&:=\int_0^\alpha
B_\phi\bigl(x\wedge\breve F(t),\breve F(t)\bigr)\,\mathrm{d}t,\\
U(y):&=\int_{1-\alpha}^1
B_\phi\bigl(y\vee\breve F(t),\breve F(t)\bigr)\,\mathrm{d}t,\\
C_\alpha(x_1,x_2)
:&=
\int_\alpha^{1-\alpha}
B_\phi\bigl((\Finv(t)\vee x_1)\wedge x_2,\Finv(t)\bigr)\,\mathrm{d} t.
\end{aligned}\]
For $\ep>0$, let
\[
\mathcal E_\alpha(\ep):=\left\{(x_1,x_2)\in\R^2:\begin{array}{l}x_1\leq\Finv(\alpha),\quad x_2\geq\Finv_+(1-\alpha),\\L(x_1)+U(x_2)=\ep\end{array}\right\},
\]
and
\[
\mathcal E_\alpha^-(\ep):=\left\{(x_1,x_2)\in\R^2:\begin{array}{l}\Finv_+(\alpha)\leq x_1\leq x_2\leq\Finv(1-\alpha),\\C_\alpha(x_1,x_2)\leq\ep\end{array}\right\}.
\]
Moreover, let
\[
c_\alpha:=(\phi')^{-1}\left(\frac{1}{1-2\alpha}\int_\alpha^{1-\alpha}\phi'(\Finv(t))\,\d t\right),\qquad \ep_0:=\int_\alpha^{1-\alpha} B_\phi(c_\alpha,\Finv(t))\,\d t.
\]

\begin{proposition}\label{prop:IQD}
Suppose $\phi$ is continuously differentiable on $\R$ and strictly convex, $\alpha\in(0,1/2)$ and $\ep>0$. Then the following statements hold.
\begin{enumerate}
\item[(i)] We have
\[
\begin{aligned}
\sup_{G\in\mathcal M_\ep(\phi)}\mathrm{IQD}_\alpha^-(G)&=\sup_{G\in\mathcal M_\ep(\phi)}\mathrm{IQD}_\alpha^+(G)\\
&=\sup\bigl\{x_2-x_1:(x_1,x_2)\in\mathcal E_\alpha(\ep)\bigr\}.
\end{aligned}
\]
The maximizing pair $(x_1^*,x_2^*)$ is unique. The unique maximizer $G^*$ of $\mathrm{IQD}_\alpha^+$ over $\mathcal M_\ep(\phi)$ has quantile function
\[
\Ginv^*(t)=\begin{cases}x_1^*\wedge\Finv(t),&0<t\leq\alpha,\\\Finv(t),&\alpha<t\leq1-\alpha,\\x_2^*\vee\Finv(t),&1-\alpha<t<1.\end{cases}
\]
The supremum of $\mathrm{IQD}_\alpha^-$ over $\mathcal M_\ep(\phi)$ is not attained.

\item[(ii)] We have
\[
\begin{aligned}
\inf_{G\in\mathcal M_\ep(\phi)}\mathrm{IQD}_\alpha^-(G)&=\inf_{G\in\mathcal M_\ep(\phi)}\mathrm{IQD}_\alpha^+(G)\\
&=\inf\bigl\{x_2-x_1:(x_1,x_2)\in\mathcal E_\alpha^-(\ep)\bigr\}.
\end{aligned}
\]
For $0<\ep\leq\ep_0$, the minimizing pair $(x_1^*,x_2^*)$ is unique, and the unique minimizer $G^*$ of $\mathrm{IQD}_\alpha^-$ over $\mathcal M_\ep(\phi)$ has quantile function
\[
\Ginv^*(t)=\begin{cases}\Finv(t),&0<t\leq\alpha,\\(\Finv(t)\vee x_1^*)\wedge x_2^*,&\alpha<t\leq1-\alpha,\\\Finv(t),&1-\alpha<t<1.\end{cases}
\] The infimum of $\mathrm{IQD}_\alpha^+$ over $\mathcal M_\ep(\phi)$ is not attained.
\end{enumerate}
\end{proposition}

\begin{remark}
If $0<\ep<\ep_0$, the minimizing pair in Proposition~\ref{prop:IQD}(ii) satisfies
\[
\Finv_+(\alpha)<x_1^*<x_2^*<\Finv(1-\alpha),\qquad C_\alpha(x_1^*,x_2^*)=\ep.
\]
The common lower bound is positive, and the infimum of $\mathrm{IQD}_\alpha^+$ is not attained. If $\ep=\ep_0>0$, the minimizing pair is $(x_1^*,x_2^*)=(c_\alpha,c_\alpha)$. The common lower bound is then zero, and the infimum of $\mathrm{IQD}_\alpha^+$ is still not attained. If $\ep>\ep_0$, both lower bounds equal zero and are attained. The minimum of $\mathrm{IQD}_\alpha^-$ is attained in all cases.
\end{remark}

\section{Intersections of Bregman--Wasserstein balls} \label{sec:two-BW-balls}

\subsection{A common reference distribution}\label{sec:common-ref}

We next extend Theorem~\ref{thm:WC-RM-BW} to the case in which the ambiguity set is the intersection of two Bregman-Wasserstein balls, generated by two possibly different convex functions. Let $\phi_1,\phi_2:\R\to\R$ be continuously differentiable and strictly convex functions, and let $\ep_1,\ep_2>0$. Consider
\begin{equation} \label{opt:WC-RM-two-balls} \argmax_{\Ginv \in\Minv} \; H_g\big(\Ginv\big) \,, \quad s.t. \quad \DB_{\phi_i}\big(G, \, F\big) \,\le \,\ep_i,\quad i=1,2. \end{equation}
For $\boldsymbol{\lambda}:=(\lambda_1,\lambda_2)\in[0,\infty)^2\setminus\{\boldsymbol{0}\}$, define
\begin{equation} \label{eq:psi-lambda-two-balls} \psi_{\boldsymbol{\lambda}}(x) := \lambda_1\phi_1(x)+\lambda_2\phi_2(x), \qquad x\in\R, \end{equation}
and
\begin{equation} \label{eq:k-lambda-two-balls} k_{\boldsymbol{\lambda}}(u) := \gamma(u) +\lambda_1\phi_1'\big(\Finv(u)\big) +\lambda_2\phi_2'\big(\Finv(u)\big), \qquad 0<u<1. \end{equation}
Since $\psi_{\boldsymbol{\lambda}}$ is continuously differentiable and strictly convex, $\psi_{\boldsymbol{\lambda}}'$ is strictly increasing. Whenever the expression is well-defined, set
\begin{equation} \label{eq:G-lambda-two-balls} \Ginv_{\boldsymbol{\lambda}}(u) := \left(\psi_{\boldsymbol{\lambda}}'\right)^{-1} \left(k_{\boldsymbol{\lambda}}^{\uparrow}(u)\right), \qquad 0<u<1. \end{equation}
\begin{theorem}[Solution to problem \eqref{opt:WC-RM-two-balls}] \label{thm:WC-RM-BW-two-balls} Let $H_g$ be a distortion risk measure with an absolutely continuous distortion function $g$, so that $\gamma\geq0$ and $\int_0^1\gamma(u)\,\d u=1$. Assume that $\phi_1$ and $\phi_2$ are continuously differentiable and strictly convex functions. Assume that $\gamma\in L^2(0,1)$ and that, for every $\boldsymbol{\lambda}\in[0,\infty)^2\setminus\{\boldsymbol{0}\}$, the function $k_{\boldsymbol{\lambda}}$ in \eqref{eq:k-lambda-two-balls} is square integrable and
$\Ginv_{\boldsymbol{\lambda}}$ in \eqref{eq:G-lambda-two-balls} is well-defined, square integrable and satisfies
\[ \DB_{\phi_i}\big(G_{\boldsymbol{\lambda}},F\big)<\infty, \qquad i=1,2. \]
Then $\Ginv^*\in\Minv$ solves problem \eqref{opt:WC-RM-two-balls} if and only if $\Ginv^*=\Ginv_{\boldsymbol{\lambda}^*}$ for some $\boldsymbol{\lambda}^*:=(\lambda_1^*,\lambda_2^*)\in[0,\infty)^2\setminus\{\boldsymbol{0}\}$ satisfying
\begin{equation} \label{eq:KKT-two-balls} \DB_{\phi_i}\big(G_{\boldsymbol{\lambda}^*},F\big)\leq \ep_i, \qquad \lambda_i^* \left( \DB_{\phi_i}\big(G_{\boldsymbol{\lambda}^*},F\big)-\ep_i \right)=0, \qquad i=1,2. \end{equation}
In that case, the solution is unique. In particular, if $\lambda_1^*>0$ and $\lambda_2^*>0$, then both constraints are binding and $\boldsymbol{\lambda}^*$ solves the system
\[ \DB_{\phi_1}\big(G_{\boldsymbol{\lambda}},F\big)=\ep_1, \qquad \DB_{\phi_2}\big(G_{\boldsymbol{\lambda}},F\big)=\ep_2. \]
If only the $i$th constraint is binding, then $\lambda_i^*>0$, $\lambda_j^*=0$ for $j\neq i$, and the solution reduces to the one-ball formula of Theorem~\ref{thm:WC-RM-BW} with $\phi=\phi_i$ and $\ep=\ep_i$, which, by Proposition~\ref{prop:signed-gamma}, is the unique solution of the one-ball problem with multiplier $\lambda_i^*$. \end{theorem}

In practice, if problem~\eqref{opt:WC-RM-two-balls} has a solution, $\boldsymbol{\lambda}^*$ can be found as follows. Let $\Ginv_{(i)}$ denote the solution of the one-ball problem with $\phi=\phi_i$ and $\ep=\ep_i$, $i=1,2$, whenever it exists, e.g., by Theorem~\ref{thm:WC-RM-BW} or Proposition~\ref{prop:signed-gamma}. If $\DB_{\phi_2}(G_{(1)},F)\leq\ep_2$, then $\Ginv^*=\Ginv_{(1)}$, and one can take $\lambda_2^*=0$; if $\DB_{\phi_1}(G_{(2)},F)\leq\ep_1$, then $\Ginv^*=\Ginv_{(2)}$, and one can take $\lambda_1^*=0$. Otherwise, both multipliers are strictly positive and $\boldsymbol{\lambda}^*$ solves the system of two equations in Theorem~\ref{thm:WC-RM-BW-two-balls}.

 \begin{remark} The theorem can be extended verbatim to $m\geq2$ BW balls. If the ambiguity set is given by \[ \DB_{\phi_i}(G,F)\leq\ep_i, \qquad i=1,\ldots,m, \] then the candidate solution is \[ \Ginv_{\boldsymbol{\lambda}}(u) = \left(\psi_{\boldsymbol{\lambda}}'\right)^{-1} \left(k_{\boldsymbol{\lambda}}^{\uparrow}(u)\right), \] where $\boldsymbol{\lambda}:=(\lambda_1,\ldots,\lambda_m)$, \[ \psi_{\boldsymbol{\lambda}}(x) = \sum_{i=1}^m\lambda_i\phi_i(x), \qquad k_{\boldsymbol{\lambda}}(u) = \gamma(u)+ \sum_{i=1}^m \lambda_i\phi_i'\big(\Finv(u)\big), \] and the multipliers satisfy the corresponding feasibility and complementary slackness conditions. \end{remark}

Theorem~\ref{thm:WC-RM-BW-two-balls} does not cover Value-at-Risk, whose distortion function is not absolutely continuous. However, the argument in the proof of Proposition~\ref{prop:VaR} (Appendix~\ref{app:general}) extends directly to the intersection of several Bregman-Wasserstein balls.

\begin{corollary}\label{cor:VaR-two-balls}
Let $\phi_1,\ldots,\phi_m$ be continuously differentiable and strictly convex on $\R$, $\alpha\in(0,1)$, $\ep_1,\ldots,\ep_m>0$, and $\mathcal M_{\boldsymbol\ep}:=\bigcap_{i=1}^m\mathcal M_{\ep_i}(\phi_i)$. For $i=1,\ldots,m$, let $x_i^*$ and $y_i^*$ be the unique solutions of
\[
\int_\alpha^1 B_{\phi_i}\big(x\vee\Finv(t),\Finv(t)\big)\,\d t=\ep_i
\quad\text{and}\quad
\int_0^\alpha B_{\phi_i}\big(y\wedge\Finv(t),\Finv(t)\big)\,\d t=\ep_i ,
\]
respectively. Then
\begin{gather*}
\sup_{G\in\mathcal M_{\boldsymbol\ep}}\Ginv(\alpha)=\sup_{G\in\mathcal M_{\boldsymbol\ep}}\Ginv_+(\alpha)=\min_{1\leq i\leq m}x_i^*
\quad\text{and}\\
\inf_{G\in\mathcal M_{\boldsymbol\ep}}\Ginv(\alpha)=\inf_{G\in\mathcal M_{\boldsymbol\ep}}\Ginv_+(\alpha)=\max_{1\leq i\leq m}y_i^*.
\end{gather*}
The worst-case and best-case distributions are given as in Proposition~\ref{prop:VaR} with $x^*=\min_{i}x_i^*$ and $y^*=\max_{i}y_i^*$, respectively.
\end{corollary}

In contrast with Theorem~\ref{thm:WC-RM-BW-two-balls}, the shape of the worst-case quantile function does not depend on the generators: the additional balls only restrict the level $x^*$, and generically only one of the constraints is binding.

\subsection{Different reference distributions}\label{sec:different-refs}
The two balls need not be centred at the same distribution. This is useful when, for instance, one reference model is credible for the body and another for the tail of the loss distribution, or when two models are each considered plausible and the ambiguity set should contain only distributions that are close to both. Let $F_1,F_2\in\mathcal M$ and consider
\begin{equation}
\label{opt:WC-RM-two-refs}
\argmax_{\Ginv\in\Minv}\;H_g\big(\Ginv\big),\quad s.t.\quad \DB_{\phi_i}\big(G,\,F_i\big)\le\ep_i,\quad i=1,2.
\end{equation}
For $\boldsymbol\lambda\in[0,\infty)^2\setminus\{\boldsymbol 0\}$, let $\psi_{\boldsymbol\lambda}$ be as in \eqref{eq:psi-lambda-two-balls}, and define
\begin{multline}
\label{eq:k-lambda-two-refs}
\tilde k_{\boldsymbol\lambda}(u):=\gamma(u)+\lambda_1\phi_1'\big(F_1^{-1}(u)\big)+\lambda_2\phi_2'\big(F_2^{-1}(u)\big)
\\
\text{and}\quad
\tilde G^{-1}_{\boldsymbol\lambda}(u):=\big(\psi_{\boldsymbol\lambda}'\big)^{-1}\big(\tilde k_{\boldsymbol\lambda}^{\uparrow}(u)\big),\qquad 0<u<1,
\end{multline}
whenever the latter is well-defined.

\begin{proposition}\label{prop:two-refs}
Let $H_g$ be as in Theorem~\ref{thm:WC-RM-BW-two-balls}. Assume that $\phi_1$ and $\phi_2$ are continuously differentiable and strictly convex, that $\gamma\in L^2(0,1)$ and that, for every $\boldsymbol\lambda\in[0,\infty)^2\setminus\{\boldsymbol 0\}$, the function $\tilde k_{\boldsymbol\lambda}$ is square integrable and $\tilde G^{-1}_{\boldsymbol\lambda}$ is well-defined, square integrable and satisfies $\DB_{\phi_i}(\tilde G_{\boldsymbol\lambda},F_i)<\infty$, $i=1,2$. Assume moreover that there exists $\Hinv\in\Minv$ with $\DB_{\phi_i}(H,F_i)<\ep_i$, $i=1,2$. Then $\Ginv^*\in\Minv$ solves problem \eqref{opt:WC-RM-two-refs} if and only if $\Ginv^*=\tilde G^{-1}_{\boldsymbol\lambda^*}$ for some $\boldsymbol\lambda^*\in[0,\infty)^2\setminus\{\boldsymbol 0\}$ satisfying
\[
\DB_{\phi_i}\big(\tilde G_{\boldsymbol\lambda^*},F_i\big)\leq\ep_i,
\qquad
\lambda_i^*\Big(\DB_{\phi_i}\big(\tilde G_{\boldsymbol\lambda^*},F_i\big)-\ep_i\Big)=0,
\qquad i=1,2.
\]
In that case, the solution is unique. If $\lambda_j^*=0$ for some $j$, then $\tilde G^{-1}_{\boldsymbol\lambda^*}$ is the solution of the one-ball problem with $\phi=\phi_i$, $F=F_i$ and $\ep=\ep_i$, $i\neq j$, by Proposition~\ref{prop:signed-gamma} with $\lambda^*=\lambda_i^*$.
\end{proposition}

\begin{remark}\label{rem:two-refs}
\begin{enumerate}[label=(\roman*),leftmargin=*]
\item For $\phi_1(x)=\phi_2(x)=x^2/2$, the worst-case quantile function is
\[
\tilde G^{-1}_{\boldsymbol\lambda^*}=\Big(\frac{\lambda_1^*F_1^{-1}+\lambda_2^*F_2^{-1}+\gamma}{\lambda_1^*+\lambda_2^*}\Big)^{\uparrow},
\]
i.e., the isotonic projection of a weighted average of the two reference quantile functions, shifted by a multiple of $\gamma$. As a weighted average of quantile functions is the quantile function of a Wasserstein barycenter, the worst case is obtained by shifting a barycenter of $F_1$ and $F_2$ according to $\gamma$ and projecting the result onto $\Minv$. Moreover, the $i$th ball consists of the quantile functions in $\Minv$ within $L^2$-distance $\sqrt{2\ep_i}$ of $F_i^{-1}$. As $\Minv$ is convex, the two balls intersect if and only if $\sqrt{2\ep_1}+\sqrt{2\ep_2}\geq d_W(F_1,F_2)$, and the Slater condition in Proposition~\ref{prop:two-refs} holds if and only if this inequality is strict.
\item As for Theorem~\ref{thm:WC-RM-BW-two-balls}, one first checks whether the solution of the one-ball problem around its own reference, whenever it exists, lies in the other ball, and does so for both balls. If neither does and problem~\eqref{opt:WC-RM-two-refs} has a solution, both constraints are binding, and $\boldsymbol\lambda^*$ minimizes the convex dual function $\boldsymbol\lambda\mapsto\sup_{\Ginv\in\Minv}\big\{H_g(\Ginv)-\sum_{i=1}^2\lambda_i\big(\DB_{\phi_i}(G,F_i)-\ep_i\big)\big\}$ over $[0,\infty)^2$; we use this in Section~\ref{sec:examples}. If the feasible set of problem~\eqref{opt:WC-RM-two-refs} is nonempty and one of the generators is quadratic, then a solution to the problem exists: the feasible set is then bounded in $L^2(0,1)$, and, as each divergence is convex and lower semicontinuous, also closed and convex, hence weakly compact, so that $H_g$ attains its maximum on it. 

\item Lower bounds are obtained by replacing $\gamma$ by $-\gamma$. In the argument excluding $\boldsymbol\lambda^*=\boldsymbol 0$, the quantile function $\Ginv^*$ is then lowered instead of raised, which uses $-\gamma\leq0$ and $\int_0^1\gamma(u)\,\d u=1$. For signed weight functions, the `if' part, including uniqueness, remains valid, but the `only if' part may fail, and solutions need not be unique: for the best-case Gini deviation and degenerate $F_1=F_2$, every degenerate distribution in the intersection is optimal.
\item The assumption that $\tilde G^{-1}_{\boldsymbol\lambda}$ is well defined for every $\boldsymbol\lambda$ is only needed for the `only if' part. The `if' part and the uniqueness only use the weak duality argument and Lemma~\ref{lem:iso} at the multiplier $\boldsymbol\lambda^*$. They therefore hold whenever $\tilde G^{-1}_{\boldsymbol\lambda^*}$ is well defined, square integrable and satisfies the conditions of Proposition~\ref{prop:two-refs}, also for generators defined on an open interval as in Section~\ref{sec:prelim}.
\end{enumerate}
\end{remark}

\begin{proposition}\label{prop:two-refs-unimodal}
Let $\phi_1,\phi_2$ be continuously differentiable and strictly convex, each on $\R$ or on an open interval as in Section~\ref{sec:prelim}, let $F_1,F_2\in\mathcal M$ take values in the respective domains, $\gamma\in L^2(0,1)$ and $\eta\in[0,1]$. Let
\begin{align*}
\mathcal S&:=\big\{\Ginv\in\Minv^\eta:\DB_{\phi_i}(G,F_i)<\infty,\ i=1,2\big\},\\
\mathcal L_{\boldsymbol\lambda}(\Ginv)&:=H_g(\Ginv)-\sum_{i=1}^2\lambda_i\,\DB_{\phi_i}(G,F_i),\qquad\Ginv\in\mathcal S,\ \boldsymbol\lambda\in[0,\infty)^2 .
\end{align*}
\begin{enumerate}[label=(\roman*)]
\item If, for some $\boldsymbol\lambda^*\in[0,\infty)^2$, $\mathcal L_{\boldsymbol\lambda^*}$ has a unique maximizer $G^{-1}_{\eta,\boldsymbol\lambda^*}$ over $\mathcal S$ and the Karush--Kuhn--Tucker conditions
\[
\DB_{\phi_i}\big(G_{\eta,\boldsymbol\lambda^*},F_i\big)\leq\ep_i,\qquad\lambda_i^*\Big(\DB_{\phi_i}\big(G_{\eta,\boldsymbol\lambda^*},F_i\big)-\ep_i\Big)=0,\qquad i=1,2,
\]
hold, then $G^{-1}_{\eta,\boldsymbol\lambda^*}$ is the unique solution to problem~\eqref{opt:WC-RM-two-refs} with $\Minv$ replaced by $\Minv^\eta$.
\item Assume, in addition, that $\phi_2$ is defined on $\R$ and $c$-strongly convex for some $c>0$, that $B_{\phi_1}(x,y)\to\infty$ as $x$ tends to a finite endpoint of the domain of $\phi_1$, for every $y$ in this domain, and that there exists $H_0^{-1}\in\Minv^\eta$ with $\DB_{\phi_i}(H_0,F_i)<\ep_i$, $i=1,2$. Then problem~\eqref{opt:WC-RM-two-refs} with $\Minv$ replaced by $\Minv^\eta$ has a solution $G_*^{-1}$, and there exists $\boldsymbol\lambda^*\in[0,\infty)^2$ such that $G_*^{-1}$ maximizes $\mathcal L_{\boldsymbol\lambda^*}$ over $\mathcal S$ and the Karush--Kuhn--Tucker conditions hold at $G_*^{-1}$. If $\boldsymbol\lambda^*\neq\boldsymbol0$, the maximizer of $\mathcal L_{\boldsymbol\lambda^*}$ over $\mathcal S$ is unique, so that $G_*^{-1}=G^{-1}_{\eta,\boldsymbol\lambda^*}$ is the unique solution.
\end{enumerate}
\end{proposition}

\begin{remark}\label{rem:two-refs-unimodal}
In contrast with Proposition~\ref{prop:two-refs}, the maximizer of $\mathcal L_{\boldsymbol\lambda^*}$ over $\mathcal S$ is in general not explicit and has to be computed numerically, e.g., under suitable additional smoothness conditions, by a Newton-type method whose steps involve weighted $L^2$-projections onto $\Minv^\eta$. Statement~(i) does not require strong convexity and provides a certificate of optimality for numerically computed multipliers. In (ii), the case $\boldsymbol\lambda^*=\boldsymbol0$ can be excluded if $g(1)>0$, e.g., for a distortion risk measure, and $\DB_{\phi_i}(G_*+1,F_i)<\infty$, $i=1,2$, where $G_*+1$ denotes the distribution with quantile function $G_*^{-1}+1$. Indeed, $\boldsymbol\lambda^*=\boldsymbol0$ means that $G_*^{-1}$ maximizes $H_g$ over $\mathcal S$, whereas $G_*^{-1}+1\in\mathcal S$, as $\Minv^\eta$ is invariant under the addition of constants, and $H_g(G_*^{-1}+1)=H_g(G_*^{-1})+g(1)$. The finiteness condition holds, e.g., for $\phi_1(x)=-\log x$ and $\phi_2(x)=x^2/2$ if $\int_0^1F_1^{-1}(u)^{-1}\,\d u<\infty$, as $B_{\phi_1}(x+1,y)\leq B_{\phi_1}(x,y)+1/y$. For lower bounds, i.e., with $g$ replaced by $-g$, the same argument uses $G_*^{-1}-1$ instead, which may take non-positive values, so that its gamma-deviance divergence may be infinite; in that case, $\boldsymbol\lambda^*\neq\boldsymbol0$ has to be verified separately, e.g., through the certificate in (i).
\end{remark}

\section{Applications}\label{sec:examples}

In this section, we illustrate our results with the insurance portfolio of \citet{bernard2022robust}. We consider the Pareto--Clayton model of Appendix~\ref{app:A1} with $a=10$, $b=1$ and $d=100$, together with the seven alternative models of Appendix~\ref{app:A2}. In contrast with Section~5.2 of \citet{bernard2022robust}, the parameters of the alternative models are chosen such that all eight distributions have exactly the same mean $100/9\approx11.11$ and the same standard deviation $4.10$; see Table~\ref{tab:ex1-models}. Although the eight distributions are almost indistinguishable in the body (their $95\%$ quantiles lie between $18.1$ and $18.8$), their tails are very different. The Inverse Weibull, Log-logistic and Inverse Gamma distributions, as well as the Pareto--Clayton model, have a power tail. The Lognormal distribution has a tail that is heavier than exponential but lighter than any power tail, while the Gamma, Inverse Gaussian and Weibull distributions have an exponential or lighter tail. All eight models are unimodal, with inflection points between $0.29$ and $0.48$.

\begin{table}[htbp]
\centering
\footnotesize
\setlength{\tabcolsep}{4pt}
\begin{tabular}{lllcc}
\hline
Model & Parameters & Tail (index) & $99.9\%$ quantile & $\eta$\\
\hline
Pareto--Clayton & $a=10$, $b=1$, $d=100$ & power ($10$) & 34.78 & 0.340\\
Inverse Weibull & $\lambda=9.299$, $k=4.423$ & power ($4.42$) & 44.33 & 0.293\\
Log-logistic & $\alpha=10.472$, $\beta=5.301$ & power ($5.30$) & 38.54 & 0.406\\
Inverse Gamma & $\alpha=9.339$, $\beta=92.661$ & power ($9.34$) & 35.34 & 0.336\\
Lognormal & $\mu=2.3441$, $\sigma=0.3574$ & lognormal & 31.45 & 0.360\\
Inverse Gaussian & $\mu=11.111$, $\lambda=81.549$ & exponential & 30.72 & 0.352\\
Gamma & $\alpha=7.339$, $\beta=1.514$ & exponential & 28.15 & 0.397\\
Weibull & $\lambda=12.452$, $k=2.949$ & lighter than exp. & 23.98 & 0.484\\
\hline
\end{tabular}
\caption{Models with the same mean and variance as the Pareto--Clayton model, with the parameterizations of Appendix~\ref{app:portfolio-risks}. All models are unimodal; $\eta=G(x_m)$ denotes the inflection point of the quantile function, with $x_m$ the mode.}
\label{tab:ex1-models}
\end{table}

We use these models for two applications: a consensus model for an insurance company and its regulator (Section~\ref{sec:example-two-refs}), and an assessment of the model risk of the Pareto--Clayton model in the spirit of \citet{bernard2022robust} (Section~\ref{sec:example-unimodal}).


\subsection{Example 1: consensus model}\label{sec:example-two-refs}

We illustrate Proposition~\ref{prop:two-refs}, in the form of Remark~\ref{rem:two-refs}(iv) for generators on an open interval, for two balls with different generators around different reference distributions. Instead of a single ball around a central model, we place balls around a heavy-tailed and a light-tailed model and intersect them. The two balls reflect the views of an insurance company and its regulator, who have to agree on a model for setting capital requirements and are sensitive to different features of the loss distribution.

The regulator is mainly concerned with tail risk and adopts a prudent attitude, but it does not have a preferred model. It specifies the set of models it finds acceptable as a ball around the heavy-tailed Inverse Weibull distribution $F_1$, which serves as the centre of the ball. The ball should contain the heavy-tailed Pareto--Clayton, Log-logistic, Inverse Gamma and Lognormal distributions. The regulator measures deviations from $F_1$ by the gamma deviance, i.e., it takes $\phi_1(x)=-\log x$, so that $B_{\phi_1}(z_1,z_2)=z_1/z_2-1-\log(z_1/z_2)$ is half the gamma deviance. This divergence compares quantiles in relative terms and penalizes a model that understates the reference quantiles more than a model that overstates them by the same proportion; a model that assigns positive probability to losses less than or equal to zero has an infinite divergence. The company looks at the distribution as a whole, including its body, and is satisfied with a model that fits the empirical distribution well. Suppose that the Gamma distribution $F_2$ is broadly consistent with the empirical distribution, that the Inverse Gaussian and Weibull distributions are deemed plausible as well, and that the company measures deviations symmetrically by the $2$-Wasserstein distance, i.e., $\phi_2(x)=x^2/2$. The models acceptable to both parties form the intersection of the two balls. The company would like to use its reference model, but acknowledges that the loss distribution is subject to model ambiguity and accepts any model in its ball. As final model for setting capital requirements, the parties adopt the distribution in the intersection with the largest $\mathrm{RVaR}_{0.99,0.9999}$, i.e., the most prudent model in the tail on which they can agree. The capital requirement is then the $\mathrm{VaR}_{0.995}$ of the final model, as in Solvency~II.

We choose each radius as small as possible such that the ball contains the plausible models of the corresponding party, i.e., $\ep_1=\max\DB_{\phi_1}(G,F_1)$ over the Pareto--Clayton, Log-logistic, Inverse Gamma and Lognormal distributions, and $\ep_2=\frac12\max d_W(F_2,G)^2$ over the Inverse Gaussian and Weibull distributions. Table~\ref{tab:ex1-distances} shows that $\ep_1=5.18\cdot10^{-3}$, determined by the Lognormal distribution, and $\ep_2=\frac12\cdot0.297=0.148$, determined by the Weibull distribution. Among the eight models, only the Lognormal distribution lies in both balls, on the boundary of the first one; the Inverse Gaussian distribution lies just outside the first ball. The quantile function $0.9\,F_{\mathrm{LN}}^{-1}+0.1\,F_1^{-1}$, where $F_{\mathrm{LN}}$ denotes the Lognormal distribution, lies strictly inside both balls, so that the Slater condition of Proposition~\ref{prop:two-refs} holds. In particular, the Pareto--Clayton model lies in the first ball but not in the second.

\begin{table}[htbp]
\centering
\footnotesize
\setlength{\tabcolsep}{3pt}
\begin{tabular}{lccccc}
\hline
& \multicolumn{2}{c}{Regulator: centre Inv.\ Weibull} & \multicolumn{2}{c}{Company: centre Gamma} & \\
Model & $10^3\,\DB_{\phi_1}(G,F_1)$ & in ball & $d_W(F_2,\cdot)^2$ & in ball & $d_W(F,\cdot)^2$\\
\hline
Pareto--Clayton & 2.80 & yes & 0.443 & no & 0\\
Inverse Weibull & 0 & centre & 2.622 & no & 0.975\\
Log-logistic & 3.89 & yes & 0.844 & no & 0.169\\
Inverse Gamma & 2.43 & yes & 0.520 & no & 0.003\\
Lognormal & \textbf{5.18} & yes (boundary) & 0.124 & yes & 0.101\\
Inverse Gaussian & 5.19 & no & 0.103 & yes & 0.134\\
Gamma & 9.44 & no & 0 & centre & 0.443\\
Weibull & 20.65 & no & \textbf{0.297} & yes (boundary) & 1.415\\
\hline
\end{tabular}
\caption{Divergence $\DB_{\phi_1}(G,F_1)$, with $\phi_1(x)=-\log x$ (half the gamma deviance), to the Inverse Weibull distribution, squared $2$-Wasserstein distances to the Gamma distribution and to the Pareto--Clayton model, and membership of the regulator's ball and the company's ball. Bold entries determine the radii.}
\label{tab:ex1-distances}
\end{table}

Table~\ref{tab:ex1-bounds} reports $\mathrm{RVaR}_{0.99,0.9999}$ and $\mathrm{VaR}_{0.995}$ for the centre of the regulator's ball, the company's reference model and the final model, together with lower and upper bounds on both risk measures over the intersection of the two balls and over a single central ball. The bounds over the intersection follow from Remark~\ref{rem:two-refs}(iv), with the Karush--Kuhn--Tucker conditions verified numerically, and the bounds on $\mathrm{VaR}_{0.995}$ over the intersection are outer bounds, obtained from bounds on Range Value-at-Risk over shrinking intervals around the level $0.995$; see Appendix~\ref{app:B} for details. In contrast with the other bounds, we have not shown that they are sharp, as Corollary~\ref{cor:VaR-two-balls} requires a common reference distribution and generators defined on all of $\R$. Here, $\psi_{\boldsymbol\lambda}'(x)=-\lambda_1/x+\lambda_2x$, and the worst-case quantile function is
\[
\tilde G^{-1}_{\boldsymbol\lambda^*}=\frac{k+\sqrt{k^2+4\lambda_1^*\lambda_2^*}}{2\lambda_2^*},
\qquad
k=\Big(\gamma-\frac{\lambda_1^*}{F_1^{-1}}+\lambda_2^*F_2^{-1}\Big)^{\uparrow}.
\]

\begin{table}[htbp]
\centering
\footnotesize
\setlength{\tabcolsep}{4pt}
\begin{tabular}{lccccc}
\hline
& Centre of & Company's & & & \\
& regulator's ball & reference & Final & & \\
Risk measure & (Inv.\ Weibull) & (Gamma) & model & Intersection & Central ball\\
\hline
$\mathrm{RVaR}_{0.99,0.9999}$ & 33.39 & 25.05 & 30.11 & $[20.76,\ 30.11]$ & $[18.87,\ 40.79]$\\
$\mathrm{VaR}_{0.995}$ & 30.79 & 24.47 & 29.54 & $[18.92,\ 33.52]$ & $[16.51,\ 48.26]$\\
\hline
\end{tabular}
\caption{Values of the risk measures for the centre of the regulator's ball, the company's reference model and the final model, i.e., the distribution in the intersection with the largest $\mathrm{RVaR}_{0.99,0.9999}$, and lower and upper bounds over the intersection of the regulator's gamma-deviance ball around the Inverse Weibull distribution and the company's $2$-Wasserstein ball around the Gamma distribution. In all cases, both constraints are binding. For $\mathrm{VaR}_{0.995}$, the interval over the intersection is an outer bound (Appendix~\ref{app:B1}). The column `Central ball' gives the bounds over the smallest $2$-Wasserstein ball around the Pareto--Clayton model that contains all eight models.}
\label{tab:ex1-bounds}
\end{table}

\begin{figure}[htbp]
\centering
\includegraphics[width=\textwidth]{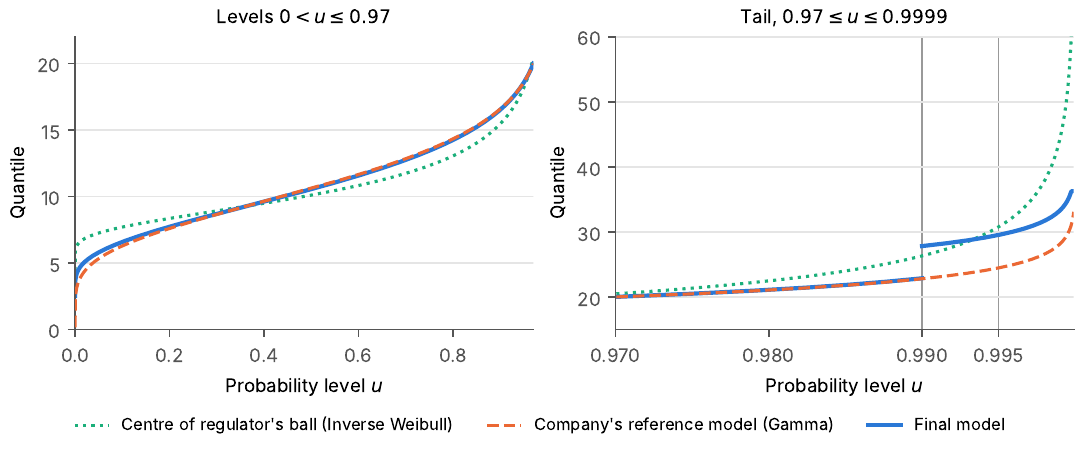}
\caption{Quantile functions of the centre of the regulator's ball (Inverse Weibull), of the company's reference model (Gamma), and of the final model, i.e., the distribution in the intersection of the two balls with the largest $\mathrm{RVaR}_{0.99,0.9999}$. Left: levels up to $0.97$; right: the tail.}
\label{fig:ex1-quantiles}
\end{figure}

We observe the following.
\begin{enumerate}[label=(\alph*)]
\item The model in the intersection with the largest $\mathrm{RVaR}_{0.99,0.9999}$ is unique (Remark~\ref{rem:two-refs}(iv)) and lies on the boundary of both balls. It stays close to the company's reference model in the body and mainly modifies its tail (Figure~\ref{fig:ex1-quantiles}): its quantile function jumps upwards at the level $0.99$, lies well above that of the company's reference model beyond, and is flat on an interval around the level $0.9999$. The company's view thus determines the body of the final model, and the regulator's view its tail. The final model does not follow the heavy tail of the centre of the regulator's ball at the highest levels, but its $\mathrm{VaR}_{0.995}$ is close to that of the centre, and much higher than that of the company's reference model.
\item The company's reference model does not belong to the regulator's ball. A consensus therefore requires the company to accept a more prudent model than its own, with a higher $\mathrm{RVaR}_{0.99,0.9999}$ and a higher capital requirement. The asymmetry of the gamma deviance contributes to the prudence of the consensus model, as the regulator's ball penalizes an understatement of the quantiles of its centre more than an overstatement of the same size.
\item Instead of intersecting their balls, the parties could agree on a single $2$-Wasserstein ball around a central model, as in \citet{bernard2022robust}. The smallest such ball around the Pareto--Clayton model that contains all models deemed plausible by either party yields much wider intervals. It must be large enough to contain the most distant plausible model and, as the $2$-Wasserstein distance is symmetric, it then also contains distributions that deviate from $F$ in the opposite direction and that neither party deems plausible.
\end{enumerate}

\subsection{Example 2: model risk assessment}\label{sec:example-unimodal}

We now assess the model risk of the Pareto--Clayton model $F$ in the spirit of Section~5.2 of \citet{bernard2022robust}. As there, we compute bounds on $\mathrm{VaR}_\alpha$ for $\alpha=0.90,0.95,0.99$, to which we add the Solvency~II level $\alpha=0.995$, over the $2$-Wasserstein ball $d_W(F,G)^2\leq1.415$ around $F$, which is the smallest ball around $F$ that contains all seven alternative models; its radius is determined by the Weibull distribution (Table~\ref{tab:ex1-distances}). In contrast with \citet{bernard2022robust}, we do not fix the mean and the variance. Instead, we use that all eight models are unimodal, and we restrict the ambiguity set to unimodal distributions whose inflection point lies between the smallest and the largest inflection point of the eight models, i.e., $\eta\in[\eta_{\min},\eta_{\max}]=[0.293,0.484]$ (Table~\ref{tab:ex1-models}). The bounds over this set are the extremes over $\eta$ of the bounds for a fixed inflection point, which follow from Proposition~\ref{prop:VaR-unimodal}; we approximate them by the extremes over the grid $\eta\in\{0.293,0.340,0.360,0.406,0.484\}$, which includes the inflection point $\eta_F=0.340$ of the reference model. Statements (i) and (ii) of Proposition~\ref{prop:VaR-unimodal} only require that the unimodal ball is nonempty, which holds for all $\eta$ considered. The identification of the approximating RVaR bounds via Corollary~\ref{thm:WC-RM-BW2} assumes that $\Finv\in\Minv^\eta$, which only holds for $\eta=\eta_F$. For $\eta\neq\eta_F$, their proofs remain valid provided that $\delta(\eta)^2/2<\ep$, where $\delta(\eta)$ is the $L^2$-distance from $\Finv$ to $\Minv^\eta$: as $\lambda\to\infty$, the divergence of the candidate solution then tends to $\delta(\eta)^2/2$ instead of $0$, and a multiplier $\lambda^*$ with a binding constraint still exists. Here, $\delta(\eta)^2/2\leq4.1\cdot10^{-6}$ for all $\eta$ considered, far below the radius. Without the shape constraint, the bounds follow from Proposition~\ref{prop:VaR}. Table~\ref{tab:ex1-unimodal} reports the results, together with the bounds based on the mean and the variance only, with and without unimodality.

\begin{table}[htbp]
\centering
\footnotesize
\setlength{\tabcolsep}{4pt}
\begin{tabular}{llccccc}
\hline
& & & \multicolumn{2}{c}{$d_W(F,G)^2\leq1.415$} & \multicolumn{2}{c}{Mean--var.}\\
 & & Ref. & Uncon. & Unimodal & Uncon. & Unimodal\\
\hline
$\mathrm{VaR}_{0.90}$ & upper & 16.29 & 22.33 & 19.65 & 23.42 & 18.72\\
 & lower & & 11.58 & 12.19 & 9.74 & --\\
$\mathrm{VaR}_{0.95}$ & upper & 18.75 & 26.66 & 23.30 & 28.99 & 22.63\\
 & lower & & 13.00 & 13.72 & 10.17 & --\\
$\mathrm{VaR}_{0.99}$ & upper & 24.79 & 40.12 & 34.11 & 51.92 & 38.14\\
 & lower & & 15.64 & 16.29 & 10.70 & --\\
$\mathrm{VaR}_{0.995}$ & upper & 27.60 & 48.26 & 40.39 & 68.97 & 49.56\\
 & lower & & 16.51 & 17.12 & 10.82 & --\\
\hline
\end{tabular}
\caption{Bounds over the ball $d_W(F,G)^2\leq1.415$ around the Pareto--Clayton model, without shape constraint (Uncon.) and over unimodal distributions with an inflection point in $[0.293,0.484]$, approximated by the extremes over $\eta\in\{0.293,0.340,0.360,0.406,0.484\}$. The last two columns give the bounds over all distributions with mean $100/9$ and standard deviation $4.10$, and the upper bounds over all unimodal such distributions of \citet{bernard2020range}.}
\label{tab:ex1-unimodal}
\end{table}

We observe the following.
\begin{enumerate}[label=(\alph*)]
\item At all levels, unimodality shortens the intervals for Value-at-Risk by $27\%$ to $31\%$, mainly by lowering the upper bound; for instance, the upper bound on $\mathrm{VaR}_{0.995}$ decreases from $48.26$ to $40.39$. Without a shape constraint, the worst case adds a jump to the quantile function at the level $\alpha$, i.e., a gap in the support, and the supremum is not attained (Proposition~\ref{prop:VaR}). Under unimodality, the quantile function is continuous, so that such a jump is excluded.
\item On the grid of inflection points considered, the upper bounds do not depend on $\eta$ at all, and the lower bounds vary by at most $0.15$, namely for $\mathrm{VaR}_{0.90}$; the bounds for $\eta=\eta_F$ therefore almost coincide with those over the whole range. The worst-case distributions modify the quantile function mainly at high levels, far from the inflection point, where unimodality only requires convexity. The best-case distributions also lower the quantile function in the body, closer to the inflection point, which explains the small effect on the lower bounds at the lowest levels.
\item At the levels $0.90$ and $0.95$, the unimodal upper bounds over the ball are slightly above the unimodal mean--variance bounds ($19.65$ against $18.72$, and $23.30$ against $22.63$). At high levels, however, they are much tighter: $34.11$ against $38.14$ for $\mathrm{VaR}_{0.99}$, and $40.39$ against $49.56$ for $\mathrm{VaR}_{0.995}$. This illustrates that, in the tail, the reference model provides more information than its first two moments. The ball-based lower bounds are also much tighter than the moment-based ones.
\item The radius of the ball is determined by the Weibull distribution, whose tail is much lighter than that of $F$. As the $2$-Wasserstein distance is symmetric, a ball that is large enough to contain this model also contains models whose tails are heavier than that of $F$ by a similar amount. Accordingly, the unimodal upper bound on $\mathrm{VaR}_{0.995}$, $40.39$, is far above the largest value among the eight models, $30.79$ for the Inverse Weibull distribution. This motivates the asymmetric divergence used by the regulator in Example~1.
\end{enumerate}

\section*{Acknowledgements}
Carole Bernard and Steven Vanduffel thank Research Foundation Flanders (FWO) for financial support with funding numbers  FWO SBO S006721N, FWO G093024N and FWO WOG W001021N. 

 \bibliographystyle{plainnat} 
 \bibliography{references}






\newpage
\appendix
\numberwithin{theorem}{subsection} 
\section{Application to a portfolio of risks}\label{app:portfolio-risks} 

This appendix collects the distributions used in Section~5.2 of \citet{bernard2022robust}. All of them are used in Section~\ref{sec:examples}, with the parameters of the alternative models in Appendix~\ref{app:A2} matched exactly to the mean and variance of the reference model (Table~\ref{tab:ex1-models}). We provide their densities or quantile functions, as well as their first two raw moments.

\subsection{Reference Pareto--Clayton model}\label{app:A1} Let $\Theta\sim\operatorname{Gamma}(a,b)$, where $a>0$ is the shape parameter and $b>0$ is the rate parameter. Conditionally on $\Theta=\theta$, let $X_1,\ldots,X_d$ be independent exponential random variables with rate $\theta$. The aggregate portfolio loss is
\[ S=\sum_{i=1}^{d}X_i. \]
Conditionally on $\Theta=\theta$, the aggregate loss satisfies
\[ S\mid\Theta=\theta\sim\operatorname{Gamma}(d,\theta), \]
where $d$ is the shape parameter and $\theta$ is the rate parameter. The scaled aggregate loss $S/b$ follows a beta-prime distribution with parameters $d$ and $a$. Equivalently, if
\[ B\sim\operatorname{Beta}(d,a), \]
then
\[ S\overset{d}{=}b\frac{B}{1-B}. \]
Consequently, the quantile function of $S$ is
\[ F_S^{-1}(u) = b\frac{F_B^{-1}(u)}{1-F_B^{-1}(u)}, \qquad 0<u<1, \]
where $F_B^{-1}$ denotes the quantile function of a $\operatorname{Beta}(d,a)$ random variable. The density of $S$ is
\[ f_S(s) = \frac{s^{d-1}b^a} {B(d,a)(s+b)^{a+d}}, \qquad s>0, \]
where $B(\cdot,\cdot)$ denotes the beta function. For every integer $k$ satisfying $k<a$, the $k$th raw moment is
\[ \mathbb{E}[S^k] = b^k\frac{B(d+k,a-k)}{B(d,a)}. \]
Equivalently,
\[ \mathbb{E}[S^k] = b^k \frac{\Gamma(d+k)\Gamma(a-k)} {\Gamma(d)\Gamma(a)}. \]
In particular, provided that $a>2$,
\[ \mathbb{E}[S] = \frac{bd}{a-1}, \]
and
\[ \mathbb{E}[S^2] = \frac{b^2d(d+1)} {(a-1)(a-2)}. \]
Therefore,
\[ \operatorname{Var}(S) = \frac{b^2d(a+d-1)} {(a-1)^2(a-2)}. \]
For the numerical illustration, we set
\[ a=10,\qquad b=1,\qquad d=100. \]
This gives
\[ \mathbb{E}[S] = \frac{100}{9} \approx 11.11, \]
\[ \mathbb{E}[S^2] = \frac{100\times101}{9\times8} \approx 140.28, \]
and
\[ \operatorname{Var}(S) = \frac{100\times109}{9^2\times8} \approx 16.82. \]

\subsection{Alternative models}\label{app:A2} We consider seven two-parameter distributions supported on $(0,+\infty)$. Unless stated otherwise, the parameters are positive. The models are parameterized as follows.
\paragraph{1. Lognormal model (LN).} The aggregate loss is represented as
\[ S=\exp(Z), \]
where
\[ Z\sim\mathcal{N}(\mu_{\mathrm{LN}},\sigma_{\mathrm{LN}}^2). \]
The density of $S$ is
\[ f_S(s) = \frac{1} {s\sigma_{\mathrm{LN}}\sqrt{2\pi}} \exp\left\{ -\frac{(\log s-\mu_{\mathrm{LN}})^2} {2\sigma_{\mathrm{LN}}^2} \right\}, \qquad s>0. \]
Its quantile function is
\[ F_S^{-1}(u) = \exp\left\{ \mu_{\mathrm{LN}} + \sigma_{\mathrm{LN}}\Phi^{-1}(u) \right\}, \qquad 0<u<1, \]
where $\Phi$ is the standard normal distribution function. The first two raw moments are
\[ \mathbb{E}[S] = \exp\left( \mu_{\mathrm{LN}} +\frac{\sigma_{\mathrm{LN}}^2}{2} \right) \]
and
\[ \mathbb{E}[S^2] = \exp\left( 2\mu_{\mathrm{LN}} +2\sigma_{\mathrm{LN}}^2 \right). \]
\paragraph{2. Gamma model ($\Gamma$).} The aggregate loss follows a Gamma distribution with shape parameter $\alpha_{\Gamma}$ and scale parameter $\beta_{\Gamma}$:
\[ S\sim\operatorname{Gamma} (\alpha_{\Gamma},\beta_{\Gamma}). \]
Its density is
\[ f_S(s) = \frac{s^{\alpha_{\Gamma}-1} \exp(-s/\beta_{\Gamma})} {\Gamma(\alpha_{\Gamma}) \beta_{\Gamma}^{\alpha_{\Gamma}}}, \qquad s>0. \]
The first two raw moments are
\[ \mathbb{E}[S] = \alpha_{\Gamma}\beta_{\Gamma} \]
and
\[ \mathbb{E}[S^2] = \alpha_{\Gamma} (\alpha_{\Gamma}+1) \beta_{\Gamma}^2. \]
Equivalently,
\[ \operatorname{Var}(S) = \alpha_{\Gamma}\beta_{\Gamma}^2. \]
\paragraph{3. Weibull model (W).} The aggregate loss follows a Weibull distribution with scale parameter $\lambda_{\mathrm{W}}$ and shape parameter $k_{\mathrm{W}}$:
\[ S\sim\operatorname{Weibull} (\lambda_{\mathrm{W}},k_{\mathrm{W}}). \]
Its distribution function is
\[ F_S(s) = 1-\exp\left\{ -\left(\frac{s}{\lambda_{\mathrm{W}}}\right)^{k_{\mathrm{W}}} \right\}, \qquad s>0. \]
Its density is
\[ f_S(s) = \frac{k_{\mathrm{W}}}{\lambda_{\mathrm{W}}} \left(\frac{s}{\lambda_{\mathrm{W}}}\right)^{k_{\mathrm{W}}-1} \exp\left\{ -\left(\frac{s}{\lambda_{\mathrm{W}}}\right)^{k_{\mathrm{W}}} \right\}, \qquad s>0. \]
Its quantile function is
\[ F_S^{-1}(u) = \lambda_{\mathrm{W}} \left[-\log(1-u)\right]^{1/k_{\mathrm{W}}}, \qquad 0<u<1. \]
The first two raw moments are
\[ \mathbb{E}[S] = \lambda_{\mathrm{W}} \Gamma\left(1+\frac{1}{k_{\mathrm{W}}}\right) \]
and
\[ \mathbb{E}[S^2] = \lambda_{\mathrm{W}}^2 \Gamma\left(1+\frac{2}{k_{\mathrm{W}}}\right). \]
\paragraph{4. Inverse Gaussian model (IG).} The aggregate loss follows an inverse Gaussian distribution with mean parameter $\mu_{\mathrm{IG}}$ and shape parameter $\lambda_{\mathrm{IG}}$:
\[ S\sim\operatorname{IG} (\mu_{\mathrm{IG}},\lambda_{\mathrm{IG}}). \]
Its density is
\[ f_S(s) = \left( \frac{\lambda_{\mathrm{IG}}} {2\pi s^3} \right)^{1/2} \exp\left\{ -\frac{\lambda_{\mathrm{IG}} (s-\mu_{\mathrm{IG}})^2} {2\mu_{\mathrm{IG}}^2s} \right\}, \qquad s>0. \]
Its distribution function is
\[
\begin{split} F_S(s) ={}& \Phi\left[ \sqrt{\frac{\lambda_{\mathrm{IG}}}{s}} \left( \frac{s}{\mu_{\mathrm{IG}}}-1 \right) \right] \\ &+ \exp\left( \frac{2\lambda_{\mathrm{IG}}}{\mu_{\mathrm{IG}}} \right) \Phi\left[ -\sqrt{\frac{\lambda_{\mathrm{IG}}}{s}} \left( \frac{s}{\mu_{\mathrm{IG}}}+1 \right) \right], \qquad s>0. \end{split}
\]
The first two raw moments are
\[ \mathbb{E}[S] = \mu_{\mathrm{IG}} \]
and
\[ \mathbb{E}[S^2] = \mu_{\mathrm{IG}}^2 + \frac{\mu_{\mathrm{IG}}^3} {\lambda_{\mathrm{IG}}}. \]
Equivalently,
\[ \operatorname{Var}(S) = \frac{\mu_{\mathrm{IG}}^3} {\lambda_{\mathrm{IG}}}. \]
\paragraph{5. Inverse Gamma model (I$\Gamma$).} The aggregate loss follows an inverse Gamma distribution with shape parameter $\alpha_{\mathrm{I}\Gamma}$ and scale parameter $\beta_{\mathrm{I}\Gamma}$:
\[ S\sim\operatorname{InvGamma} (\alpha_{\mathrm{I}\Gamma}, \beta_{\mathrm{I}\Gamma}). \]
Its density is
\[ f_S(s) = \frac{\beta_{\mathrm{I}\Gamma}^{\alpha_{\mathrm{I}\Gamma}}} {\Gamma(\alpha_{\mathrm{I}\Gamma})} s^{-\alpha_{\mathrm{I}\Gamma}-1} \exp\left( -\frac{\beta_{\mathrm{I}\Gamma}}{s} \right), \qquad s>0. \]
Provided that $\alpha_{\mathrm{I}\Gamma}>1$,
\[ \mathbb{E}[S] = \frac{\beta_{\mathrm{I}\Gamma}} {\alpha_{\mathrm{I}\Gamma}-1}. \]
Provided that $\alpha_{\mathrm{I}\Gamma}>2$,
\[ \mathbb{E}[S^2] = \frac{\beta_{\mathrm{I}\Gamma}^2} {(\alpha_{\mathrm{I}\Gamma}-1) (\alpha_{\mathrm{I}\Gamma}-2)}. \]
Consequently,
\[ \operatorname{Var}(S) = \frac{\beta_{\mathrm{I}\Gamma}^2} {(\alpha_{\mathrm{I}\Gamma}-1)^2 (\alpha_{\mathrm{I}\Gamma}-2)}. \]
\paragraph{6. Inverse Weibull model (IW).} The aggregate loss follows an inverse Weibull distribution with scale parameter $\lambda_{\mathrm{IW}}$ and shape parameter $k_{\mathrm{IW}}$. Its distribution function is
\[ F_S(s) = \exp\left\{ -\left( \frac{\lambda_{\mathrm{IW}}}{s} \right)^{k_{\mathrm{IW}}} \right\}, \qquad s>0. \]
Its density is
\[ f_S(s) = k_{\mathrm{IW}} \lambda_{\mathrm{IW}}^{k_{\mathrm{IW}}} s^{-k_{\mathrm{IW}}-1} \exp\left\{ -\left( \frac{\lambda_{\mathrm{IW}}}{s} \right)^{k_{\mathrm{IW}}} \right\}, \qquad s>0. \]
Its quantile function is
\[ F_S^{-1}(u) = \lambda_{\mathrm{IW}} \left[-\log(u)\right]^{-1/k_{\mathrm{IW}}}, \qquad 0<u<1. \]
Provided that $k_{\mathrm{IW}}>1$,
\[ \mathbb{E}[S] = \lambda_{\mathrm{IW}} \Gamma\left( 1-\frac{1}{k_{\mathrm{IW}}} \right). \]
Provided that $k_{\mathrm{IW}}>2$,
\[ \mathbb{E}[S^2] = \lambda_{\mathrm{IW}}^2 \Gamma\left( 1-\frac{2}{k_{\mathrm{IW}}} \right). \]
\paragraph{7. Log-logistic model (LL).} The aggregate loss follows a log-logistic distribution with scale parameter $\alpha_{\mathrm{LL}}$ and shape parameter $\beta_{\mathrm{LL}}$. Its distribution function is
\[ F_S(s) = \frac{1} {1+\left(\alpha_{\mathrm{LL}}/s\right)^{\beta_{\mathrm{LL}}}}, \qquad s>0. \]
Its density is
\[ f_S(s) = \frac{\beta_{\mathrm{LL}}}{\alpha_{\mathrm{LL}}} \frac{ \left(s/\alpha_{\mathrm{LL}}\right)^{\beta_{\mathrm{LL}}-1} }{ \left[ 1+\left(s/\alpha_{\mathrm{LL}}\right)^{\beta_{\mathrm{LL}}} \right]^2 }, \qquad s>0. \]
Its quantile function is
\[ F_S^{-1}(u) = \alpha_{\mathrm{LL}} \left( \frac{u}{1-u} \right)^{1/\beta_{\mathrm{LL}}}, \qquad 0<u<1. \]
Provided that $\beta_{\mathrm{LL}}>1$,
\[ \mathbb{E}[S] = \alpha_{\mathrm{LL}} \frac{\pi/\beta_{\mathrm{LL}}} {\sin(\pi/\beta_{\mathrm{LL}})}. \]
Provided that $\beta_{\mathrm{LL}}>2$,
\[ \mathbb{E}[S^2] = \alpha_{\mathrm{LL}}^2 \frac{2\pi/\beta_{\mathrm{LL}}} {\sin(2\pi/\beta_{\mathrm{LL}})}. \]


\section{Technical details for Example 1}\label{app:B}

\subsection{Value-at-Risk over the intersection}\label{app:B1}
As $\mathrm{RVaR}_{\alpha-h,\alpha}(G)\leq\Ginv(\alpha)\leq\mathrm{RVaR}_{\alpha,\alpha+h}(G)$ for every distribution $G$ and every $h>0$, the lower bound on $\mathrm{RVaR}_{0.995-h,0.995}$ and the upper bound on $\mathrm{RVaR}_{0.995,0.995+h}$ over the intersection are outer bounds on $\mathrm{VaR}_{0.995}$ for every $h$. In Example~1, these bounds do not change for $h\leq10^{-3}$, and we report their common values.

\subsection{Validity of the worst-case quantile function}\label{app:B2}
The bounds over the intersection follow from Remark~\ref{rem:two-refs}(iv), with the Karush--Kuhn--Tucker conditions verified numerically. By Remark~\ref{rem:two-refs}(iv), it suffices that the candidate solution is well defined at the multiplier found numerically, which is the case as both multipliers are positive, so that $\psi_{\boldsymbol\lambda}'(x)=-\lambda_1/x+\lambda_2x$ maps $(0,\infty)$ onto $\R$. Moreover, $\tilde k_{\boldsymbol\lambda}$ is square integrable, as $\int_0^1F_1^{-1}(u)^{-2}\,\d u<\infty$ for the Inverse Weibull distribution, and the bounds $\lambda_1/(|y|+\sqrt{\lambda_1\lambda_2})\leq(\psi_{\boldsymbol\lambda}')^{-1}(y)\leq|y|/\lambda_2+\sqrt{\lambda_1/\lambda_2}$, $y\in\R$, imply that the candidate solution is square integrable and has finite divergences.

\section{Proofs}\label{app:general}

This appendix contains the proofs of all results in Sections~\ref{sec:WC-DRM}--\ref{sec:two-BW-balls}. Sections~\ref{app:C1}--\ref{app:C3} contain the proofs of Theorem~\ref{thm:general}, Propositions~\ref{prop:general-multiplier} and~\ref{prop:general-stability} and Corollary~\ref{cor:general-ac}, together with the consequences of \eqref{eq:strong-smooth} used in Section~\ref{sec:shape}. Section~\ref{app:C4} contains the remaining proofs for Sections~\ref{sec:ac} and~\ref{sec:shape}, including that of Lemma~\ref{lem:iso}, and Sections~\ref{app:C5} and~\ref{app:C6} the proofs for Sections~\ref{sec:VaR} and~\ref{sec:two-BW-balls}, respectively.

In Sections~\ref{app:C1}--\ref{app:C3}, unless stated otherwise, the standing assumptions of Section~\ref{sec:general-g} hold: $1<p<\infty$, $p'=p/(p-1)$, $\phi\in C^1(\R)$ is strictly convex and satisfies \eqref{eq:general-p-growth}, $F\in\mathcal M_p$, and $g$ is of bounded variation with $g(0)=0$ and satisfies \eqref{eq:endpoint} with $\nu>1/p$. We write $\lVert\cdot\rVert_p$ for the norm of $L^p(0,1)$ and $C$ for a finite constant that may change from line to line. For a non-decreasing, left-continuous function $h$ on $(0,1)$, we denote by $\d h$ its Lebesgue--Stieltjes measure on $(0,1)$, so that $\d h([s,t))=h(t)-h(s)$ for $0<s<t<1$ and $\d h(\{u\})=h(u+)-h(u)$; in particular, $\d\Ginv(\{u\})=\Delta_u(G)$. For $\Ginv\in\Minv_p$ and $N>0$, we write $\Ginv_{(N)}:=\max\{\min\{\Ginv,N\},-N\}$ for the truncated quantile function and $G_{(N)}$ for its distribution function.

\subsection{Preliminaries for Section~\ref{sec:general-g}}\label{app:C1}

\begin{lemma}\label{lem:C-growth}
Let $\phi\in C^1(\R)$ be strictly convex and satisfy \eqref{eq:general-p-growth}. Then there exists a constant $C_\phi$ such that the following hold.
\begin{enumerate}[label=(\alph*),leftmargin=*]
\item $|\phi'(x)|\leq C_\phi(1+|x|^{p-1})$ for $x\in\R$.
\item $\phi'$ is a strictly increasing bijection of $\R$, and $|(\phi')^{-1}(y)|\leq C_\phi(1+|y|^{p'-1})$ for $y\in\R$.
\item The convex conjugate $\phi^*$ is finite on $\R$, $|\phi^*(y)|\leq C_\phi(1+|y|^{p'})$, and $\phi^*(\phi'(x))=x\phi'(x)-\phi(x)$ for $x\in\R$.
\item For $G,H\in\mathcal M_p$, $\phi(\Ginv)\in L^1(0,1)$, $\phi'(\Hinv)\in L^{p'}(0,1)$ and $\DB_\phi(G,H)<\infty$, and $\Ginv\mapsto\DB_\phi(G,H)$ is continuous on $\Minv_p$ with respect to $\lVert\cdot\rVert_p$.
\item For every $H\in\mathcal M_p$, there exists $C_H<\infty$ such that
\begin{equation}\label{eq:general-coercivity}
\DB_\phi(G,H)\geq\frac{a_\phi}2\,\lVert\Ginv\rVert_p^p-C_H,\qquad G\in\mathcal M_p .
\end{equation}
In particular, divergence balls are bounded in $L^p(0,1)$.
\item For $H\in\mathcal M_p$ and $G_n\in\mathcal M_p$ satisfying $\lim_{n\to\infty}\DB_\phi(G_n,H)=0$, we have
\begin{equation}\label{eq:general-fixed-base}
\lim_{n\to\infty}\lVert\Ginv_n-\Hinv\rVert_p=0 
\end{equation}
\end{enumerate}
\end{lemma}

\begin{proof}
(a) For $h>0$, convexity gives $\phi'(x)\leq(\phi(x+h)-\phi(x))/h$ and $\phi'(x)\geq(\phi(x)-\phi(x-h))/h$. With $h=1+|x|$, \eqref{eq:general-p-growth} bounds both difference quotients in absolute value by $\big(A_\phi(1+(2|x|+1)^p)+b_\phi\big)/(1+|x|)\leq C_\phi(1+|x|^{p-1})$ for some $C_\phi>0$. We establish the claim in (a).

(b) As $\phi$ is strictly convex, $\phi'$ is strictly increasing and continuous. By convexity, $\phi(0)\geq\phi(x)-x\phi'(x)$, implying that $x\phi'(x)\geq a_\phi|x|^p-b_\phi-\phi(0)$ by \eqref{eq:general-p-growth}. Hence $\phi'(x)\to\pm\infty$ as $x\to\pm\infty$, implying that $\phi'(\R)=\R$. Moreover, if $y=\phi'(x)$ with $|x|\geq1$, then for some $C>0$, $a_\phi|x|^{p}-C\leq|x||y|$, which gives $|x|^{p-1}\leq \frac{C(1+|y|)}{a_\phi}$. Hence, we have $|(\phi')^{-1}(y)|\leq C_\phi(1+|y|^{p'-1}), y\in\R$ for some $C_\phi>0$. 

(c) By \eqref{eq:general-p-growth}, $\phi^*(y)\leq\sup_{x\in\R}\{xy-a_\phi|x|^p\}+b_\phi\leq C(1+|y|^{p'})$, and $\phi^*(y)\geq-\phi(0)$. Hence, we have $|\phi^*(y)|\leq C_\phi(1+|y|^{p'}),~y\in\R$ for some $C_\phi>0$.  For $y=\phi'(x)$, the concave function $z\mapsto zy-\phi(z)$ is maximized at $z=x$, which implies $\phi^*(\phi'(x))=x\phi'(x)-\phi(x)$ for $x\in\R$.

(d) As $-b_\phi\leq\phi(x)\leq A_\phi(1+|x|^p)$, we have $|\phi(\Ginv)|\leq A_\phi(1+|\Ginv|^p) +b_\phi$, implying $\phi(\Ginv)\in L^1(0,1)$. Moreover, it follows from (a) that $|\phi'(\Hinv)|^{p'}\leq C(1+|\Hinv|^p)$ for some $C>0$, implying  $\phi'(\Hinv)\in L^{p'}(0,1)$.  Hence, using H\"older's inequality,  all terms of $\DB_\phi(G,H)=\int_0^1\big(\phi(\Ginv)-\phi(\Hinv)-\phi'(\Hinv)(\Ginv-\Hinv)\big)\,\d u$ are integrable, which implies $\DB_\phi(G,H)<\infty$.  By (a) and the mean value theorem, $|\phi(x)-\phi(y)|\leq C_\phi|x-y|(1+|x|^{p-1}+|y|^{p-1})$, so that, by H\"older's inequality, $\Ginv\mapsto\int_0^1\phi(\Ginv)\,\d u$ is continuous on $L^p(0,1)$.  Using the fact that $\phi'(\Hinv)\in L^{p'}(0,1)$ and H\"older's inequality, $\Ginv\mapsto\int_0^1\phi'(\Hinv)\Ginv\,\d u$ is continuous on $L^p(0,1)$. Therefore, $\Ginv\mapsto\DB_\phi(G,H)$ is continuous on $\Minv_p$ with respect to $\lVert\cdot\rVert_p$.

(e) By \eqref{eq:general-p-growth} and using H\"older's inequality, we have
\[
\DB_\phi(G,H)\geq a_\phi\lVert\Ginv\rVert_p^p-b_\phi-\int_0^1\phi(\Hinv)\,\d u-\lVert\phi'(\Hinv)\rVert_{p'}\big(\lVert\Ginv\rVert_p+\lVert\Hinv\rVert_p\big).
\]
Moreover, using Young's inequality, we have 
\[\lVert\phi'(\Hinv)\rVert_{p'}\lVert\Ginv\rVert_p\leq \frac{a_\phi}{2}\lVert\Ginv\rVert_p^p+\frac{1}{p'}\left(\frac{2}{p a_\phi}\right)^{1/(p-1)}\lVert\phi'(\Hinv)\rVert_{p'}^{p'}.\]
Combining the above two inequalities yields, for some $C_H\in\R$,
\[
\DB_\phi(G,H)\geq\frac{a_\phi}2\lVert\Ginv\rVert_p^p-C_H,~G\in\mathcal M_p.
\]

(f) By (e), the sequence $\{\Ginv_n\}_{n\geq 1}$ is bounded in $L^p(0,1)$. For $y\in\R$, $x\mapsto B_\phi(x,y)$ is convex, vanishes only at $x=y$ by strict convexity of $\phi$, and is therefore non-increasing on $(-\infty,y]$ and non-decreasing on $[y,\infty)$. Hence, using the monotonicity and  continuity of $B_{\phi}$, we have, for $M,\delta>0$, $m_{M,\delta}:=\inf\{B_\phi(x,y):|y|\leq M,\ |x-y|\geq\delta\}=\min_{|y|\leq M}\min\{B_\phi(y-\delta,y),B_\phi(y+\delta,y)\}>0$, implying that $B_\phi(x,y)/m_{M,\delta}\geq 1$ whenever $|y|\leq M,\ |x-y|\geq\delta$. Thus the Lebesgue measure of $\{t\in (0,1):|\Ginv_n(t)-\Hinv(t)|\geq\delta,\ |\Hinv(t)|\leq M\}$ is bounded by $\DB_\phi(G_n,H)/m_{M,\delta}$, which tends to $0$ as $n\to\infty$. Clearly, the Lebesgue measure of $\{t\in (0,1):|\Hinv(t)|>M\}$ tends to $0$ as $M\to\infty$. Consequently, $\Ginv_n\to\Hinv$  in Lebesgue measure as $n\to\infty$. As $\{\Ginv_n\}_{n\geq 1}$ is bounded in $L^p(0,1)$ and $p>1$, it follows that $\Ginv_n\to\Hinv$ weakly in $L^p(0,1)$. By (d), we have $\phi'(\Hinv)\in L^{p'}(0,1)$. Therefore,
\[
\int_0^1\phi(\Ginv_n)\,\d u=\DB_\phi(G_n,H)+\int_0^1\phi(\Hinv)\,\d u+\int_0^1\phi'(\Hinv)\big(\Ginv_n-\Hinv\big)\,\d u\to\int_0^1\phi(\Hinv)\,\d u .
\]
There exist subsequences such that the non-negative functions $\phi(\Ginv_{n_k})+b_\phi$ converge \ to $\phi(\Hinv)+b_\phi$ a.e., and their integrals converge. Hence, $\lVert\phi(\Ginv_{n_k})-\phi(\Hinv)\rVert_1\to 0$, which further implies that $\{\phi(\Ginv_{n_k})\}_{k\geq 1}$  is uniformly integrable. Since $a_\phi|\Ginv_n|^p\leq\phi(\Ginv_n)+b_\phi$, $\{|\Ginv_{n_k}|^p\}_{k\geq 1}$  is uniformly integrable. Using $|\Ginv_{n_k}-\Hinv|^p\leq 2^p(|\Ginv_{n_k}|^p+|\Hinv|^p)$, we have $\lVert\Ginv_{n_k}-\Hinv\rVert_p\to0$.  As every subsequence has an a.e.\ convergent further subsequence, the claim in (f) holds. We complete the proof.
\end{proof}

\begin{lemma}\label{lem:C-basic}
Let $b\colon[0,1]\to\R$ be of bounded variation with $b(0)=0$ and satisfy \eqref{eq:endpoint} with some constants $K_b>0$, $t_0\in(0,1/2]$ and $\nu>1/p$.
\begin{enumerate}[label=(\alph*),leftmargin=*]
\item Both integrals in \eqref{eq:signed-choquet} are absolutely convergent for $H_b(G)$, $G\in\mathcal M_p$, and there exists a constant $C_b$ such that
\begin{equation}\label{eq:general-H-bound}
|H_b(G)|\leq C_b\lVert\Ginv\rVert_p,\qquad G\in\mathcal M_p .
\end{equation}
\item $H_b(G_{(N)})\to H_b(G)$ as $N\to\infty$, uniformly on $L^p$-bounded subsets of $\mathcal M_p$.
\item If $b$ is upper semicontinuous, $G_n,G\in\mathcal M_p$, $\Ginv_n\to\Ginv$ a.e.\ and $\sup_{n\geq 1}\lVert\Ginv_n\rVert_p<\infty$, then $\limsup_{n\to\infty}H_b(G_n)\leq H_b(G)$.
\item For $\Ginv\in\Minv_p$, $|\Ginv(u)|\leq\lVert\Ginv\rVert_p\min(u,1-u)^{-1/p}$ for $0<u<1$. Every $L^p$-bounded sequence in $\Minv_p$ has a subsequence that converges a.e.\ to an element of $\Minv_p$.
\item The envelope $\hat b$ is upper semicontinuous and of bounded variation, $\hat b(0)=b(0)$, $\hat b(1)=b(1)$, and $\hat b$ satisfies \eqref{eq:endpoint} with the same $\nu$, the constant $2K_b$ and possibly different $t_0$. The set $\mathcal E_b$ is at most countable.
\end{enumerate}
\end{lemma}

\begin{proof}
(a) Let $R:=\lVert\Ginv\rVert_p>0$; the case $R=0$ is trivial. By Markov's inequality, $1-G(x)\leq R^p/x^p$ and $G(-x)\leq R^p/x^p$ for $x>0$. Hence, for $x\geq x_R:=R\,t_0^{-1/p}$, \eqref{eq:endpoint} gives
\begin{equation}\label{eq:C-tails}
|b(1-G(x))|\leq K_bR^{p\nu}x^{-p\nu},\qquad|b(1-G(-x))-b(1)|\leq K_bR^{p\nu}x^{-p\nu}.
\end{equation}
As $p\nu>1$, both tails are integrable, and, with $M_b:=\sup_{t\in[0,1]}|b(t)|<\infty$,
\[
|H_b(G)|\leq(2M_b+|b(1)|)\,x_R+2K_bR^{p\nu}\int_{x_R}^\infty x^{-p\nu}\,\d x=\Big(\frac{2M_b+|b(1)|}{t_0^{1/p}}+\frac{2K_b\,t_0^{\nu-1/p}}{p\nu-1}\Big)R .
\]

(b) The distribution functions of $G$ and $G_{(N)}$ coincide on $(-N,N)$, so that $H_b(G)-H_b(G_{(N)})$ only involves the integrands of \eqref{eq:signed-choquet} on $\{|x|\geq N\}$. By \eqref{eq:C-tails}, these are bounded by $2K_bR^{p\nu}\int_N^\infty x^{-p\nu}\,\d x$ for $N\geq x_R$, uniformly in $\lVert\Ginv\rVert_p\leq R$.

(c) Let $U\sim U(0,1)$. Then $\Ginv_n(U)\to\Ginv(U)$ a.s., implying that $G_n(x)\to G(x)$ at every continuity point $x$ of $G$, hence for a.e.\ $x$. At these points, upper semicontinuity gives $\limsup_{n\to\infty}b(1-G_n(x))\leq b(1-G(x))$. On every bounded interval, the integrands in \eqref{eq:signed-choquet} are bounded by $M_b+|b(1)|$ with $M_b=\sup_{t\in[0,1]}|b(t)|<\infty$. Using Fatou lemma for nonnegative functions $M_b+|b(1)|-b(1 - G_n(x))$, we have
\begin{align*}&\limsup_{n\to\infty}\int_0^{N}b(1 - G_n(x))\,\mathrm{d}x+\int_{-N}^0 \left(b(1 -  G_n(x))-b(1)\right) \,\mathrm{d}x\\
&\leq \int_0^{N}b(1 - G(x))\,\mathrm{d}x+\int_{-N}^0 \left(b(1 -  G(x))-b(1)\right)\,\mathrm{d}x.
\end{align*}
By \eqref{eq:C-tails}, the tails are small uniformly in $n$, which yields the claim.

(d) If $\Ginv(u)\geq0$, then $\int_u^1|\Ginv(v)|^p\,\d v\geq(1-u)\Ginv(u)^p$, and if $\Ginv(u)<0$, then $\int_0^u|\Ginv(v)|^p\,\d v\geq u|\Ginv(u)|^p$, which proves the bound. Hence an $L^p$-bounded sequence $\{\Ginv_n(u)\}_{n\geq 1}$ in $\Minv_p$ is uniformly bounded on every compact subinterval of $(0,1)$. Helly's selection theorem and a diagonal argument yield a subsequence $\{\Ginv_{n_k}(u)\}_{k\geq 1}$ that converges at all continuity points of a non-decreasing limit, whose left-continuous version is denoted by $\Ginv(u)$.  By Fatou's lemma, we have
$$\int_0^1|\Ginv(u)|^p\d u\leq \liminf_{k\to\infty}\int_0^1|\Ginv_{n_k}(u)|^p\d u<\infty.$$
Hence, $\Ginv\in\Minv_p$.

(e) As $b$ is of bounded variation, it is bounded and has one-sided limits, and $\hat b(t)=\lim_{\delta\downarrow0}\sup_{|s-t|<\delta}b(s)$ for $0<t<1$, so that $\hat b$ is upper semicontinuous on $(0,1)$. By \eqref{eq:endpoint}, $b$ is continuous at $0$ and $1$, so that $\hat b$ is upper semicontinuous on $[0,1]$. Given a partition $0\leq t_1<\cdots<t_n\leq1$, approximate each $\hat b(t_i)$ by $b(s_i)$ with $s_i\to t_i$ from the appropriate side, keeping the order of the $s_i$; letting $s_i\to t_i$ in the variation sums shows that the total variation of $\hat b$ does not exceed that of $b$. Moreover, \eqref{eq:endpoint} for $\hat b$ follows by taking one-sided limits in \eqref{eq:endpoint} for $b$, after reducing $t_0$ if necessary, which gives $|\hat b(t)|\leq K_bt^\nu$ and $|\hat b(1-t)-\hat b(1)|\leq K_bt^\nu$ separately. Finally, $\mathcal E_b:=\{t\in(0,1):\hat b(t)>b(t)\}$ is contained in the set of discontinuity points of $b$, which is at most countable.
\end{proof}

\begin{lemma}\label{lem:C-identity}
Let $b\colon[0,1]\to\R$ be bounded and Borel measurable with $b(0)=0$.
\begin{enumerate}[label=(\alph*),leftmargin=*]
\item For every $G\in\mathcal M_p$, the image of the Lebesgue measure on $\{x\in\R:0<G(x)<1\}$ under $G$ is $\d\Ginv$, i.e., $\mathrm{Leb}\{x\in\R:G(x)\in A\}=\d\Ginv(A)$ for every Borel set $A\subseteq(0,1)$. If $G$ has a bounded quantile function, with $x_0:=\Ginv(0+)$, then
\[
H_b(G)=b(1)\,x_0+\int_{(0,1)}b(1-u)\,\d\Ginv(u).
\]
\item If $b$ is of bounded variation and satisfies \eqref{eq:endpoint} with $\nu>1/p$, then, for every $G\in\mathcal M_p$,
\begin{equation}\label{eq:C-anchor}
H_b(G)=b(1)\,\Ginv(1/2)+\int_{(0,1/2)}\big(b(1-u)-b(1)\big)\,\d\Ginv(u)+\int_{[1/2,1)}b(1-u)\,\d\Ginv(u),
\end{equation}
where both integrals converge absolutely.
\item Under the assumptions of (b), $H_b\big((1-\theta)\Ginv+\theta\Hinv\big)=(1-\theta)H_b(\Ginv)+\theta H_b(\Hinv)$ for all $G,H\in\mathcal M_p$ and $\theta\in[0,1]$.
\item If $b$ is absolutely continuous with $b'\in L^{p'}(0,1)$, then, for every $G\in\mathcal M_p$, both integrals in \eqref{eq:signed-choquet} converge absolutely and
\begin{equation}\label{eq:C-ac-identity}
H_b(G)=\int_0^1b'(1-u)\,\Ginv(u)\,\d u .
\end{equation}
\end{enumerate}
\end{lemma}

\begin{proof}
(a) For $0<s<t<1$, the set $\{x\in\R:s<G(x)\leq t\}$ is, up to its endpoints, the interval $[\Ginv_+(s),\Ginv_+(t))$, and thus has Lebesgue measure $\Ginv_+(t)-\Ginv_+(s)=\d\Ginv((s,t])$. As both measures are finite on compact subsets of $(0,1)$ and coincide on these intervals, they coincide on the Borel sets of $(0,1)$. If $\Ginv$ is bounded, let $x_1:=\Ginv(1-)$. By \eqref{eq:signed-choquet}, as $G=0$ on $(-\infty,x_0)$ and $G=1$ on $[x_1,\infty)$, $H_b(G)=b(1)x_0+\int_{x_0}^{x_1}b(1-G(x))\,\d x$, and the image property yields the identity.

(b) For bounded $\Ginv$, \eqref{eq:C-anchor} follows from (a), as $\d\Ginv((0,1/2))=\Ginv(1/2)-x_0$. For general $G\in\mathcal M_p$, the bound in Lemma~\ref{lem:C-basic}(d), which also holds for $\Ginv_+$, and integration by parts show that $\int_{(0,t_0)}u^\nu\,\d\Ginv(u)<\infty$: the boundary term $u^\nu|\Ginv_+(u)|\leq\lVert\Ginv\rVert_pu^{\nu-1/p}$ vanishes as $u\downarrow0$, and $\int_0^{t_0}u^{\nu-1}|\Ginv(u)|\,\d u\leq\lVert\Ginv\rVert_p\big(\int_0^{t_0}u^{(\nu-1)p'}\,\d u\big)^{1/p'}<\infty$ by H\"older's inequality, as $(\nu-1)p'>-1$ is equivalent to $\nu>1/p$. Likewise, $\int_{(1-t_0,1)}(1-u)^\nu\,\d\Ginv(u)<\infty$. By \eqref{eq:endpoint} and the boundedness of $b$, both integrals in \eqref{eq:C-anchor} converge absolutely. The measures $\d\Ginv_{(N)}$ are dominated by $\d\Ginv$ and coincide with $\d\Ginv$ on every compact subinterval of $(0,1)$ for $N$ large enough. Applying \eqref{eq:C-anchor} to $G_{(N)}$ and letting $N\to\infty$, with dominated convergence on the right-hand side and Lemma~\ref{lem:C-basic}(b) on the left-hand side, proves \eqref{eq:C-anchor} for $G$.

(c) As $\d\big((1-\theta)\Ginv+\theta\Hinv\big)=(1-\theta)\,\d\Ginv+\theta\,\d\Hinv$, this follows from the linearity of the right-hand side of \eqref{eq:C-anchor} in $\Ginv$.

(d) As $b(0)=0$, $b(1-G(x))=\int_{G(x)}^1b'(1-u)\,\d u$ for $x\geq0$, and $b(1-G(x))-b(1)=-\int_0^{G(x)}b'(1-u)\,\d u$ for $x<0$. Moreover, $G(x)<u$ if and only if $x<\Ginv(u)$, so that $\mathrm{Leb}\{x\geq0:G(x)<u\}=(\Ginv(u))_+$ and $\mathrm{Leb}\{x<0:G(x)\geq u\}=(\Ginv(u))_-$ for $0<u<1$. By Tonelli's theorem,
\begin{align*}
\int_0^\infty\big|b(1-G(x))\big|\,\d x+\int_{-\infty}^0\big|b(1-G(x))-b(1)\big|\,\d x&\leq\int_0^1|b'(1-u)|\,|\Ginv(u)|\,\d u\\
&\leq\lVert b'\rVert_{p'}\lVert\Ginv\rVert_p<\infty,
\end{align*}
by H\"older's inequality, and Fubini's theorem yields \eqref{eq:C-ac-identity}.
\end{proof}

\begin{lemma}\label{lem:C-gap}
For every $G\in\mathcal M_p$,
\begin{equation}\label{eq:general-gap-Lp}
H_{\hat g}(G)-H_g(G)=\mathcal J_g(G)=\sum_{t\in\mathcal E_g}\big(\hat g(t)-g(t)\big)\Delta_{1-t}(G)<\infty .
\end{equation}
\end{lemma}

\begin{proof}
As $\hat g(1)=g(1)$, $H_{\hat g}(G)-H_g(G)=\int_\R(\hat g-g)\big(1-G(x)\big)\,\d x$, where the integrand is non-negative and vanishes unless $1-G(x)\in\mathcal E_g$. For $0<u<1$, the set $\{x\in\R:G(x)=u\}$ is, up to its endpoints, the interval $[\Ginv(u),\Ginv_+(u))$, and thus has length $\Delta_u(G)$. Hence, the identity in \eqref{eq:general-gap-Lp} holds, and finiteness follows from Lemma~\ref{lem:C-basic}(a) and (e).
\end{proof}

\begin{lemma}\label{lem:C-patch}
Let $G\in\mathcal M_p$ and let $S\subseteq\mathcal E_g$ be finite. For $t\in S$, let  
\[
m_t:=\begin{cases}\Ginv_+(1-t),&g(t+)\geq g(t-),\\ \Ginv(1-t),&g(t+)<g(t-),\end{cases}
\]
and, for $\varrho>0$ so small that the intervals $(1-t-\varrho,1-t+\varrho]$, $t\in S$, are disjoint and contained in $(0,1)$, let $\Ginv_\varrho:=m_t$ on $(1-t-\varrho,1-t+\varrho]$, $t\in S$, and $\Ginv_\varrho:=\Ginv$ elsewhere. Then $\Ginv_\varrho\in\Minv_p$, and, as $\varrho\downarrow0$,
\[
\begin{gathered}
\lVert\Ginv_\varrho-\Ginv\rVert_p\to0,\qquad\DB_\phi(G_\varrho,F)\to\DB_\phi(G,F),\\
H_g(G_\varrho)\to H_g(G)+\sum_{t\in S}\big(\hat g(t)-g(t)\big)\Delta_{1-t}(G).
\end{gathered}
\]
\end{lemma}

\begin{proof}
 We choose $\varrho>0$  small enough such that the intervals $(1-t-\varrho,1-t+\varrho]$, $t\in S$, are disjoint and contained in $(0,1)$.
As $\Ginv(1-t)\leq m_t\leq\Ginv_+(1-t)$, $\Ginv_\varrho$ is non-decreasing, and it is left-continuous by the choice of half-open intervals. It differs from $\Ginv$ only on a set of measure at most $2|S|\varrho$, contained in a fixed compact subinterval of $(0,1)$ on which $\Ginv$ is bounded. This yields the $L^p$-convergence, and Lemma~\ref{lem:C-growth}(d) yields the convergence of the divergences.

Let $t\in S$. As $t\in\mathcal E_g$, $\hat g(t)=\max\{g(t-),g(t+)\}>g(t)$. Let $x\in\big(\Ginv(1-t),\Ginv_+(1-t)\big)$, so that $G(x)=1-t$ and $g(1-G(x))=g(t)$. Direct computation shows that $1-G_\varrho(x)=t+\varrho$ if $m_t=\Ginv_+(1-t)$ and $1-G_\varrho(x)=t-\varrho$ if $m_t=\Ginv(1-t)$. In both cases, $g(1-G_\varrho(x))\to\hat g(t)$. If
$x\notin\bigcup_{t\in S}
[\Ginv(1-t),\Ginv_+(1-t)]$,
then $G_\varrho(x)=G(x)$
for all sufficiently small $\varrho>0$. 
Hence $g(1-G_\varrho(x))$ converges for every $x$ outside the finite set $\{\Ginv(1-t),\Ginv_+(1-t):t\in S\}$, to $\hat g(t)$ on $(\Ginv(1-t),\Ginv_+(1-t))$ and to $g(1-G(x))$ elsewhere. Moreover, $G_\varrho$ and $G$ only differ on a fixed bounded interval, on which the difference of the integrands in \eqref{eq:signed-choquet} is bounded by $2\sup|g|$. As the integrands converge a.e., dominated convergence yields the claim.
\end{proof}

\begin{lemma}\label{lem:C-recovery}
Let $\ep>0$. For every $G\in\mathcal M_{\ep,p}(\phi)$, there exist $G_n\in\mathcal M_{\ep,p}(\phi)$ such that
\begin{equation}\label{eq:general-recovery-Lp}
\lVert\Ginv_n-\Ginv\rVert_p\to0\qquad\text{and}\qquad H_g(G_n)\to H_{\hat g}(G).
\end{equation}
\end{lemma}

\begin{proof}
Let $\theta_n\in(0,1)$ with $\theta_n\downarrow0$, and $R_n^{-1}:=(1-\theta_n)\Ginv+\theta_n\Finv\in\Minv_p$. As $B_\phi(\cdot,y)$ is convex with $B_\phi(y,y)=0$, $\DB_\phi(R_n,F)\leq(1-\theta_n)\ep<\ep$. Moreover, $R_n^{-1}\to\Ginv$ in $L^p(0,1)$, and $H_{\hat g}(R_n)\to H_{\hat g}(G)$ by Lemma~\ref{lem:C-identity}(c) and Lemma~\ref{lem:C-basic}(e). By Lemma~\ref{lem:C-gap}, there is a finite $S_n\subseteq\mathcal E_g$ with $0\leq\mathcal J_g(R_n)-\sum_{t\in S_n}(\hat g(t)-g(t))\Delta_{1-t}(R_n)<1/n$. As $R_n$ lies strictly inside the ball, Lemma~\ref{lem:C-patch}, applied to $R_n$ and $S_n$ with $\varrho$ small enough, yields $G_n\in\mathcal M_{\ep,p}(\phi)$ with $\lVert\Ginv_n-R_n^{-1}\rVert_p<1/n$ and $\big|H_g(G_n)-H_g(R_n)-\sum_{t\in S_n}(\hat g(t)-g(t))\Delta_{1-t}(R_n)\big|<1/n$; if $S_n=\emptyset$, let $G_n:=R_n$. Hence $|H_g(G_n)-H_{\hat g}(R_n)|<2/n$ by Lemma~\ref{lem:C-gap}, and the claim follows.
\end{proof}

\subsection{Upper semicontinuous distortion functions}\label{app:C2}

In this subsection, $T_F$, $g_\lambda$, $\overline g_\lambda$, $\bar k_\lambda$, $\Ginv_\lambda$ and $W(\lambda)$ are as in Section~\ref{sec:general-g}. 

\begin{proposition}\label{prop:C-usc}
Under the standing assumptions, let $g$ be upper semicontinuous. Then $\bar k_\lambda\in L^{p'}(0,1)$ and $\Ginv_\lambda\in\Minv_p$ for $\lambda>0$, and $d(\lambda):=\DB_\phi(G_\lambda,F)$ is finite, continuous and non-increasing on $(0,\infty)$, tends to $0$ as $\lambda\to\infty$, and is strictly decreasing if $g(1)\neq0$. Moreover, $\lambda\mapsto\Ginv_\lambda$ is continuous in $L^p(0,1)$, and:
\begin{enumerate}[label=(\roman*),leftmargin=*]
\item If $g(1)\neq0$ or $g(t)>0$ for some $t\in(0,1)$, then $d(\lambda)\to\infty$ as $\lambda\downarrow0$. For every $\ep>0$, there exists $\lambda^*>0$ with $d(\lambda^*)=\ep$, $\Ginv_{\lambda^*}$ is the unique maximizer of $H_g$ over $\mathcal M_{\ep,p}(\phi)$, and $V_g(\ep)=\lambda^*\ep+W(\lambda^*)$. The maximizer does not depend on the choice of $\lambda^*$, and $\lambda^*$ is unique if $g(1)\neq0$.
\item If $g(1)=0$ and $g\leq0$, then $H_g\leq0$ on $\Minv_p$, $\bar\ep:=\min\{\DB_\phi(G,F):\Ginv\in\Minv_p,\ H_g(\Ginv)=0\}<\infty$ is attained, and $d(\lambda)\to\bar\ep$ as $\lambda\downarrow0$. For $0<\ep<\bar\ep$, there exists $\lambda^*>0$ with $d(\lambda^*)=\ep$, and the conclusions of (i) on the maximizer, its uniqueness and the value $V_g(\ep)$ hold. For $\ep\geq\bar\ep$, $V_g(\ep)=0$, and the maximizers are exactly the $G\in\mathcal M_{\ep,p}(\phi)$ with $H_g(G)=0$.
\item For every $\ep>0$, $V_g(\ep)=\inf_{\lambda>0}\{\lambda\ep+W(\lambda)\}$, and every $\lambda^*>0$ with $d(\lambda^*)=\ep$ attains the infimum.
\item If $d(\lambda^*)=\ep$ for some $\lambda^*>0$, in either case (i) or (ii), including $\ep=\bar\ep$ in case (ii), then $\Ginv_{\lambda^*}$ is the unique maximizer of $H_g$ over $\mathcal M_{\ep,p}(\phi)$, $V_g(\ep)=\lambda^*\ep+W(\lambda^*)$, and every $\lambda'>0$ with $d(\lambda')=\ep$ satisfies $\Ginv_{\lambda'}=\Ginv_{\lambda^*}$.
\end{enumerate}
\end{proposition}

\begin{lemma}\label{lem:C-majorant}
Let $\lambda>0$. The least concave majorant $\overline g_\lambda$ of $g_\lambda$ exists, is continuous and absolutely continuous on $[0,1]$, and $\partial_+\overline g_\lambda\in L^{p'}(0,1)$. Moreover, $\overline g_\lambda(0)=0$ and $\overline g_\lambda(1)=g(1)+\lambda\int_0^1\phi'(\Finv(v))\,\d v$, and the least concave majorants of $g+\lambda T_F$ and $\hat g+\lambda T_F$ coincide. If $g$ is upper semicontinuous, then $\overline g_\lambda$ is affine on every open interval contained in $\{t\in(0,1):\overline g_\lambda(t)>g_\lambda(t)\}$.
\end{lemma}

\begin{proof}
By Lemma~\ref{lem:C-growth}(d), $\phi'(\Finv)$ is non-decreasing and in $L^{p'}(0,1)$. Hence, $T_F$ is concave and absolutely continuous, with $T_F'(t)=\phi'(\Finv(1-t))$ a.e. Choose $\rho$ with $1/p<\rho<\min\{\nu,1\}$ and let $\varsigma(t):=t^\rho+(1-t)^\rho-1$, which is concave, positive on $(0,1)$ and vanishes at $0$ and $1$. By the boundedness of $g$ and by \eqref{eq:endpoint}, as $\rho<\nu$ and $\rho<1$, the function $(g(t)-g(1)t)/\varsigma(t)$ is bounded from above on $(0,1)$, so that $\beta(t):=g(1)t+K_\beta\varsigma(t)\geq g(t)$ on $[0,1]$ for some constant $K_\beta\geq0$.  Note that $\rho>1/p$ implies $p'(1-\rho)<1$. Hence, we have $\beta'\in L^{p'}(0,1)$. Consequently, $R_\lambda:=\beta+\lambda T_F$ is a continuous concave majorant of $g_\lambda$ with $R_\lambda'\in L^{p'}(0,1)$, which coincides with $g_\lambda$ at $0$ and $1$.

By definition,  $g_\lambda\leq\overline g_\lambda\leq R_\lambda$, which together with the concavity of $\overline{g}_\lambda$ implies that $\overline{g}_\lambda$ is continuous over $[0,1]$ and  is absolutely continuous over $(0,1)$, with $\overline g_\lambda(x)-\overline g_\lambda(0)=\int_0^x\chi(t)\,\d t$, where $\chi:=\partial_+\overline g_\lambda$ is non-increasing. Let $\chi_R:=R_\lambda'$. For every $z\in\R$, monotonicity and the common endpoint values give
\begin{align*}
\int_0^1(\chi-z)_+\,\d t&=\sup_{0\leq x\leq1}\big\{\overline g_\lambda(x)-\overline g_\lambda(0)-zx\big\}\\
&\leq\sup_{0\leq x\leq1}\big\{R_\lambda(x)-R_\lambda(0)-zx\big\}=\int_0^1(\chi_R-z)_+\,\d t,\\
\int_0^1(z-\chi)_+\,\d t&=\sup_{0\leq x\leq1}\big\{z(1-x)-\overline g_\lambda(1)+\overline g_\lambda(x)\big\}\\
&\leq\sup_{0\leq x\leq1}\big\{z(1-x)-R_\lambda(1)+R_\lambda(x)\big\}=\int_0^1(z-\chi_R)_+\,\d t .
\end{align*}
As $\psi(x):=|x|^{p'}$ satisfies $\psi(x)=\int_0^\infty\big((x-z)_++(-x-z)_+\big)\,\psi''(z)\,\d z$, with $\psi''(z)=p'(p'-1)z^{p'-2}$ integrable near $0$, and as $(-x-z)_+=((-z)-x)_+$, Fubini’s theorem yields $\int_0^1|\chi|^{p'}\,\d t\leq\int_0^1|\chi_R|^{p'}\,\d t<\infty$.

 Obviously,  the least concave majorants of $g+\lambda T_F$ and $\hat g+\lambda T_F$ coincide.

If $g$ is upper semicontinuous, so is $g_\lambda$, which further implies that  $\overline g_\lambda-g_\lambda$ is lower semicontinuous.  Hence, $O:=\{\overline g_\lambda>g_\lambda\}\cap(0,1)$ is open. Every point of $O$ has a closed neighbourhood $I=[u_1,u_2]\subseteq O$ on which $\overline g_\lambda-g_\lambda$ is bounded below by some $\delta>0$ and on which, by continuity, $\overline g_\lambda$ differs from the line segment joining $(u_1,\overline g_\lambda(u_1))$ and $(u_2,\overline g_\lambda(u_2))$ by less than $\delta$. Replacing $\overline g_\lambda$ on $I$ by this line segment yields a concave majorant of $g_\lambda$, so that, by minimality, $\overline g_\lambda$ is affine on $I$, and hence on every connected component of $O$. We complete the proof.
\end{proof}

\begin{lemma}\label{lem:C-penalized}
Let $g$ be upper semicontinuous and $\lambda>0$. Then $\bar k_\lambda\in L^{p'}(0,1)$, $\Ginv_\lambda\in\Minv_p$, $W(\lambda)$ is finite, and, for every $G\in\mathcal M_p$,
\begin{equation}\label{eq:general-majorization}
H_g(G)+\lambda\big\langle \phi'(\Finv),\Ginv\big\rangle\leq\lambda\big\langle\bar k_\lambda,\Ginv\big\rangle ,
\end{equation}
with equality for every $\Ginv\in\Minv_p$ that is constant on every connected component of $\{u\in(0,1):\overline g_\lambda(1-u)>g_\lambda(1-u)\}$, in particular for $\Ginv_\lambda$. Moreover, $\Ginv_\lambda$ is the unique maximizer of $H_g(\Ginv)-\lambda\DB_\phi(G,F)$ over $\Minv_p$,
\[
W(\lambda)=\sup_{\Ginv\in\Minv_p}\big\{H_g(\Ginv)-\lambda\DB_\phi(G,F)\big\}=H_g(G_\lambda)-\lambda\DB_\phi(G_\lambda,F),
\]
and, for every $G\in\mathcal M_p$,
\begin{equation}\label{eq:general-penalty-gap}
W(\lambda)-\big(H_g(G)-\lambda\DB_\phi(G,F)\big)\geq\lambda\,\DB_\phi(G,G_\lambda) .
\end{equation}
\end{lemma}

\begin{proof}
By Lemma~\ref{lem:C-majorant}, $\bar k_\lambda\in L^{p'}(0,1)$.  It follows from Lemma~\ref{lem:C-growth}(b) that $|\Ginv_\lambda|^p\leq C(1+|\bar k_\lambda|^{p'})$, and hence $\Ginv_\lambda\in\Minv_p$. By Lemma~\ref{lem:C-growth}(c) and (d), $W(\lambda)$ is finite.

Let $G\in\mathcal M_p$. The functions $T_F$ and $\overline g_\lambda$ vanish at $0$ and are absolutely continuous, with derivatives $T_F'(1-u)=\phi'(\Finv(u))$ and $\overline g_\lambda'(1-u)=\lambda\bar k_\lambda(u)$ for a.e.\ $u$, which belong to $L^{p'}(0,1)$. By Lemma~\ref{lem:C-identity}(d), $H_{T_F}(G)=\langle \phi'(\Finv),\Ginv\rangle$ and $H_{\overline g_\lambda}(G)=\lambda\langle\bar k_\lambda,\Ginv\rangle$, with absolutely convergent integrals in \eqref{eq:signed-choquet}, and the same holds for $H_g(G)$ by Lemma~\ref{lem:C-basic}(a). Hence the signed Choquet integrals are additive in the distortion function, and, as $\overline g_\lambda$ and $g_\lambda$ coincide at $0$ and $1$, we have
\begin{align*}
\lambda\big\langle\bar k_\lambda,\Ginv\big\rangle-H_g(G)-\lambda\big\langle \phi'(\Finv),\Ginv\big\rangle&=H_{\overline g_\lambda}(G)-H_{g_\lambda}(G)=\int_\R\big(\overline g_\lambda-g_\lambda\big)\big(1-G(x)\big)\,\d x\\
&=\int_{(0,1)}\big(\overline g_\lambda-g_\lambda\big)(1-u)\,\d\Ginv(u),
\end{align*}
where the last equality follows from Lemma~\ref{lem:C-identity}(a), as the integrand vanishes if $G(x)\in\{0,1\}$. The right-hand side is non-negative, which proves \eqref{eq:general-majorization}, and it vanishes if $\d\Ginv$ puts no mass on the open set $\{u\in(0,1):\overline g_\lambda(1-u)>g_\lambda(1-u)\}$, in particular if $\Ginv$ is constant on its connected components. By Lemma~\ref{lem:C-majorant}, $\bar k_\lambda$, and hence $\Ginv_\lambda$, is constant on these connected components.

By Lemma~\ref{lem:C-growth}(c), we have $\DB_\phi(G,F)=\int_0^1\big(\phi(\Ginv)-\phi'(\Finv)\Ginv+\phi^*(\phi'(\Finv))\big)\,\d u$.  Hence, using \eqref{eq:general-majorization} and the Fenchel--Young inequality, we have
\begin{align*}
H_g(G)-\lambda\DB_\phi(G,F)&\leq\lambda\int_0^1\big(\bar k_\lambda\Ginv-\phi(\Ginv)-\phi^*(\phi'(\Finv))\big)\,\d u\\
&\leq\lambda\int_0^1\big(\phi^*(\bar k_\lambda)-\phi^*(\phi'(\Finv))\big)\,\d u=W(\lambda),
\end{align*}
with equality in both inequalities for $G=G_\lambda$, as $\phi'(\Ginv_\lambda)=\bar k_\lambda$. This proves the formula for $W(\lambda)$. Moreover,  by Lemma~\ref{lem:C-growth}(c), the difference between $W(\lambda)$ and the first upper bound equals $\lambda\int_0^1\big(\phi^*(\bar k_\lambda)-\bar k_\lambda\Ginv+\phi(\Ginv)\big)\,\d u=\lambda\int_0^1\big(\bar k_\lambda\Ginv_\lambda-\phi(\Ginv_\lambda)-\bar k_\lambda\Ginv+\phi(\Ginv)\big)\,\d u=\lambda\DB_\phi(G,G_\lambda)$, which proves \eqref{eq:general-penalty-gap}. If the left-hand side of \eqref{eq:general-penalty-gap} equals zero, then $B_\phi(\Ginv,\Ginv_\lambda)=0$, which implies  $\Ginv=\Ginv_\lambda$ a.e.\ by strict convexity of $\phi$, and then everywhere by left-continuity of $\Ginv$ and $\Ginv_\lambda$. This proves uniqueness. We complete the proof.
\end{proof}

\begin{proof}[Proof of Proposition~\ref{prop:C-usc}]
The claims of $\bar k_\lambda\in L^{p'}(0,1)$ and $\Ginv_\lambda\in\Minv_p$  follow from Lemma~\ref{lem:C-penalized}, and the finiteness of $d(\lambda)$ follows from Lemma \ref{lem:C-growth} (d).

\emph{Step 1: monotonicity.} Applying \eqref{eq:general-penalty-gap} at $\lambda$ with $G=G_\mu$ and at $\mu$ with $G=G_\lambda$ and adding the two inequalities gives
\begin{align}\label{eq:general-multiplier-gap}
&W(\lambda)-\big(H_g(G_\mu)-\lambda\DB_\phi(G_\mu,F)\big)+W(\mu)-\big(H_g(G_\lambda)-\mu\DB_\phi(G_\lambda,F)\big)\nonumber\\
&=(\mu-\lambda)\big(d(\lambda)-d(\mu)\big)\geq\lambda\,\DB_\phi(G_\mu,G_\lambda)+\mu\,\DB_\phi(G_\lambda,G_\mu)\geq0 .
\end{align}
Hence $d$ is non-increasing, and strictly decreasing between any two values of $\lambda$ for which the quantile functions $\Ginv_\lambda$ differ. By Lemma~\ref{lem:C-majorant},
\begin{equation}\label{eq:C-mean}
\int_0^1\phi'\big(\Ginv_\lambda(u)\big)\,\d u=\int_0^1\bar k_\lambda(u)\,\d u=\frac{\overline g_\lambda(1)-\overline g_\lambda(0)}\lambda=\int_0^1\phi'(\Finv(u))\,\d u+\frac{g(1)}\lambda ,
\end{equation}
so that, if $g(1)\neq0$, then $\Ginv_\lambda$ are pairwise different and $d$ is strictly decreasing.

\emph{Step 2: continuity and the limit at $\infty$.} Using Lemma~\ref{lem:C-penalized} at $G=F$, we have 
\[H_g(G_\lambda)-H_g(F)-\lambda d(\lambda)=W(\lambda)-H_g(F)\geq \lambda\DB_\phi(F,G_\lambda)\geq0 ,\]
which together with \eqref{eq:general-H-bound} gives
\[
\lambda\,d(\lambda)\leq H_g(G_\lambda)-H_g(F)\leq C_g\lVert\Ginv_\lambda\rVert_p+|H_g(F)| .
\]
Together with \eqref{eq:general-coercivity} for $H=F$, we have
\[
\lambda\,d(\lambda)\leq C_g\left(\frac{2(C_F+d(\lambda_0))}{a_\phi}\right)^{1/p}+|H_g(F)|
\]
for $\lambda$ over $[\lambda_0,\infty)$ with any $\lambda_0>0$, which implies $d(\lambda)\to0$ as $\lambda\to\infty$. For $\mu\to\lambda>0$, \eqref{eq:general-multiplier-gap} gives $\DB_\phi(G_\mu,G_\lambda)\to0$, so that $\Ginv_\mu\to\Ginv_\lambda$ in $L^p(0,1)$ by \eqref{eq:general-fixed-base}, and $d$ is continuous by Lemma~\ref{lem:C-growth}(d).

\emph{Step 3: the limit at $0$ in case (i).} If $g(1)\neq0$, then $H_g(x)=x\,g(1)$ for constant quantile functions $x\in\R$; if $g(t)>0$ for some $t\in(0,1)$, then $H_g(x\Id_{(1-t,1)})=x\,g(t)$ for quantile functions $x\Id_{(1-t,1)}$ with $x>0$. In both cases, $H_g$ is unbounded from above on bounded quantile functions, which have finite divergence by Lemma~\ref{lem:C-growth}(d). Suppose that $d(\lambda)$ has a finite limit as $\lambda\downarrow0$. Then, by \eqref{eq:general-coercivity} and \eqref{eq:general-H-bound}, \[H_g(G_\lambda)\leq C_g\left(\frac{2(C_F+d(\lambda))}{a_\phi}\right)^{1/p}\leq M\] for some $M<\infty$ and all $\lambda>0$. Choose a bounded $\Hinv\in\Minv_p$ with $H_g(H)>M+1$. By Lemma~\ref{lem:C-penalized}, $H_g(G_\lambda)\geq H_g(G_\lambda)-\lambda d(\lambda)\geq H_g(H)-\lambda\DB_\phi(H,F)$, which implies $H_g(G_\lambda)>M$ for small $\lambda$, leading to a contradiction. Hence $d(\lambda)\to\infty$ as $\lambda\downarrow0$, and, by Steps~1 and~2 and the intermediate value theorem,  we have the existence of  $\lambda^*>0$ satisfying $d(\lambda^*)=\ep$ for every $\ep>0$.

\emph{Step 4: optimality.} If $d(\lambda^*)=\ep$, then by Lemma \ref{lem:C-penalized}, $W(\lambda^*)=H_g(G_{\lambda^*})-\lambda^*\epsilon$. Hence, for every $G\in\mathcal M_{\ep,p}(\phi)$, by \eqref{eq:general-penalty-gap},
\[
\begin{split}
H_g(G_{\lambda^*})-H_g(G)&=W(\lambda^*)-\big(H_g(G)-\lambda^*\DB_\phi(G,F)\big)+\lambda^*\big(\ep-\DB_\phi(G,F)\big)\\
&\geq\lambda^*\,\DB_\phi(G,G_{\lambda^*}).
\end{split}
\]
Hence $\Ginv_{\lambda^*}$ is the unique maximizer, by strict convexity of $\phi$, and $V_g(\ep)=H_g(G_{\lambda^*})=\lambda^*\ep+W(\lambda^*)$. If $d(\lambda')=d(\lambda^*)$, then  by \eqref{eq:general-multiplier-gap}, we have $\Ginv_{\lambda'}=\Ginv_{\lambda^*}$. If additionally $g(1)\neq0$,  then by \eqref{eq:C-mean}, we further have $\lambda'=\lambda^*$. As this argument applies to every $\lambda^*>0$ with $d(\lambda^*)=\ep$, this proves (iv) and, with Step~3, (i).

\emph{Step 5: the case $g(1)=0$ and $g\leq0$.} By Lemma~\ref{lem:C-identity}(a), $H_g(G)=\int_{(0,1)}g(1-u)\,\d\Ginv(u)\leq0$ for bounded $\Ginv$, and Lemma~\ref{lem:C-basic}(b) extends this to $\Minv_p$. Constant quantile functions have objective $0$ and finite divergence, so that $\bar\ep$ is finite. Let $\{G_n\}_{n\geq 1}$ be a sequence in $\mathcal M_p$ such that $H_g(G_n)=0$ and $\lim_{n\to\infty}\DB_\phi(G_n,F)=\bar\ep$. By \eqref{eq:general-coercivity}, it is $L^p$-bounded, and by Lemma~\ref{lem:C-basic}(d), there exists a subsequence converging  a.e.\ to some $\Ginv_0\in\Minv_p$. By Fatou's lemma, we have $\DB_\phi(G_0,F)\leq\bar\ep$. Moreover, it follows from Lemma~\ref{lem:C-basic}(c) that $H_g(G_0)\geq0$, implying that $H_g(G_0)=0$ and the minimum is attained at $G_0$, i.e., $\DB_\phi(G_0,F)=\bar\ep$. Comparing $G_\lambda$ with $G_0$ in Lemma~\ref{lem:C-penalized}, we have $H_g(G_\lambda)-\lambda d(\lambda)\geq H_g(G_0)-\lambda\DB_\phi(G_0,F)$. Using $H_g(G_\lambda)\leq0$, $H_g(G_0)=0$ and $\DB_\phi(G_0,F)\leq\bar\ep$, we have
\begin{equation}\label{eq:C-exceptional}
d(\lambda)\leq\bar\ep,\qquad0\leq-H_g(G_\lambda)\leq\lambda\,\bar\ep,
\end{equation}
which implies $d_0:=\lim_{\lambda\downarrow0}d(\lambda)\leq\bar\ep$. By \eqref{eq:general-coercivity} and Lemma~\ref{lem:C-basic}(d), a subsequence of $\{\Ginv_{\lambda_n}\}_{n\geq 1}$ with $\lambda_n\downarrow 0$ converges a.e.\ to some $\Ginv_\infty\in\Minv_p$. Moreover,  by Lemma~\ref{lem:C-basic}(c) and \eqref{eq:C-exceptional}, we have $H_g(G_\infty)=0$; and  by Fatou's lemma, we have $\DB_\phi(G_\infty,F)\leq d_0$.  Hence, $\bar\ep\leq d_0$, implying $d_0=\bar\ep$. For $0<\ep<\bar\ep$, Step~2 and the intermediate value theorem yield a $\lambda^*>0$ with $d(\lambda^*)=\ep$, and Step~4 applies. For $\ep\geq\bar\ep$, $G_0\in\mathcal M_{\ep,p}(\phi)$ attains the upper bound $0$, which proves (ii).

\emph{Step 6: the dual formula.} For every $G\in\mathcal M_{\ep,p}(\phi)$ and $\lambda>0$, $H_g(G)\leq H_g(G)+\lambda(\ep-\DB_\phi(G,F))\leq\lambda\ep+W(\lambda)$, so that $V_g(\ep)\leq\inf_{\lambda>0}\{\lambda\ep+W(\lambda)\}$. If $d(\lambda^*)=\ep$, equality holds at $\lambda^*$ by Step~4. Otherwise, $g(1)=0$, $g\leq0$ and $\ep\geq\bar\ep$. Using $G_0$ given in Step 5,  we have  $W(\lambda)\geq H_g(G_0)-\lambda \DB_\phi(G_0,F)\geq -\lambda \bar\ep$, implying  $-\lambda\bar\ep\leq W(\lambda)\leq0$. Hence,  $\lambda\ep+W(\lambda)\to0=V_g(\ep)$ as $\lambda\downarrow0$. This proves (iii). We complete the proof.
\end{proof}

\subsection{Proofs of the results in Section~\ref{sec:general-g} and of Corollary~\ref{cor:general-ac}}\label{app:C3}

\begin{proof}[Proof of Theorem~\ref{thm:general}]
By Lemma~\ref{lem:C-basic}(e), $\hat g$ satisfies the assumptions of Proposition~\ref{prop:C-usc}. It follows from  Proposition~\ref{prop:C-usc}(iii) that $V_{\hat{g}}(\ep)=\inf_{\lambda>0}\{\lambda\ep+W(\lambda)\}$. Moreover, by  Proposition~\ref{prop:C-usc}(i) or (ii), we have the existence of $G^*\in \mathcal M_{\ep,p}(\phi)$ such that $H_{\hat g}(G^*)=V_{\hat{g}}(\epsilon)$.  As $H_g\leq H_{\hat g}$, we have $V_g(\ep)\leq V_{\hat{g}}(\ep)$. Moreover, applying Lemma~\ref{lem:C-recovery} for $G=G^*$, we have $V_g(\ep)=V_{\hat{g}}(\ep)$. This proves (i). 

If $\DB_\phi(G_{\lambda^*},F)=\ep$, Proposition~\ref{prop:C-usc}(iv), applied to $\hat g$, shows that $\Ginv_{\lambda^*}$ is the unique maximizer of $H_{\hat g}$ over $\mathcal M_{\ep,p}(\phi)$ and that $V_g(\ep)=H_{\hat{g}}(G_{\lambda^*})=\lambda^*\ep+W(\lambda^*)$. As $\hat g(1)=g(1)$, and as $\hat g(t)>0$ for some $t\in(0,1)$ if and only if $g(t')>0$ for some $t'\in(0,1)$, such a $\lambda^*$ exists for every $\ep>0$ under the stated condition, by Proposition~\ref{prop:C-usc}(i).

For every $G\in\mathcal M_{\ep,p}(\phi)$, Lemma~\ref{lem:C-gap} and the first equality in \eqref{eq:general-value} give
\[
V_g(\ep)-H_g(G)=\Big(\max_{H\in\mathcal M_{\ep,p}(\phi)}H_{\hat g}(H)-H_{\hat g}(G)\Big)+\mathcal J_g(G),
\]
where both terms on the right-hand side are non-negative. Hence $G$ maximizes $H_g$ over $\mathcal M_{\ep,p}(\phi)$ if and only if both terms in the right-hand side vanish. Under the assumption of (ii), the first term vanishes only for $G=G_{\lambda^*}$, and, as $\hat g(t)-g(t)>0$ for every $t\in\mathcal E_g$, $\mathcal J_g(G_{\lambda^*})=0$ is equivalent to \eqref{eq:general-attain}. This proves (ii).
\end{proof}

\begin{proof}[Proof of Proposition~\ref{prop:general-multiplier}]
As $\Ginv_\lambda$ is the same for $g$ and $\hat g$ (Lemma~\ref{lem:C-majorant}), the statements about  $d$ follow from Proposition~\ref{prop:C-usc}, applied to $\hat g$, as the conditions on $g(1)$ and on the sign of $g$ are the same for $g$ and $\hat g$; if $g(1)=0$ and $g\leq0$, then $H_g\leq H_{\hat g}\leq0$ on $\Minv_p$. In that case, $V_g(\ep)=\max_{G\in\mathcal M_{\ep,p}(\phi)}H_{\hat g}(G)=0$ for $\ep\geq\bar\ep$, by Theorem~\ref{thm:general}(i). We complete the proof.
\end{proof}

\begin{proof}[Proof of Proposition~\ref{prop:general-stability}]
By Theorem~\ref{thm:general}(ii) and Lemma~\ref{lem:C-gap},
\[
V_g(\ep)-H_g(G)=W(\lambda^*)-\big(H_{\hat g}(G)-\lambda^*\DB_\phi(G,F)\big)+\lambda^*\big(\ep-\DB_\phi(G,F)\big)+\mathcal J_g(G).
\]
Applying \eqref{eq:general-penalty-gap} to $\hat g$ gives 
\begin{equation*}\label{eq:general-p-stability}
V_g(\ep)-H_g(G)\geq\lambda^*\DB_\phi(G,G_{\lambda^*})
+\lambda^*\big(\ep-\DB_\phi(G,F)\big)+\mathcal J_g(G).
\end{equation*}
The three terms on the right-hand side are non-negative for $G\in\mathcal M_{\ep,p}(\phi)$. For a maximizing sequence, they tend to $0$, so that $\DB_\phi(G_n,F)\to\ep$, $\mathcal J_g(G_n)\to0$ and $\DB_\phi(G_n,G_{\lambda^*})\to0$, and the latter implies $\Ginv_n\to\Ginv_{\lambda^*}$ in $L^p(0,1)$ by \eqref{eq:general-fixed-base}. We complete the proof.
\end{proof}

\begin{proof}[Proof of Corollary~\ref{cor:general-ac}]
As $g$ is continuous, $\hat g=g$ and $\mathcal E_g=\emptyset$. As $\gamma\in L^2(0,1)$, $H_g(\Ginv)=\langle\gamma,\Ginv\rangle$ for $\Ginv\in\Minv$ (Section~\ref{sec:prelim}), so that, with $p=2$, \eqref{eq:general-majorization} reads $\langle k_\lambda-\bar k_\lambda,\Ginv\rangle\leq0$ for all $\Ginv\in\Minv$. By Lemma~\ref{lem:C-penalized}, $\bar k_\lambda\in\Minv$, and, as $\bar k_\lambda$ is constant on every connected component of $\{u\in(0,1):\overline g_\lambda(1-u)>g_\lambda(1-u)\}$, equality holds in \eqref{eq:general-majorization} for $\Ginv=\bar k_\lambda$, so that $\langle k_\lambda-\bar k_\lambda,\bar k_\lambda\rangle=0$. Hence $\langle k_\lambda-\bar k_\lambda,h-\bar k_\lambda\rangle\leq0$ for every $h\in\Minv$. For the projection $k_\lambda^{\uparrow}$, it follows that $\langle k_\lambda- k_\lambda^{\uparrow},h- k_\lambda^{\uparrow}\rangle\leq 0$ for every $h\in\Minv$. Combining those two inequalities gives $\langle \overline{k}_\lambda- k_\lambda^{\uparrow}, \overline{k}_\lambda- k_\lambda^{\uparrow}\rangle\leq 0$, implying  $\bar k_\lambda=k_\lambda^\uparrow$. The remaining claims follow from Theorem~\ref{thm:general} and Proposition~\ref{prop:general-multiplier} with $p=2$, as $\mathcal E_g=\emptyset$.
\end{proof}

\subsection{Remaining proofs for Sections~\ref{sec:ac} and~\ref{sec:shape}}\label{app:C4}

\begin{proof}[Proof of Lemma~\ref{lem:iso}]
Let $\mathcal C$ denote the closed convex cone of non-decreasing functions in $L^2(0,1)$, let $I:=\psi'(J)$, which is an open interval, and let $r:=k-k^\uparrow$. As $k^\uparrow$ is the $L^2$-projection of $k$ onto $\mathcal C$, it holds that $\langle r,h-k^\uparrow\rangle\leq0$ for all $h\in\mathcal C$. Choosing $h=k^\uparrow+h_0$ with $h_0\in\mathcal C$, and $h\in\{0,2k^\uparrow\}$, we obtain
\begin{equation}\label{eq:iso-proj}
\text{(a)}\ \ \langle r,h_0\rangle\leq0\ \text{ for all } h_0\in\mathcal C,
\qquad
\text{(b)}\ \ \langle r,k^\uparrow\rangle=0 .
\end{equation}

\emph{Step 1.} We show that $\langle r,f(k^\uparrow)\rangle=0$ for every continuous non-decreasing $f:I\to\R$ with $f(k^\uparrow)\in L^2(0,1)$. First, let $f$ be bounded and Lipschitz continuous with constant $L$. As $f(k^\uparrow)\in\mathcal C$, (a) gives $\langle r,f(k^\uparrow)\rangle\leq0$. For $0<t<1/L$, the map $x\mapsto x-tf(x)$ is non-decreasing, so that $k^\uparrow-tf(k^\uparrow)\in\mathcal C$, and (a) and (b) give $-t\langle r,f(k^\uparrow)\rangle\leq0$. Hence $\langle r,f(k^\uparrow)\rangle=0$. Next, let $f$ be bounded. The functions $f_n(x):=\inf_{y\in I}\{f(y)+n|x-y|\}$, $x\in I$, are bounded, non-decreasing and $n$-Lipschitz, and increase pointwise to $f$ by continuity of $f$. Since $r\in L^1(0,1)$, dominated convergence yields $\langle r,f(k^\uparrow)\rangle=\lim_n\langle r,f_n(k^\uparrow)\rangle=0$. Finally, for general $f$, apply the previous case to the truncations $f_N:=(f\wedge N)\vee(-N)$ and use dominated convergence with the integrable bound $|r|\,|f(k^\uparrow)|$.

\emph{Step 2.} The function $(\psi')^{-1}:I\to\R$ is continuous and strictly increasing. As $k^\uparrow$ is non-decreasing and left-continuous, so is $\Ginv_*$. Since $\Ginv_*\in L^2(0,1)$, we have $\Ginv_*\in\Minv$. Let $\Ginv\in\Minv$. By convexity of $\psi$,
\[
\psi(\Ginv)-\psi(\Ginv_*)-\psi'(\Ginv_*)(\Ginv-\Ginv_*)\geq0,
\qquad\text{where}\quad \psi'(\Ginv_*)=k^\uparrow .
\]
Hence $\ell_{\Ginv}\geq-r\,(\Ginv-\Ginv_*)$, which is integrable as $r,\Ginv,\Ginv_*\in L^2(0,1)$, and
\[
\int_0^1\ell_{\Ginv}(u)\,\d u\ \geq\ -\langle r,\Ginv\rangle+\langle r,\Ginv_*\rangle\ \geq\ 0,
\]
by (a) and  Step~1 with $f=(\psi')^{-1}$. If $\int_0^1\ell_{\Ginv}(u)\,\d u=0$, then the non-negative function $\psi(\Ginv)-\psi(\Ginv_*)-\psi'(\Ginv_*)(\Ginv-\Ginv_*)$ has zero integral and thus vanishes a.e. By strict convexity of $\psi$, $\Ginv=\Ginv_*$ a.e., and since both functions are non-decreasing and left-continuous, $\Ginv=\Ginv_*$ on $(0,1)$.
\end{proof}

\begin{proof}[Proof of Theorem~\ref{thm:WC-RM-BW}]
Let $\lambda>0$. Since
\[
\DB_\phi(G,F)=\int_0^1\Big(\phi\big(\Ginv(u)\big)-\phi\big(\Finv(u)\big)-\phi'\big(\Finv(u)\big)\big(\Ginv(u)-\Finv(u)\big)\Big)\,\d u,
\]
maximizing the Lagrangian $H_g(\Ginv)-\lambda\DB_\phi(G,F)$ over $\Minv$ is, up to terms that do not depend on $\Ginv$, equivalent to
\begin{equation}
\label{eq1aa}
    \argmin_{\Ginv \in \Minv} \;
        \textstyle\int_0^1
     \left(\phi\big(\Ginv(u)\big)
    - k_\lambda(u)\Ginv(u)\right)\,\d u\,.
\end{equation}
This reduction is justified as follows. Let $\Ginv\in\Minv$ with $\DB_\phi(G,F)<\infty$. As $\Finv,\Ginv,\phi'(\Finv),\gamma\in L^2(0,1)$, the functions $\phi'(\Finv)\Finv$, $\phi'(\Finv)\Ginv$ and $\gamma\Ginv$ are integrable by the Cauchy--Schwarz inequality. By convexity of $\phi$, $\phi(0)+\phi'(0)x\leq\phi(x)\leq\phi(0)+x\phi'(x)$ for $x\in\R$, where the second inequality follows from $\phi(0)\geq\phi(x)-x\phi'(x)$. Hence $|\phi(\Finv)|\leq|\phi(0)|+|\phi'(0)|\,|\Finv|+|\Finv|\,|\phi'(\Finv)|$, so that $\phi(\Finv)\in L^1(0,1)$. For a generator defined on an open interval $J$, the same argument uses a fixed $a\in J$ and the bounds $\phi(a)+\phi'(a)(x-a)\leq\phi(x)\leq\phi(a)+\phi'(x)(x-a)$ for $x\in J$. The pointwise identity
\[
\phi(\Ginv)=B_\phi(\Ginv,\Finv)+\phi(\Finv)+\phi'(\Finv)(\Ginv-\Finv)
\]
then shows that $\phi(\Ginv)\in L^1(0,1)$. Consequently, all integrals in the above representation of $\DB_\phi(G,F)$ are finite, and
\begin{equation}\label{eq:lagr-identity}
\begin{split}
H_g(\Ginv)-\lambda\DB_\phi(G,F)={}&-\lambda\int_0^1\Big(\phi\big(\Ginv(u)\big)-k_\lambda(u)\Ginv(u)\Big)\,\d u\\
&+\lambda\int_0^1\Big(\phi\big(\Finv(u)\big)-\phi'\big(\Finv(u)\big)\Finv(u)\Big)\,\d u,
\end{split}
\end{equation}
where the last integral is finite and does not depend on $\Ginv$. If $\DB_\phi(G,F)=\infty$, the Lagrangian equals $-\infty$, and, by the same pointwise identity, $\phi(\Ginv)$ is the sum of a non-negative function with infinite integral and an integrable function, so that the objective in \eqref{eq1aa} equals $+\infty$. Such $\Ginv$ are therefore irrelevant for both problems, and maximizing the Lagrangian over $\Minv$ is equivalent to \eqref{eq1aa}, provided that the solution of \eqref{eq1aa} has a finite divergence, which holds for $\Ginv_\lambda$ below by assumption.

By Lemma~\ref{lem:iso}, applied with $\psi=\phi$ and $k=k_\lambda$, the unique solution to \eqref{eq1aa} is
\begin{equation*}
    \Ginv_{\lambda}(u)
    :=
        (\phi')^{-1}({k^{\uparrow}_{\lambda}(u)}),\qquad 0<u<1.
\end{equation*}
Note that the assumptions on $\gamma(u)$ and $\Ginv_{\lambda}(u)$ imply that $H_g(\Ginv_{\lambda})< \infty$. Next we show that there exists $\lambda^*>0$ such that $\DB_\phi(G_{\lambda^*},F)=\ep$.

To this end, we first show that $\lambda \rightarrow  k^{\uparrow}_\lambda$ is decreasing and continuous w.r.t. the $L^2-$ topology. The property of being decreasing follows from the fact that  $$k^{\uparrow}_\lambda(u) = \inf_{s \geq u} \sup_{r \leq u}  \frac{1}{s - r} \int_r^s k_\lambda(t) \ dt,$$ see  \citet{mammen1991estimating}. 
As for continuity, let us consider any sequence $e_n$ with $e_n \to 0$, as $n \to +\infty$ and denote $1/\lambda_n = 1/\lambda + e_n$. 
From Theorem 2.3 in \citet{Brunk1965AMS}, 
\begin{align*}
\lVert k^\uparrow_{\lambda_n} - k^\uparrow_\lambda\rVert^2 
& \le
\lVert k_{\lambda_n} - k_\lambda \rVert^2
=
{|e_n|^2}\, \lVert \gamma \rVert^2 
\end{align*}
and thus 

\begin{align*}
\lVert k^\uparrow_{\lambda_n} - k^\uparrow_\lambda\rVert^2 
\to 0, \quad \text{for} \quad n \to +\infty, 
\end{align*}
i.e., $\lambda \rightarrow k^{\uparrow}_\lambda$ is continuous. 

Next we show that for every monotone sequence $\lambda_n$ with $\lambda_n\to\lambda$ as $n\to+\infty$, it holds for almost all $t$ that 
${k^\uparrow_{\lambda_n}(t)}
\to k^\uparrow_{\lambda}(t), \quad \text{for} \quad n \to +\infty.$ 
Assume first that $\lambda_n$ is {a decreasing} sequence. $L^2-$ continuity of $k^{\uparrow}_\lambda$ implies there exists a subsequence $\lambda_{n_j}$ with $\lambda_{n_j}\to\lambda$ as $j\to+\infty$ such that for almost every $t$,  
\begin{align*}
{k^\uparrow_{\lambda_{n_j}}(t)}
\to k^\uparrow_{\lambda}(t), \quad \text{for} \quad j \to +\infty. 
\end{align*}
As the $k^\uparrow_{\lambda_{n}}(t)$ form an increasing sequence, it follows that also 
\begin{align*}
{k^\uparrow_{\lambda_n}(t)}
\to k^\uparrow_{\lambda}(t), \quad \text{for} \quad n \to +\infty. 
\end{align*}
A similar reasoning applies when $\lambda_n$ is an increasing sequence. 

As a consequence, it follows that also 
for every monotone sequence $\lambda_n$ with $\lambda_n \to \lambda$ as $n \to +\infty$, it holds for almost every $t$ that
\begin{align*}
B_{\phi}(\Ginv_{\lambda_n}(t),\Finv(t))
\to B_{\phi}(\Ginv_\lambda(t),\Finv(t)), \quad \text{for} \quad n \to +\infty. 
\end{align*}

Moreover, since $k_\lambda\geq \phi'(\Finv)$ a.e. and the isotonic projection is order-preserving, $k_\lambda^\uparrow\geq \phi'(\Finv)$ and hence $\Ginv_\lambda\geq\Finv$. Likewise, if $\lambda_1<\lambda_2$, then $\Ginv_{\lambda_1}\geq\Ginv_{\lambda_2}$ a.e. Therefore $B_\phi(\Ginv_\lambda,\Finv)$ is pointwise non-increasing in $\lambda$. 

As the ${\DB_\phi( G_{\lambda},F)}$ are finite and decreasing in $\lambda$ (note that $\gamma(u)$ is non-negative), it follows that 
$$ {\DB_\phi( G_{\lambda},F)} =\lim_{\lambda_n \to \lambda } {\DB_\phi( G_{\lambda_n},F)} $$ 
i.e., 
$ \lambda \rightarrow \DB_\phi( G_{\lambda},F)$ is continuous; as this function is monotone in $\lambda$, continuity along monotone sequences suffices.  To see this, if $\lambda_n\downarrow\lambda$, one can
apply the monotone convergence theorem and if $\lambda_n\uparrow\lambda$,
one can invoke  dominated convergence. 


 Moreover, $k_\lambda ={\phi^\prime (\Finv)} +\frac{1}{\lambda}\gamma \rightarrow {\phi^\prime (\Finv)}$ for $\lambda \rightarrow \infty$  and since ${\phi^\prime (\Finv)}$ is non-decreasing, it is its own isotonic projection. As the isotonic projection is non-expansive, $k^\uparrow_\lambda \rightarrow \phi^\prime (\Finv)$ in $L^2$ for $\lambda \rightarrow \infty$, and since $k^\uparrow_\lambda$ is monotone in $\lambda$, this convergence also holds a.e. Hence, by dominated convergence (using any fixed finite $\lambda_0$ as a dominating divergence for $\lambda\geq\lambda_0$),  
$$ {\lim_{\lambda \to \infty }{\DB_\phi( G_{\lambda},F)} = \DB_\phi( F,F) =0}.$$ Furthermore, as per assumption, we also have that $$\lim_{\lambda \to 0 } {\DB_\phi( G_{\lambda},F)} = \infty.$$ Hence, existence of $\lambda^*$ follows. 


Furthermore, $\lambda_1 < \lambda_2$ implies that   
$H_g(\Ginv_{\lambda_1}) \ge H_g(\Ginv_{\lambda_2})$ and  $\DB_\phi( G_{\lambda_1},F) > \DB_\phi( G_{\lambda_2},F)$. To show this, note that since the isotonic projection preserves integrals it holds that  
\[
\begin{aligned}
\int_0^1 \left(k_{\lambda_1}^{\uparrow}(u)
      - k_{\lambda_2}^{\uparrow}(u)\right)\,du
&=
\int_0^1 \left(k_{\lambda_1}(u)-k_{\lambda_2}(u)\right)\,du  \\
&=
\left(\frac{1}{\lambda_1}-\frac{1}{\lambda_2}\right)
\int_0^1 \gamma(u)\,du  \\
&>0.
\end{aligned}
\]
Therefore, 
\[
k_{\lambda_1}^{\uparrow}(u)>k_{\lambda_2}^{\uparrow}(u)
\]
on a set of strictly positive measure. Since \((\phi')^{-1}\) is strictly increasing, it follows that
\[
{\Ginv_{\lambda_1}(u)>\Ginv_{\lambda_2}(u)}
\]
on a set of strictly positive measure and thus  $H_g(\Ginv_{\lambda_1}) \ge H_g(\Ginv_{\lambda_2}).$

{In fact, this inequality is strict. Put $t_i:=1/\lambda_i$ and, for $\Ginv\in\Minv$ with $\DB_\phi(G,F)<\infty$, let
\[
J_t(\Ginv):=\DB_\phi(G,F)-t\,H_g(\Ginv),
\]
which is finite. For such $\Ginv$, $J_{t_i}(\Ginv)-J_{t_i}(\Ginv_{\lambda_i})=\int_0^1\ell_{\Ginv}(u)\,\d u$, where $\ell_{\Ginv}$ is defined in Lemma~\ref{lem:iso} with $\psi=\phi$ and $k=k_{\lambda_i}=\phi'(\Finv)+t_i\gamma$. Hence $\Ginv_{\lambda_i}$ is the unique minimizer of $J_{t_i}$. As $\Ginv_{\lambda_1}\neq\Ginv_{\lambda_2}$, we have
$J_{t_1}(\Ginv_{\lambda_1})<J_{t_1}(\Ginv_{\lambda_2})$ and
$J_{t_2}(\Ginv_{\lambda_2})<J_{t_2}(\Ginv_{\lambda_1})$. Adding these inequalities yields
\[
(t_1-t_2)\bigl(H_g(\Ginv_{\lambda_1})-H_g(\Ginv_{\lambda_2})\bigr)>0,
\]
and hence $H_g(\Ginv_{\lambda_1})>H_g(\Ginv_{\lambda_2})$.}

Furthermore, both \({\Ginv_{\lambda_1}}\) and \({\Ginv_{\lambda_2}}\) are bigger than \({\Finv}\) and thus since \(\phi\) is strictly convex, it follows that a.e.
\[
B_\phi\bigl({\Ginv_{\lambda_1}(u)},{\Finv(u)}\bigr)
\ge
B_\phi\bigl({\Ginv_{\lambda_2}(u)},{\Finv(u)}\bigr)
\]
with strict inequality on a set with strictly positive measure. Hence,
\[
{\DB_\phi(G_{\lambda_1},F)}
>
{\DB_\phi(G_{\lambda_2},F)}.
\]

This allows us to conclude that $G_{\lambda^*}$ is an optimal solution to problem \eqref{opt:WC-RMtoto}, where $\lambda^*>0$ is such that  $\DB_\phi(G_{\lambda^*},F)=\ep$. Indeed, since $\Ginv_{\lambda^*}$ uniquely maximizes $H_g(\Ginv)-\lambda^*\DB_\phi(G,F)$ over $\Minv$, it holds for every feasible $\Hinv\in\Minv$ that $H_g(\Hinv)\leq H_g(\Hinv)-\lambda^*\bigl(\DB_\phi(H,F)-\ep\bigr)\leq H_g(\Ginv_{\lambda^*})-\lambda^*\bigl(\DB_\phi(G_{\lambda^*},F)-\ep\bigr)=H_g(\Ginv_{\lambda^*})$; in particular, an optimal solution exists. Moreover, if a feasible $\Hinv\in\Minv$ satisfies $H_g(\Hinv)=H_g(\Ginv_{\lambda^*})$, then both inequalities above are equalities, so $\Hinv$ also maximizes $H_g(\Ginv)-\lambda^*\DB_\phi(G,F)$ over $\Minv$, and thus $\Hinv=\Ginv_{\lambda^*}$. Hence the solution is unique, and $\lambda^*$ is unique as $\DB_\phi(G_\lambda,F)$ is strictly decreasing in $\lambda$.

\end{proof}

\begin{proof}[Verification of Remark~\ref{rem:sufficient}]
Since $\Finv\in L^\infty(0,1)$ and $\phi'$ is continuous, it follows that
$\phi'(\Finv)\in L^\infty(0,1)$. Hence, for every $\lambda>0$,
\[
k_\lambda
=
\phi'(\Finv)+\frac1\lambda\gamma
\in L^\infty(0,1).
\]
The isotonic projection of a bounded function is again bounded, and thus we also have that
$k_\lambda^\uparrow\in L^\infty(0,1)$. Since $\phi'$ has range $\mathbb R$,
$(\phi')^{-1}$ is well-defined on $\mathbb R$, and thus
\[
\Ginv_\lambda
=
(\phi')^{-1}(k_\lambda^\uparrow)
\]
is well-defined. Moreover, since $k_\lambda^\uparrow$ is bounded and
$(\phi')^{-1}$ is continuous on bounded intervals, we obtain
$\Ginv_\lambda\in L^\infty(0,1)\subset L^2(0,1)$.

Furthermore, since $\Ginv_\lambda$ and $\Finv$ are bounded and $\phi$ and
$\phi'$ are continuous on bounded intervals, it holds that
\[
\DB_\phi(G_\lambda,F)<\infty .
\]

Let $\Gamma:=\|\gamma\|_\infty$ and let $C>0$ be such that
$|\phi'(\Finv(u))|\le C$ a.e. Since isotonic projections preserve bounds, we have
\[
-C\le{} k_\lambda^\uparrow(u)\le C+\frac{\Gamma}{\lambda}
\qquad\text{a.e.}
\]
Moreover, since isotonic projection preserves integrals,
\[
\int_0^1 k_\lambda^\uparrow(u)\,du
=
\int_0^1 \phi'(\Finv(u))\,du+\frac1\lambda .
\]
Note that $\Gamma\geq\int_0^1\gamma(u)\,\d u=1$. Let $\mu$ denote the Lebesgue measure on $(0,1)$ and $A_\lambda:=\{u:k_\lambda^\uparrow(u)\geq 1/(2\lambda)\}$. By the bounds above,
\[
-C+\frac1\lambda\leq\int_0^1 k_\lambda^\uparrow(u)\,\d u\leq\Big(C+\frac{\Gamma}{\lambda}\Big)\mu(A_\lambda)+\frac{1}{2\lambda}\big(1-\mu(A_\lambda)\big)\leq\Big(C+\frac{\Gamma}{\lambda}\Big)\mu(A_\lambda)+\frac{1}{2\lambda},
\]
so that
\[
\mu(A_\lambda)\geq\frac{1-2C\lambda}{2\Gamma+2C\lambda}\geq\frac{1}{4\Gamma}
\qquad\text{for all sufficiently small }\lambda>0.
\]
On $A_\lambda$, it holds that $\Ginv_\lambda(u)\geq x_\lambda:=(\phi')^{-1}\big(1/(2\lambda)\big)$, and $x_\lambda\to+\infty$ as $\lambda\downarrow0$, because $\phi'$ is strictly increasing with range $\R$.

Let $M\geq0$ be such that $|\Finv(u)|\leq M$ a.e., and put $\kappa:=\phi'(M+1)-\phi'(M)$, which is strictly positive as $\phi$ is strictly convex. For $|y|\leq M$ and $x\geq M+1$, since $\phi'$ is non-decreasing,
\[
B_\phi(x,y)=\int_y^x\big(\phi'(s)-\phi'(y)\big)\,\d s\geq\int_{M+1}^x\big(\phi'(s)-\phi'(M)\big)\,\d s\geq\kappa\,(x-M-1).
\]
As $B_\phi\geq0$, it follows that, for all sufficiently small $\lambda$ such that $x_\lambda\geq M+1$,
\[
\DB_\phi(G_\lambda,F)\geq\int_{A_\lambda}B_\phi\big(\Ginv_\lambda(u),\Finv(u)\big)\,\d u\geq\frac{\kappa\,(x_\lambda-M-1)}{4\Gamma}\to\infty
\quad\text{as }\lambda\downarrow0.
\]
Thus all assumptions of Theorem~\ref{thm:WC-RM-BW} are satisfied.
\end{proof}

\begin{proof}[Proof of Proposition~\ref{prop:signed-gamma}]
Let $\Ginv\in\Minv$ with $\DB_\phi(G,F)<\infty$. With $\psi=\phi$ and $k=k_{\lambda^*}$ in Lemma~\ref{lem:iso}, and since $k_{\lambda^*}=\phi'(\Finv)+\gamma/\lambda^*$,
\[
\int_0^1\ell_{\Ginv}(u)\,\d u=\DB_\phi(G,F)-\DB_\phi(G_{\lambda^*},F)-\frac{1}{\lambda^*}\big(H_g(\Ginv)-H_g(\Ginv_{\lambda^*})\big)\geq0,
\]
with equality if and only if $\Ginv=\Ginv_{\lambda^*}$. Hence $\Ginv_{\lambda^*}$ is the unique maximizer of $H_g(\Ginv)-\lambda^*\DB_\phi(G,F)$, and the claim follows by the weak duality argument at the end of the proof of Theorem~\ref{thm:WC-RM-BW}.
\end{proof}

\begin{lemma}\label{lem:C-strong-smooth}
Let $\phi$ be continuously differentiable and satisfy \eqref{eq:strong-smooth}. Then $\phi'$ is strictly increasing with $\phi'(\R)=\R$, $(\phi')^{-1}$ is $1/c$-Lipschitz continuous, $|\phi'(x)|\leq|\phi'(0)|+L|x|$,
\begin{equation}\label{eq:B-sandwich}
\tfrac c2(x-y)^2\leq B_\phi(x,y)\leq\tfrac L2(x-y)^2,\qquad x,y\in\R ,
\end{equation}
and, for all $G,H\in\mathcal M$,
\begin{gather}
\tfrac c2\lVert\Ginv-\Finv\rVert^2\leq\DB_\phi(G,F)\leq\tfrac L2\lVert\Ginv-\Finv\rVert^2,\label{eq:D-sandwich}\\
\big|\DB_\phi(G,F)-\DB_\phi(H,F)\big|\leq L\big(\lVert\Ginv-\Finv\rVert+\lVert\Hinv-\Finv\rVert\big)\lVert\Ginv-\Hinv\rVert .\label{eq:D-lip}
\end{gather}
Moreover, the convex conjugate $\phi^*$ is finite on $\R$, $|\phi^*(y)|\leq C_\phi(1+y^2)$ for some constant $C_\phi$, and $\phi^*(\phi'(x))=x\phi'(x)-\phi(x)$ for $x\in\R$. In particular, $\phi$ is strictly convex and satisfies \eqref{eq:general-p-growth} with $p=2$.
\end{lemma}

\begin{proof}
By \eqref{eq:strong-smooth} with $x=s$, $\phi'(s)-\phi'(y)$ has the sign of $s-y$ and $c|s-y|\leq|\phi'(s)-\phi'(y)|\leq L|s-y|$. This yields the claims on $\phi'$ and $(\phi')^{-1}$, and, as $B_\phi(x,y)=\int_y^x\big(\phi'(s)-\phi'(y)\big)\,\d s$, \eqref{eq:B-sandwich}. Integrating \eqref{eq:B-sandwich} with $x=\Ginv(u)$ and $y=\Finv(u)$ gives \eqref{eq:D-sandwich}. Moreover, $|B_\phi(x,y)-B_\phi(x',y)|=|\phi(x)-\phi(x')-\phi'(y)(x-x')|\leq L\big(|x-y|+|x'-y|\big)|x-x'|$, and the Cauchy--Schwarz inequality yields \eqref{eq:D-lip}. By \eqref{eq:B-sandwich} with $y=0$, $\phi(0)+\phi'(0)x+\frac c2x^2\leq\phi(x)\leq\phi(0)+\phi'(0)x+\frac L2x^2$, so that $\frac1{2L}(y-\phi'(0))^2-\phi(0)\leq\phi^*(y)\leq\frac1{2c}(y-\phi'(0))^2-\phi(0)$. Finally, for $y=\phi'(x)$, the concave function $z\mapsto zy-\phi(z)$ is maximized at $z=x$. The two-sided bound on $\phi$ yields \eqref{eq:general-p-growth} with $p=2$, e.g., with $a_\phi=c/4$, $A_\phi=\max\{L,|\phi(0)|+|\phi'(0)|^2/c\}$ and a suitable $b_\phi$, as $|\phi'(0)x|\leq\frac c4x^2+\frac{\phi'(0)^2}{c}$.
\end{proof}

\begin{lemma}\label{lem:bregman-proj}
Let $\phi$ be continuously differentiable and satisfy \eqref{eq:strong-smooth}, let $\eta\in[0,1]$ and $k\in L^2(0,1)$. Then the minimum in \eqref{eq:bregman-proj} is finite and attained at a unique $\Ginv\in\Minv^\eta$, i.e., $\Pi^\eta_\phi(k)$ is well defined.
\end{lemma}

\begin{proof}
By Lemma~\ref{lem:C-strong-smooth}, $\phi'(\R)=\R$ and $(\phi')^{-1}$ is $1/c$-Lipschitz continuous, so that $K:=(\phi')^{-1}(k)$ satisfies $|K|\leq|(\phi')^{-1}(0)|+|k|/c$ and belongs to $L^2(0,1)$. For $\Ginv\in\Minv^\eta$, let $\Phi(\Ginv):=\int_0^1B_\phi\big(\Ginv(u),K(u)\big)\,\d u$. We use three properties of $\Phi$.
\begin{enumerate}[label=(\roman*),leftmargin=*]
\item By \eqref{eq:B-sandwich}, $\frac c2\lVert\Ginv-K\rVert^2\leq\Phi(\Ginv)\leq\frac L2\lVert\Ginv-K\rVert^2<\infty$.
\item As in the proof of \eqref{eq:D-lip}, which does not use the monotonicity of the second argument, $|\Phi(\Ginv)-\Phi(\Hinv)|\leq L\big(\lVert\Ginv-K\rVert+\lVert\Hinv-K\rVert\big)\lVert\Ginv-\Hinv\rVert$, so that $\Phi$ is continuous on $\Minv^\eta$ with respect to $\lVert\cdot\rVert$.
\item As $x\mapsto\phi(x)-\frac c2x^2$ is convex, $B_\phi(x,y)-\frac c2(x-y)^2$ is the Bregman divergence of a convex function and hence convex in $x$. Therefore, for $\Ginv,\Hinv\in\Minv^\eta$,
\[
\Phi\big(\tfrac12\Ginv+\tfrac12\Hinv\big)\leq\tfrac12\Phi(\Ginv)+\tfrac12\Phi(\Hinv)-\tfrac c8\lVert\Ginv-\Hinv\rVert^2 .
\]
\end{enumerate}
Let $m:=\inf_{\Ginv\in\Minv^\eta}\Phi(\Ginv)\in[0,\infty)$ and let $(\Ginv_n)$ be a minimizing sequence in $\Minv^\eta$. As $\Minv^\eta$ is convex, $\frac12(\Ginv_n+\Ginv_m)\in\Minv^\eta$, and (iii) yields
\[
\lVert\Ginv_n-\Ginv_m\rVert^2\leq\frac4c\big(\Phi(\Ginv_n)+\Phi(\Ginv_m)-2m\big)\to0\qquad\text{as }n,m\to\infty .
\]
Hence $(\Ginv_n)$ is a Cauchy sequence in $L^2(0,1)$. As $\Minv^\eta$ is closed in $L^2(0,1)$ (Section~\ref{sec:prelim}), it converges to some $\Ginv_*\in\Minv^\eta$, and $\Phi(\Ginv_*)=m$ by (ii). If $\Ginv_1,\Ginv_2\in\Minv^\eta$ are both minimizers, then (iii) gives $m\leq m-\frac c8\lVert\Ginv_1-\Ginv_2\rVert^2$, so that $\Ginv_1=\Ginv_2$ a.e., and, as both are non-decreasing and left-continuous, on $(0,1)$.
\end{proof}

\begin{proof}[Proof of Theorem~\ref{thm:WC-RM-BW2-general}]
Throughout, $\lVert\cdot\rVert$ denotes the norm of $L^2(0,1)$, $\lambda>0$ and $t:=1/\lambda$.

\emph{Step 1: preliminaries.} By Lemma~\ref{lem:C-strong-smooth}, $\phi'(\R)=\R$, $(\phi')^{-1}$ is $1/c$-Lipschitz continuous, and \eqref{eq:B-sandwich}, \eqref{eq:D-sandwich} and \eqref{eq:D-lip} hold.

\emph{Step 2: the Lagrangian.} As in the proof of Theorem~\ref{thm:WC-RM-BW}, maximizing the Lagrangian $H_g(\Ginv)-\lambda\DB_\phi(G,F)$ over $\Minv^\eta$ is equivalent to minimizing
\[
J_t(\Ginv):=\DB_\phi(G,F)-t\,H_g(\Ginv)
\]
over $\Minv^\eta$. Since $k_\lambda\in L^2(0,1)$ and $(\phi')^{-1}$ is Lipschitz continuous, $K_\lambda:=(\phi')^{-1}(k_\lambda)\in L^2(0,1)$, and a direct computation shows that
\[
J_t(\Ginv)=\int_0^1B_\phi\big(\Ginv(u),K_\lambda(u)\big)\,\d u+\kappa_\lambda ,
\]
where $\kappa_\lambda$ is finite by \eqref{eq:B-sandwich} and does not depend on $\Ginv$. Hence the minimizers of $J_t$ over $\Minv^\eta$ are the Bregman projections in \eqref{eq:bregman-proj} with $k=k_\lambda$.

By \eqref{eq:D-lip}, $J_t$ is continuous on $\Minv^\eta$. Since $\phi(x)-\frac c2x^2$ is convex, $J_t(\Ginv)-\frac c2\lVert\Ginv\rVert^2$ is convex, i.e., for $\Ginv,\Hinv\in\Minv^\eta$ and $\theta\in[0,1]$,
\begin{equation}\label{eq:strong}
J_t\big((1-\theta)\Ginv+\theta\Hinv\big)\leq(1-\theta)J_t(\Ginv)+\theta J_t(\Hinv)-\tfrac c2\,\theta(1-\theta)\lVert\Hinv-\Ginv\rVert^2 .
\end{equation}
By Lemma~\ref{lem:bregman-proj}, applied with $k=k_\lambda$, $J_t$ attains its minimum over $\Minv^\eta$ at a unique $\Ginv_\lambda$. Letting $\theta\downarrow0$ in \eqref{eq:strong} with $\Ginv=\Ginv_\lambda$ yields
\begin{equation}\label{eq:growth}
J_t(\Hinv)-J_t(\Ginv_\lambda)\geq\tfrac c2\lVert\Hinv-\Ginv_\lambda\rVert^2\qquad\text{for all }\Hinv\in\Minv^\eta .
\end{equation}
Finally, as $|\phi'(x)|\leq|\phi'(0)|+L|x|$, dominated convergence shows that the derivative of $\theta\mapsto J_t\big(\Ginv+\theta(\Hinv-\Ginv)\big)$ at $\theta=0$ equals the left-hand side of \eqref{eq:VI}, as $k_\lambda=\phi'(\Finv)+t\gamma$. Thus \eqref{eq:VI} is the first-order condition for minimizing the convex functional $J_t$ over the convex set $\Minv^\eta$, which is necessary and sufficient.

\emph{Step 3: monotonicity in $\lambda$.} Let $0<\lambda_2<\lambda_1$, $t_i:=1/\lambda_i$, so that $t_1<t_2$, and write $\Ginv_i:=\Ginv_{\lambda_i}$. Applying \eqref{eq:growth} to $J_{t_1}$ at $\Ginv_2$ and to $J_{t_2}$ at $\Ginv_1$ and adding the two inequalities, we obtain
\begin{equation}\label{eq:add}
\begin{split}
c\,\lVert\Ginv_2-\Ginv_1\rVert^2&\leq(t_2-t_1)\big(H_g(\Ginv_2)-H_g(\Ginv_1)\big)\\
&\leq(t_2-t_1)\lVert\gamma\rVert\,\lVert\Ginv_2-\Ginv_1\rVert .
\end{split}
\end{equation}
Hence $\lVert\Ginv_{\lambda_1}-\Ginv_{\lambda_2}\rVert\leq|1/\lambda_1-1/\lambda_2|\,\lVert\gamma\rVert/c$, so that $\lambda\mapsto\Ginv_\lambda$ is continuous in $L^2(0,1)$, and $\lambda\mapsto H_g(\Ginv_\lambda)$ is non-increasing. Moreover, $J_{t_1}(\Ginv_1)\leq J_{t_1}(\Ginv_2)$ and \eqref{eq:add} yield
\begin{align*}
\DB_\phi(G_2,F)-\DB_\phi(G_1,F)&\geq t_1\big(H_g(\Ginv_2)-H_g(\Ginv_1)\big)\\
&\geq\frac{c\,t_1}{t_2-t_1}\lVert\Ginv_2-\Ginv_1\rVert^2\geq0 ,
\end{align*}
with strict inequality if $\Ginv_1\neq\Ginv_2$.

\emph{Step 4: strict monotonicity and continuity of the divergence.} As $\Minv^\eta$ is invariant under the addition of constants, we may choose $\Hinv=\Ginv_\lambda\pm1$ in \eqref{eq:VI}, which gives
\begin{equation}\label{eq:mean-cond}
\int_0^1\Big(\phi'\big(\Ginv_\lambda(u)\big)-\phi'\big(\Finv(u)\big)\Big)\,\d u=\frac1\lambda\int_0^1\gamma(u)\,\d u=\frac{m_\gamma}{\lambda} .
\end{equation}
Hence $\Ginv_{\lambda_1}\neq\Ginv_{\lambda_2}$ whenever $\lambda_1\neq\lambda_2$, and by Step~3, $\lambda\mapsto\DB_\phi(G_\lambda,F)$ is strictly decreasing. By \eqref{eq:D-lip} and the continuity of $\lambda\mapsto\Ginv_\lambda$, it is also continuous.

\emph{Step 5: limits.} Since $\Finv\in\Minv^\eta$, we have $J_t(\Ginv_\lambda)\leq J_t(\Finv)$, so that, by \eqref{eq:D-sandwich},
\begin{align*}
\DB_\phi(G_\lambda,F)&\leq\frac1\lambda\big(H_g(\Ginv_\lambda)-H_g(\Finv)\big)\leq\frac1\lambda\lVert\gamma\rVert\,\lVert\Ginv_\lambda-\Finv\rVert\\
&\leq\frac1\lambda\lVert\gamma\rVert\sqrt{2\DB_\phi(G_\lambda,F)/c},
\end{align*}
that is, $\DB_\phi(G_\lambda,F)\leq2\lVert\gamma\rVert^2/(c\lambda^2)\to0$ as $\lambda\to\infty$. Conversely, by \eqref{eq:mean-cond}, \eqref{eq:strong-smooth} and \eqref{eq:D-sandwich},
\begin{align*}
\frac{|m_\gamma|}\lambda&\leq\int_0^1\Big|\phi'\big(\Ginv_\lambda(u)\big)-\phi'\big(\Finv(u)\big)\Big|\,\d u\leq L\int_0^1\big|\Ginv_\lambda(u)-\Finv(u)\big|\,\d u\\
&\leq L\lVert\Ginv_\lambda-\Finv\rVert\leq L\sqrt{2\DB_\phi(G_\lambda,F)/c},
\end{align*}
where the second inequality holds as \eqref{eq:strong-smooth} and the monotonicity of $\phi'$ imply $|\phi'(x)-\phi'(y)|\leq L|x-y|$. Hence $\DB_\phi(G_\lambda,F)\geq c\,m_\gamma^2/(2L^2\lambda^2)\to\infty$ as $\lambda\downarrow0$. By Step~4 and the intermediate value theorem, there exists a unique $\lambda^*>0$ such that $\DB_\phi(G_{\lambda^*},F)=\ep$.

\emph{Step 6: optimality.} By Step~2, $\Ginv_{\lambda^*}$ uniquely maximizes $H_g(\Ginv)-\lambda^*\DB_\phi(G,F)$ over $\Minv^\eta$. Hence, for every feasible $\Hinv\in\Minv^\eta$,
\begin{align*}
H_g(\Hinv)
&\leq H_g(\Hinv)-\lambda^*\bigl(\DB_\phi(H,F)-\ep\bigr)\\
&\leq H_g(\Ginv_{\lambda^*})-\lambda^*\bigl(\DB_\phi(G_{\lambda^*},F)-\ep\bigr)\\
&=H_g(\Ginv_{\lambda^*}),
\end{align*}
and if $H_g(\Hinv)=H_g(\Ginv_{\lambda^*})$, both inequalities are equalities, so that $\Hinv$ maximizes the Lagrangian over $\Minv^\eta$ and $\Hinv=\Ginv_{\lambda^*}$.
\end{proof}

\begin{proof}[Proof of Corollary~\ref{thm:WC-RM-BW2}]
For $\phi(x)=x^2/2$, \eqref{eq:strong-smooth} holds with $c=L=1$, and $\int_0^1B_\phi(\Ginv(u),k(u))\,\d u=\frac12\lVert\Ginv-k\rVert_2^2$, so that the Bregman projection onto $\mathcal C$ is the $L^2$-projection. The claim follows from Theorem~\ref{thm:WC-RM-BW2-general} and Remark~\ref{rem:WC-RM-BW2-general}(i).
\end{proof}

\begin{proof}[Proof of Proposition~\ref{prop:general-phi-unimodal}]
For every $\Hinv\in\Minv^\eta$ with $\DB_\phi(H,F)\leq\ep$,
\begin{align*}
H_g(\Hinv)&\leq H_g(\Hinv)-\lambda^*\big(\DB_\phi(H,F)-\ep\big)\\
&\leq H_g(\Ginv_{\lambda^*})-\lambda^*\big(\DB_\phi(G_{\lambda^*},F)-\ep\big)=H_g(\Ginv_{\lambda^*}).
\end{align*}
If $H_g(\Hinv)=H_g(\Ginv_{\lambda^*})$, then both inequalities are equalities, so that $\Hinv$ maximizes $\mathcal L_{\lambda^*}$, and hence $\Hinv=\Ginv_{\lambda^*}$.
\end{proof}

\subsection{Proofs for Section~\ref{sec:VaR}}\label{app:C5}

\begin{proof}[Proof of Proposition~\ref{prop:VaR}]
Recall that $z\mapsto B_\phi(z,y)$ vanishes only at $z=y$, is strictly decreasing on $(-\infty,y]$ and strictly increasing on $[y,\infty)$.

(i) For $x\geq \Finv_+(\alpha)$, let $G_x\in\M$ be given by $\Ginv_x(t)=\Finv(t)$ for $t\leq\alpha$ and $\Ginv_x(t)=x\vee \Finv(t)$ for $t>\alpha$; it is non-decreasing and left-continuous, and $(\Ginv_x)_+(\alpha)=x$. Let $G\in\M$ with $x:=\Ginv_+(\alpha)\geq\Finv_+(\alpha)$. For $t\leq\alpha$, $\Ginv_x(t)=\Finv(t)$. For $t>\alpha$, $\Ginv(t)\geq x$, so that either $\Ginv(t)<\Finv(t)$, in which case $\Ginv_x(t)=\Finv(t)$, or $\Finv(t)\leq\Ginv_x(t)\leq\Ginv(t)$. In both cases,
\begin{equation*}
B_\phi\big(\Ginv(t),\Finv(t)\big)\geq B_\phi\big(\Ginv_x(t),\Finv(t)\big),\qquad 0<t<1, \tag{$*$}
\end{equation*}
and in particular $\DB_\phi(G_x,F)\leq\DB_\phi(G,F)$.

Let $G\in \mathcal M_\ep(\phi)$. If $\Ginv_+(\alpha)<\Finv_+(\alpha)$, then for $x=\Finv_+(\alpha)$, $G_x=F$, so that $G_x\in \mathcal M_\ep(\phi)$ and $(\Ginv_x)_+(\alpha)>\Ginv_+(\alpha)$. If $\Ginv_+(\alpha)\geq \Finv_+(\alpha)$, then for $x=\Ginv_+(\alpha)$, ($*$) gives $\DB_\phi(G_x,F)\leq\DB_\phi(G,F)\leq \ep$, so that $G_x\in \mathcal M_\ep(\phi)$ and $(\Ginv_x)_+(\alpha)=\Ginv_+(\alpha)$. Therefore,
\begin{align*}
\sup_{G\in \mathcal M_\ep(\phi)}\Ginv_+(\alpha)
&=\sup\big\{(\Ginv_x)_+(\alpha):x\geq \Finv_+(\alpha),\ G_x\in \mathcal M_\ep(\phi)\big\}\\
&=\sup\Big\{x:\int_{\alpha}^{1}B_{\phi}\big(x\vee \Finv(t),\Finv(t)\big)\,\d t\leq \ep\Big\}.
\end{align*}
The integral vanishes for $x\leq\Finv_+(\alpha)$. On $[\Finv_+(\alpha),\infty)$ it is finite, since its integrand vanishes unless $\Finv_+(\alpha)\leq\Finv(t)<x$, and continuous by dominated convergence, since the integrands are uniformly bounded when $x$ ranges over a compact interval. It is strictly increasing, since $\Finv(t)<x$ on a right neighbourhood of $\alpha$ and $B_\phi(\cdot,y)$ is strictly increasing on $[y,\infty)$. Moreover, it tends to $\infty$ as $x\to\infty$, since strict convexity implies $B_\phi(x,z)\to\infty$ for every fixed $z\in\R$, and Fatou's lemma applies. Hence the supremum above equals the unique $x^*$ with $\int_{\alpha}^{1}B_{\phi}(x^*\vee \Finv(t),\Finv(t))\,\d t=\ep$, and it is attained by $G^*=G_{x^*}$.

Next, we show $\sup_{G\in \mathcal M_\ep(\phi)}\Ginv(\alpha)=\sup_{G\in \mathcal M_\ep(\phi)}\Ginv_+(\alpha)$. Clearly, the left-hand side is at most the right-hand side, which equals $x^*>\Finv_+(\alpha)$. For $0<\kappa<x^*-\Finv_+(\alpha)$ and $n\geq 2$, let $H_n\in\M$ satisfy $\Hinv_n(t)=\Finv(t)$ for $t\leq \alpha-\alpha/n$ and $\Hinv_n(t)=(x^*-\kappa)\vee \Finv(t)$ for $t>\alpha-\alpha/n$. Then $\lim_{n\to\infty}{\DB_\phi}(H_n,F)={\DB_\phi}(G_{x^*-\kappa},F)<\ep$, so that $H_n\in \mathcal M_\ep(\phi)$ for all sufficiently large $n$, and $\Hinv_n(\alpha)=x^*-\kappa$. Hence $\sup_{G\in \mathcal M_\ep(\phi)}\Ginv(\alpha)\geq x^*-\kappa$, and, as $\kappa>0$ was arbitrary, the two suprema coincide.

Finally, let $G\in\mathcal M_\ep(\phi)$ with $x:=\Ginv_+(\alpha)\geq x^*$. By ($*$) and the monotonicity of the above integral, $\ep\geq\DB_\phi(G,F)\geq\DB_\phi(G_x,F)\geq\DB_\phi(G_{x^*},F)=\ep$. Thus $x=x^*$, and ($*$) is an equality a.e. Hence $\Ginv=\Finv$ a.e.\ on $(0,\alpha)$, and $\Ginv=x^*\vee\Finv$ a.e.\ on $(\alpha,1)$: where $\Ginv(t)\geq\Finv(t)$, we have $\Ginv(t)\geq x^*\vee\Finv(t)$ and $B_\phi(\cdot,\Finv(t))$ is strictly increasing on $[\Finv(t),\infty)$, while $\Ginv(t)<\Finv(t)$ would give $B_\phi(\Ginv(t),\Finv(t))>0=B_\phi(\Ginv^*(t),\Finv(t))$. By left-continuity, $\Ginv=\Ginv^*$, so that $G^*$ is the unique maximizer of $\Ginv_+(\alpha)$. If some $G\in\mathcal M_\ep(\phi)$ satisfied $\Ginv(\alpha)=x^*$, then $\Ginv_+(\alpha)\geq x^*$ and thus $G=G^*$, which contradicts $\Ginv^*(\alpha)=\Finv(\alpha)\leq\Finv_+(\alpha)<x^*$. Hence $\sup_{G\in\mathcal M_\ep(\phi)}\Ginv(\alpha)$ is not attained.

(ii) For $y\leq \Finv(\alpha)$, let $G_y\in\M$ be given by $\Ginv_y(t)=y\wedge \Finv(t)$ for $t\leq \alpha$ and $\Ginv_y(t)=\Finv(t)$ for $t>\alpha$; it is non-decreasing and left-continuous, and $\Ginv_y(\alpha)=y$. Let $G\in\M$ with $y:=\Ginv(\alpha)\leq\Finv(\alpha)$. For $t>\alpha$, $\Ginv_y(t)=\Finv(t)$. For $t\leq\alpha$, $\Ginv(t)\leq y$, so that either $\Ginv(t)>\Finv(t)$, in which case $\Ginv_y(t)=\Finv(t)$, or $\Ginv(t)\leq\Ginv_y(t)\leq\Finv(t)$. In both cases,
\begin{equation*}
B_\phi\big(\Ginv(t),\Finv(t)\big)\geq B_\phi\big(\Ginv_y(t),\Finv(t)\big),\qquad 0<t<1, \tag{$**$}
\end{equation*}
and in particular $\DB_\phi(G_y,F)\leq\DB_\phi(G,F)$.

Let $G\in \mathcal M_\ep(\phi)$. If $\Ginv(\alpha)>\Finv(\alpha)$, then for $y=\Finv(\alpha)$, $G_y=F\in \mathcal M_\ep(\phi)$ and $\Ginv_y(\alpha)<\Ginv(\alpha)$. If $\Ginv(\alpha)\leq \Finv(\alpha)$, then for $y=\Ginv(\alpha)$, ($**$) gives $G_y\in \mathcal M_\ep(\phi)$ and $\Ginv_y(\alpha)=\Ginv(\alpha)$. Hence
\[
\inf_{G\in \mathcal M_\ep(\phi)}\Ginv(\alpha)=\inf\Big\{y:\int_{0}^{\alpha}B_{\phi}\big(y\wedge \Finv(t),\Finv(t)\big)\,\d t\leq \ep\Big\}.
\]
The integral vanishes for $y\geq\Finv(\alpha)$. On $(-\infty,\Finv(\alpha)]$ it is finite, since its integrand vanishes unless $y<\Finv(t)\leq\Finv(\alpha)$, and continuous by dominated convergence, since the integrands are uniformly bounded when $y$ ranges over a compact interval. It is strictly decreasing on $(-\infty,\Finv(\alpha))$, since $\Finv(t)>y$ on a left neighbourhood of $\alpha$ by left-continuity of $\Finv$. Moreover, it tends to $\infty$ as $y\to-\infty$, since strict convexity implies $B_\phi(y,z)\to\infty$ for every fixed $z\in\R$, and Fatou's lemma applies. Hence the infimum equals the unique $y^*$ with $\int_{0}^{\alpha}B_{\phi}(y^*\wedge \Finv(t),\Finv(t))\,\d t=\ep$, and it is attained by $G^*=G_{y^*}$.

If $G\in\mathcal M_\ep(\phi)$ is another minimizer, then $\Ginv(\alpha)=y^*$, and, by ($**$),
\[
\ep\geq\DB_\phi(G,F)\geq\DB_\phi(G_{y^*},F)=\ep.
\]
Hence ($**$) is an equality a.e., and, arguing as at the end of (i), $\Ginv=\Finv$ a.e.\ on $(\alpha,1)$ and $\Ginv=y^*\wedge\Finv$ a.e.\ on $(0,\alpha)$, hence everywhere by left-continuity. Thus $G^*$ is the unique minimizer.

Finally, $\inf_{G\in \mathcal M_\ep(\phi)}\Ginv(\alpha)\leq\inf_{G\in \mathcal M_\ep(\phi)}\Ginv_+(\alpha)$. Conversely, for $0<\kappa<\Finv(\alpha)-y^*$ and $n\geq2$, let $\Hinv_n(t)=(y^*+\kappa)\wedge\Finv(t)$ for $t\leq\alpha+(1-\alpha)/n$ and $\Hinv_n(t)=\Finv(t)$ otherwise. Then $\lim_{n\to\infty}\DB_\phi(H_n,F)=\DB_\phi(G_{y^*+\kappa},F)<\ep$. Moreover, since $y^*+\kappa<\Finv(\alpha)\leq\Finv_+(\alpha)$ and $\Finv$ is non-decreasing, $\Hinv_n(t)=y^*+\kappa$ immediately to the right of $\alpha$, so that $(\Hinv_n)_+(\alpha)=y^*+\kappa$. Hence $H_n\in\mathcal M_\ep(\phi)$ for all sufficiently large $n$ and
\[
\inf_{G\in\mathcal M_\ep(\phi)}\Ginv_+(\alpha)
\leq y^*+\kappa.
\]
Letting $\kappa\downarrow0$ shows that the two infima coincide.
\end{proof}

\begin{proof}[Proof of Proposition~\ref{prop:VaR-unimodal}]
Let $G\in\mathcal M_\ep^\eta(\phi)$. Then $\lVert\Ginv\rVert_2\leq\lVert\Finv\rVert_2+\lVert\Ginv-\Finv\rVert_2=\lVert\Finv\rVert_2+\sqrt{2\DB_\phi(G,F)}\leq R$. As $\Ginv$ is non-decreasing, $\Ginv(u)\geq0$ implies $\int_u^1\Ginv(t)^2\,\d t\geq(1-u)\Ginv(u)^2$, and $\Ginv(u)<0$ implies $\int_0^u\Ginv(t)^2\,\d t\geq u\,\Ginv(u)^2$. Hence $|\Ginv(u)|\leq R/\sqrt{\min(u,1-u)}$ for all $0<u<1$. In particular, with $a:=(1+\alpha)/2$ and $K:=R/\sqrt{\min(\alpha/2,1-a)}$, it holds that $|\Ginv(u)|\leq K$ for all $u\in[\alpha/2,a]$.

Next, we show that $\Ginv$ is Lipschitz continuous at $\alpha$, uniformly in $G\in\mathcal M_\ep^\eta(\phi)$. If $\alpha\geq\eta$, then $\Ginv$ is convex on $[\eta,1)$, and for $0<h\leq(1-\alpha)/2$,
\[
\frac{\Ginv(\alpha+h)-\Ginv(\alpha)}{h}\leq\frac{\Ginv(a)-\Ginv(\alpha)}{a-\alpha}\leq\frac{2K}{a-\alpha}.
\]
If $\alpha<\eta$, then $\Ginv$ is concave on $(0,\eta]$, and for $0<h\leq\eta-\alpha$,
\[
\frac{\Ginv(\alpha+h)-\Ginv(\alpha)}{h}\leq\frac{\Ginv(\alpha)-\Ginv(\alpha/2)}{\alpha/2}\leq\frac{4K}{\alpha}.
\]
Similarly, if $\alpha\leq\eta$, concavity yields $\big(\Ginv(\alpha)-\Ginv(\alpha-h)\big)/h\leq\big(\Ginv(\alpha-h)-\Ginv(\alpha/2)\big)/(\alpha/2-h)\leq8K/\alpha$ for $0<h\leq\alpha/4$, and if $\alpha>\eta$, convexity yields $\big(\Ginv(\alpha)-\Ginv(\alpha-h)\big)/h\leq\big(\Ginv(a)-\Ginv(\alpha)\big)/(a-\alpha)\leq2K/(a-\alpha)$ for $0<h\leq\alpha-\eta$. Hence there exist $h_0>0$ and $\Lambda<\infty$, depending only on $\alpha$, $\eta$ and $R$, such that for all $G\in\mathcal M_\ep^\eta(\phi)$ and $0<h\leq h_0$,
\[
\Ginv(\alpha+h)-\Ginv(\alpha)\leq \Lambda h\qquad\text{and}\qquad\Ginv(\alpha)-\Ginv(\alpha-h)\leq \Lambda h.
\]

Finally, since $\Ginv$ is non-decreasing, it follows that for all $G\in\mathcal M_\ep^\eta(\phi)$ and $0<h\leq h_0$,
\[
\Ginv(\alpha)\leq\mathrm{RVaR}_{\alpha,\alpha+h}(G)\leq\Ginv(\alpha+h)\leq\Ginv(\alpha)+\Lambda h
\]
and
\[
\Ginv(\alpha)-\Lambda h\leq\Ginv(\alpha-h)\leq\mathrm{RVaR}_{\alpha-h,\alpha}(G)\leq\Ginv(\alpha).
\]
Taking the supremum, respectively the infimum, over $\mathcal M_\ep^\eta(\phi)$ and letting $h\downarrow0$ proves (i) and (ii). The last statement follows from Corollary~\ref{thm:WC-RM-BW2} and \eqref{eq:best-case}, as both weight functions belong to $L^2(0,1)$ and integrate to one.
\end{proof}

\begin{proof}[Proof of Corollary~\ref{cor:VaR-unimodal-general}]
Since $B_\phi(x,y)-\frac c2(x-y)^2$ is the Bregman divergence of the convex function $\phi(x)-cx^2/2$, it is non-negative. Hence $\lVert\Ginv-\Finv\rVert_2^2\leq 2\DB_\phi(G,F)/c\leq 2\ep/c$ for all $G\in\mathcal M_\ep^\eta(\phi)$, so that $\lVert\Ginv\rVert_2\leq R$. The proof of Proposition~\ref{prop:VaR-unimodal} uses $\phi$ only through this bound and therefore carries over.
\end{proof}

\begin{proof}[Proof of Proposition~\ref{prop:IQD}]
Recall that $B_\phi(\cdot,y)$ is strictly decreasing on $(-\infty,y]$ and strictly increasing on $[y,\infty)$.

(i) For $x_1\leq\Finv(\alpha)$ and $x_2\geq\Finv_+(1-\alpha)$, let $G_{x_1,x_2}$ have the quantile function stated in (i). Then
\[
\mathrm{IQD}_\alpha^+(G_{x_1,x_2})=x_2-x_1,\qquad \DB_\phi(G_{x_1,x_2},F)=L(x_1)+U(x_2).
\]
For any $G\in\mathcal M_\ep(\phi)$, set $x_1:=\Ginv(\alpha)\wedge\Finv(\alpha)$ and $x_2:=\Ginv_+(1-\alpha)\vee\Finv_+(1-\alpha)$. Applying the pointwise clipping inequalities from the proof of Proposition~\ref{prop:VaR} to the two tails gives
\[
\DB_\phi(G_{x_1,x_2},F)\leq\DB_\phi(G,F),\qquad \mathrm{IQD}_\alpha^+(G)\leq x_2-x_1.
\]
Consequently,
\[
\sup_{G\in\mathcal M_\ep(\phi)}\mathrm{IQD}_\alpha^+(G)=\max\bigl\{x_2-x_1:x_1\leq\Finv(\alpha),\ x_2\geq\Finv_+(1-\alpha),\ L(x_1)+U(x_2)\leq\ep\bigr\}.
\]
Indeed, as established in the VaR proof, $L$ and $U$ are continuous on their respective domains and tend to infinity at the unbounded endpoints, so the feasible set is compact. Every maximizing pair exhausts the budget, since otherwise $x_2$ could be increased. Thus the maximum equals the stated maximum over $\mathcal E_\alpha(\ep)$.

Strict convexity of $\phi$ implies strict convexity of $L$ and $U$ on their respective domains. Hence the maximizing pair $(x_1^*,x_2^*)$ is unique: the midpoint of two distinct maximizing pairs would have the same objective value and leave unused budget. If $G$ also maximizes $\mathrm{IQD}_\alpha^+$, the preceding reduction preserves its objective value and
\[
\ep\geq\DB_\phi(G,F)\geq L(x_1^*)+U(x_2^*)=\ep.
\]
Equality in the pointwise clipping inequalities forces $\Ginv=\Ginv_{x_1^*,x_2^*}$ a.e., and hence everywhere by left-continuity. This proves uniqueness of $G^*$.

To compare the suprema, for $s\in(0,1)$ set
\[
x_{1,s}:=(1-s)\Finv(\alpha)+sx_1^*,\qquad x_{2,s}:=(1-s)\Finv_+(1-\alpha)+sx_2^*.
\]
Convexity gives $L(x_{1,s})+U(x_{2,s})\leq s\ep<\ep$. Extend the lower-tail clipping of $G_{x_{1,s},x_{2,s}}$ to $(0,\alpha+h]$ and the upper-tail clipping to $(1-\alpha-h,1)$, where $0<h<(1-2\alpha)/2$. The resulting distribution has $\mathrm{IQD}_\alpha^-=x_{2,s}-x_{1,s}$, and its divergence tends to $L(x_{1,s})+U(x_{2,s})$ as $h\downarrow0$, because $\Finv$ is locally bounded. It is therefore feasible for sufficiently small $h$. Letting $s\uparrow1$ and using $\mathrm{IQD}_\alpha^-\leq\mathrm{IQD}_\alpha^+$ proves equality of the suprema.

The common supremum is strictly larger than $\Finv_+(1-\alpha)-\Finv(\alpha)$, whereas
\[
\mathrm{IQD}_\alpha^-(G^*)=\Finv(1-\alpha)-\Finv_+(\alpha)\leq\Finv_+(1-\alpha)-\Finv(\alpha).
\]
Any maximizer of $\mathrm{IQD}_\alpha^-$ would also maximize $\mathrm{IQD}_\alpha^+$ and hence equal $G^*$, a contradiction. Thus its supremum is not attained.

(ii) Write $a:=\Finv_+(\alpha)$ and $b:=\Finv(1-\alpha)$. For $a\leq x_1\leq x_2\leq b$, let $G_{x_1,x_2}$ now have the quantile function stated in (ii). Then
\[
\mathrm{IQD}_\alpha^-(G_{x_1,x_2})=x_2-x_1,\qquad \DB_\phi(G_{x_1,x_2},F)=C_\alpha(x_1,x_2).
\]
For arbitrary $G\in\mathcal M_\ep(\phi)$, project $\Ginv_+(\alpha)$ and $\Ginv(1-\alpha)$ onto $[a,b]$, obtaining $x_1$ and $x_2$. This does not increase their difference. For $\alpha<t<1-\alpha$, the clipped value $(\Finv(t)\vee x_1)\wedge x_2$ lies between $\Finv(t)$ and $\Ginv(t)$, so
\[
C_\alpha(x_1,x_2)\leq\DB_\phi(G,F),\qquad x_2-x_1\leq\mathrm{IQD}_\alpha^-(G).
\]
Therefore,
\[
\min_{G\in\mathcal M_\ep(\phi)}\mathrm{IQD}_\alpha^-(G)=\min\bigl\{x_2-x_1:(x_1,x_2)\in\mathcal E_\alpha^-(\ep)\bigr\}.
\]
The minimum on the right exists by continuity of $C_\alpha$ and compactness of $\mathcal E_\alpha^-(\ep)$, and the corresponding clipped distribution attains it on the left.

The function
\[
c\longmapsto C_\alpha(c,c)=\int_\alpha^{1-\alpha}B_\phi(c,\Finv(t))\,\d t
\]
has unique minimizer $c_\alpha$ and minimum $\ep_0$, as follows by differentiation. Suppose $0<\ep<\ep_0$. Every minimizing pair has positive width and exhausts the budget, since otherwise its first coordinate could be increased slightly. Moreover,
\[
C_\alpha(x_1,x_2)=\int_\alpha^{1-\alpha}B_\phi(x_1\vee\Finv(t),\Finv(t))\,\d t+\int_\alpha^{1-\alpha}B_\phi(x_2\wedge\Finv(t),\Finv(t))\,\d t
\]
is strictly convex on $\{a\leq x_1\leq x_2\leq b\}$. Thus the midpoint argument from (i) proves uniqueness of the minimizing pair $(x_1^*,x_2^*)$.

If $G$ is any minimizing distribution, its projected pair must equal $(x_1^*,x_2^*)$, and the projection must preserve its positive width. Hence its two quantiles already lie in $[a,b]$ and equal $x_1^*$ and $x_2^*$. Therefore,
\[
\ep\geq\DB_\phi(G,F)\geq C_\alpha(x_1^*,x_2^*)=\ep.
\]
Equality in the pointwise divergence comparison forces $\Ginv$ to equal the stated clipped quantile a.e., and hence everywhere by left-continuity. This proves uniqueness of the minimizing distribution.

If $\ep=\ep_0>0$, the unique minimizing pair is $(c_\alpha,c_\alpha)$. Any distribution with zero $\mathrm{IQD}_\alpha^-$ has a constant quantile $c$ on $(\alpha,1-\alpha)$ and satisfies
\[
\DB_\phi(G,F)\geq\int_\alpha^{1-\alpha}B_\phi(c,\Finv(t))\,\d t\geq\ep_0.
\]
Equality forces $c=c_\alpha$ and $\Ginv=\Finv$ a.e. outside that interval, proving uniqueness also in this case.

It remains to compare the two lower bounds. For $0<\ep\leq\ep_0$, set
\[
x_{1,s}:=(1-s)a+sx_1^*,\qquad x_{2,s}:=(1-s)b+sx_2^*,\qquad 0<s<1.
\]
Since $C_\alpha(a,b)=0$, convexity gives $C_\alpha(x_{1,s},x_{2,s})\leq s\ep<\ep$. Extend the central clipping to $(\alpha-h,1-\alpha+h]$, leaving $\Finv$ unchanged elsewhere. The resulting distribution has $\mathrm{IQD}_\alpha^+=x_{2,s}-x_{1,s}$, and its divergence tends to $C_\alpha(x_{1,s},x_{2,s})$ as $h\downarrow0$. Thus it is feasible for sufficiently small $h$. Letting $s\uparrow1$ and using $\mathrm{IQD}_\alpha^-\leq\mathrm{IQD}_\alpha^+$ proves equality of the lower bounds.

If $\ep>\ep_0$, replace $\Finv$ by $c_\alpha$ on $(\alpha-h,1-\alpha+h]$ and leave it unchanged elsewhere. Since $a\leq c_\alpha\leq b$, this defines a non-decreasing, left-continuous quantile. Its divergence tends to $\ep_0<\ep$ as $h\downarrow0$, and both interquantile differences are zero. Hence the two lower bounds coincide also in this case.

Finally, suppose $0<\ep\leq\ep_0$, and write $m:=x_2^*-x_1^*$. If some $G\in\mathcal M_\ep(\phi)$ attained the infimum of $\mathrm{IQD}_\alpha^+$, then
\[
m\leq\mathrm{IQD}_\alpha^-(G)\leq\mathrm{IQD}_\alpha^+(G)=m.
\]
Thus $G$ would also minimize $\mathrm{IQD}_\alpha^-$, and uniqueness would imply $G=G^*$. However, $C_\alpha(x_1^*,x_2^*)=\ep>0=C_\alpha(a,b)$ implies $(x_1^*,x_2^*)\neq(a,b)$. Since $a\leq x_1^*\leq x_2^*\leq b$, we obtain
\[
m=x_2^*-x_1^*<b-a\leq\Finv_+(1-\alpha)-\Finv(\alpha)=\mathrm{IQD}_\alpha^+(G^*),
\]
a contradiction. Hence the infimum of $\mathrm{IQD}_\alpha^+$ is not attained.
\end{proof}
\subsection{Proofs for Section~\ref{sec:two-BW-balls}}\label{app:C6}

\begin{proof}[Proof of Theorem~\ref{thm:WC-RM-BW-two-balls}]
Throughout, $\boldsymbol e_i$ denotes the $i$th unit vector and
\[
\mathcal X:=\big\{\Ginv\in\Minv:\DB_{\phi_i}(G,F)<\infty,\ i=1,2\big\}.
\]
Since each Bregman divergence is convex in its first argument, both $\mathcal X$ and the feasible set of \eqref{opt:WC-RM-two-balls} are convex, and the latter is contained in $\mathcal X$.

\emph{Step 1: Lagrangian maximizer.} Let $\boldsymbol{\lambda}\in[0,\infty)^2\setminus\{\boldsymbol{0}\}$. Up to constants that do not depend on $\Ginv$, the Lagrangian associated with \eqref{opt:WC-RM-two-balls} is
\begin{equation} \label{eq:Lagrangian-two-balls}
\mathcal L_{\boldsymbol{\lambda}}(\Ginv) = \int_0^1 \gamma(u)\Ginv(u)\,\d u - \sum_{i=1}^2 \lambda_i \int_0^1 B_{\phi_i}\big(\Ginv(u),\Finv(u)\big)\,\d u .
\end{equation}
Expanding $B_{\phi_i}$ as in Section~\ref{sec:prelim}, maximizing \eqref{eq:Lagrangian-two-balls} over $\mathcal X$ is equivalent to minimizing
\begin{equation} \label{eq:min-two-balls}
\int_0^1 \Big( \psi_{\boldsymbol{\lambda}}\big(\Ginv(u)\big) - k_{\boldsymbol{\lambda}}(u)\Ginv(u) \Big)\,\d u
\end{equation}
over $\Ginv\in\mathcal X$, with $\psi_{\boldsymbol{\lambda}}$ and $k_{\boldsymbol{\lambda}}$ as in \eqref{eq:psi-lambda-two-balls} and \eqref{eq:k-lambda-two-balls}. This reduction is justified as in the proof of Theorem~\ref{thm:WC-RM-BW}: since $\phi_i'(\Finv)=k_{\boldsymbol e_i}-\gamma\in L^2(0,1)$, all integrals involved are finite for every $\Ginv\in\mathcal X$. Since $\psi_{\boldsymbol{\lambda}}$ is strictly convex, Lemma~\ref{lem:iso}, applied with $\psi=\psi_{\boldsymbol{\lambda}}$ and $k=k_{\boldsymbol{\lambda}}$, shows that the unique minimizer of \eqref{eq:min-two-balls} over $\Minv$ is $\Ginv_{\boldsymbol{\lambda}}$ given in \eqref{eq:G-lambda-two-balls}. As $\Ginv_{\boldsymbol{\lambda}}\in\mathcal X$ by assumption, it is also the unique minimizer over $\mathcal X$, and hence the unique maximizer of the Lagrangian over $\mathcal X$.

\emph{Step 2: ``only if''.} Let $\Ginv^*$ solve \eqref{opt:WC-RM-two-balls}. On $\mathcal X$, the objective $H_g$ is linear and real-valued, the constraint functionals are convex and real-valued, and the optimal value is finite. Slater's condition holds since $\Finv\in\mathcal X$ and $\DB_{\phi_i}(F,F)=0<\ep_i$, $i=1,2$. By the Lagrange multiplier theorem for convex problems (see, e.g., \citet[Section~8.6, Theorem~1]{luenberger1969optimization}), there exists $\boldsymbol{\lambda}^*\in[0,\infty)^2$ such that
\[
\lambda_i^* \big( \DB_{\phi_i}(G^*,F)-\ep_i \big)=0, \qquad i=1,2,
\]
and $\Ginv^*$ maximizes $\mathcal L_{\boldsymbol{\lambda}^*}$ over $\mathcal X$.

We show that $\boldsymbol{\lambda}^*\neq\boldsymbol{0}$. Otherwise, $\Ginv^*$ maximizes $H_g$ over $\mathcal X$. As $\gamma\geq0$ and $\int_0^1\gamma(u)\,\d u=1$, we can choose $b\in(0,1)$ with $\int_0^b\gamma(u)\,\d u>0$, and we choose $x_0>\Ginv^*(b)$. Define
\[
\Hinv(u):=\max\big\{\Ginv^*(u),x_0\big\},\qquad 0<u<1,
\]
which is non-decreasing and left-continuous, so that $\Hinv\in\Minv$. By convexity of $\phi_i$,
\[
0\leq B_{\phi_i}\big(x_0,\Finv(u)\big)\leq\big(\Finv(u)-x_0\big)\big(\phi_i'(\Finv(u))-\phi_i'(x_0)\big),
\]
and the right-hand side is integrable by the Cauchy--Schwarz inequality, as $\Finv\in L^2(0,1)$ since $F\in\mathcal M$, and $\phi_i'(\Finv)\in L^2(0,1)$. As $B_{\phi_i}(\Hinv(u),\Finv(u))$ equals either $B_{\phi_i}(\Ginv^*(u),\Finv(u))$ or $B_{\phi_i}(x_0,\Finv(u))$, and both are non-negative,
\[
\DB_{\phi_i}(H,F)\leq\DB_{\phi_i}(G^*,F)+\int_0^1 B_{\phi_i}\big(x_0,\Finv(u)\big)\,\d u<\infty,\qquad i=1,2,
\]
that is, $\Hinv\in\mathcal X$. Moreover, since $\gamma\geq0$ and $\Ginv^*\leq\Ginv^*(b)$ on $(0,b]$,
\[
H_g(\Hinv)-H_g(\Ginv^*)=\int_0^1\gamma(u)\big(x_0-\Ginv^*(u)\big)^+\,\d u\geq\big(x_0-\Ginv^*(b)\big)\int_0^b\gamma(u)\,\d u>0,
\]
contradicting the optimality of $\Ginv^*$ over $\mathcal X$. Hence $\boldsymbol{\lambda}^*\neq\boldsymbol{0}$.

By Step~1, $\Ginv_{\boldsymbol\lambda^*}$ is the unique maximizer of $\mathcal L_{\boldsymbol\lambda^*}$ over $\mathcal X$, and therefore $\Ginv^* = \Ginv_{\boldsymbol{\lambda}^*}$.

\emph{Step 3: ``if''.} Suppose that $\boldsymbol{\lambda}^*\in[0,\infty)^2\setminus\{\boldsymbol{0}\}$ satisfies \eqref{eq:KKT-two-balls}. For any feasible $\Ginv\in\Minv$, we have $\Ginv\in\mathcal X$ and $\DB_{\phi_i}(G,F)\leq\ep_i$, $i=1,2$, and therefore
\begin{align*}
H_g(\Ginv) &\leq H_g(\Ginv) - \sum_{i=1}^2 \lambda_i^* \big( \DB_{\phi_i}(G,F)-\ep_i \big) \\
&\leq H_g(\Ginv_{\boldsymbol{\lambda}^*}) - \sum_{i=1}^2 \lambda_i^* \big( \DB_{\phi_i}(G_{\boldsymbol{\lambda}^*},F)-\ep_i \big)
= H_g(\Ginv_{\boldsymbol{\lambda}^*}),
\end{align*}
by Step~1 and complementary slackness. Since $\Ginv_{\boldsymbol{\lambda}^*}$ is feasible by \eqref{eq:KKT-two-balls}, it is optimal. If a feasible $\Hinv\in\Minv$ satisfies $H_g(\Hinv)=H_g(\Ginv_{\boldsymbol{\lambda}^*})$, then both inequalities are equalities, so $\Hinv$ maximizes $\mathcal L_{\boldsymbol{\lambda}^*}$ over $\mathcal X$ and thus $\Hinv=\Ginv_{\boldsymbol{\lambda}^*}$ by Step~1.

\emph{Step 4: one positive multiplier.} Let $\lambda_2^*=0$, so that $\lambda_1^*>0$; the case $\lambda_1^*=0$ is symmetric. Then $\psi_{\boldsymbol{\lambda}^*}'=\lambda_1^*\phi_1'$ and $k_{\boldsymbol{\lambda}^*}=\lambda_1^*\big(\phi_1'(\Finv)+\gamma/\lambda_1^*\big)$. Since the isotonic projection is positively homogeneous, i.e., $(s\,k)^\uparrow=s\,k^\uparrow$ for $s>0$,
\[
\Ginv_{\boldsymbol{\lambda}^*}=(\phi_1')^{-1}\Big(\big(\phi_1'(\Finv)+\tfrac{1}{\lambda_1^*}\gamma\big)^{\uparrow}\Big).
\]
As the first constraint is binding by complementary slackness, Proposition~\ref{prop:signed-gamma}, applied with $\phi=\phi_1$, $\ep=\ep_1$ and $\lambda^*=\lambda_1^*$, shows that this is the unique solution of the one-ball problem. It coincides with the formula of Theorem~\ref{thm:WC-RM-BW}, without requiring the global assumptions of that theorem.
\end{proof}

\begin{proof}[Proof of Corollary~\ref{cor:VaR-two-balls}]
We prove the statement for the supremum; the proof for the infimum is similar. Let $G\in\mathcal M_{\boldsymbol\ep}$ with $x:=\Ginv_+(\alpha)\geq\Finv_+(\alpha)$, and let $G_x$ be as in the proof of Proposition~\ref{prop:VaR}. For $t\leq\alpha$, $B_{\phi_i}(\Ginv_x(t),\Finv(t))=0$. For $t>\alpha$, either $\Ginv(t)<\Finv(t)$, in which case $\Ginv_x(t)=\Finv(t)$, or $\Finv(t)\leq\Ginv_x(t)\leq\Ginv(t)$. Since $z\mapsto B_{\phi_i}(z,y)$ is non-decreasing on $[y,\infty)$, in both cases $B_{\phi_i}(\Ginv_x(t),\Finv(t))\leq B_{\phi_i}(\Ginv(t),\Finv(t))$. Hence, we obtain that  $\DB_{\phi_i}(G_x,F)\leq\DB_{\phi_i}(G,F)\leq\ep_i$ for $i=1,\ldots,m$, and thus that  $G_x\in\mathcal M_{\boldsymbol\ep}$. The case $\Ginv_+(\alpha)<\Finv_+(\alpha)$ is treated as in the proof of Proposition~\ref{prop:VaR}. Hence,
\begin{align*}
\sup_{G\in\mathcal M_{\boldsymbol\ep}}\Ginv_+(\alpha)
&=\sup\Big\{x:\int_\alpha^1 B_{\phi_i}\big(x\vee\Finv(t),\Finv(t)\big)\,\d t\leq\ep_i,\ i=1,\ldots,m\Big\}\\
&=\min_{1\leq i\leq m}x_i^*,
\end{align*}
where we used that each of these integrals is continuous and strictly increasing in $x$ on $(\Finv_+(\alpha),\infty)$, as shown in the proof of Proposition~\ref{prop:VaR}. Finally, the equality with $\sup_{G\in\mathcal M_{\boldsymbol\ep}}\Ginv(\alpha)$ follows from the approximation argument in that proof, as $\DB_{\phi_i}(G_{x^*-\kappa},F)<\ep_i$ for all $i$ and every sufficiently small $\kappa>0$. Let $i^*$ be an index with $x_{i^*}^*=x^*$. If $G\in\mathcal M_{\boldsymbol\ep}$ satisfies $\Ginv_+(\alpha)\geq x^*$, the argument at the end of the proof of Proposition~\ref{prop:VaR}(i), applied to $\phi_{i^*}$ and $\ep_{i^*}$, yields $G=G^*$. Hence $G^*$ is the unique maximizer of $\Ginv_+(\alpha)$ over $\mathcal M_{\boldsymbol\ep}$, and $\sup_{G\in\mathcal M_{\boldsymbol\ep}}\Ginv(\alpha)$ is not attained. For the infimum, $G_{y^*}$ belongs to $\mathcal M_{\boldsymbol\ep}$, as $y^*\geq y_i^*$ for all $i$, and therefore attains it.
\end{proof}

\begin{proof}[Proof of Proposition~\ref{prop:two-refs}]
Up to terms that do not depend on $\Ginv$, the Lagrangian of problem \eqref{opt:WC-RM-two-refs} equals
\begin{align*}
&\int_0^1\gamma(u)\Ginv(u)\,\d u-\sum_{i=1}^2\lambda_i\int_0^1\Big(\phi_i\big(\Ginv(u)\big)-\phi_i'\big(F_i^{-1}(u)\big)\Ginv(u)\Big)\,\d u\\
&\qquad=-\int_0^1\Big(\psi_{\boldsymbol\lambda}\big(\Ginv(u)\big)-\tilde k_{\boldsymbol\lambda}(u)\Ginv(u)\Big)\,\d u .
\end{align*}
This has the same form as in the proof of Theorem~\ref{thm:WC-RM-BW-two-balls}, with $k_{\boldsymbol\lambda}$ replaced by $\tilde k_{\boldsymbol\lambda}$. Lemma~\ref{lem:iso} therefore yields that $\tilde G^{-1}_{\boldsymbol\lambda}$ is the unique minimizer of the corresponding integral functional over $\Minv$. Since this candidate has finite divergence under both generators by assumption, it is the unique maximizer of the Lagrangian over the set of quantile functions with both divergences finite. The remainder of that proof applies with $F$ replaced by $F_i$ in the $i$th constraint and with two changes in Step~2. First, Slater's condition holds at the strictly feasible quantile function assumed to exist in Proposition~\ref{prop:two-refs}, instead of at $\Finv$. Second, in the argument excluding $\boldsymbol\lambda^*=\boldsymbol 0$, the integrability of $B_{\phi_i}(x_0,F_i^{-1})$ follows from $F_i^{-1}\in L^2(0,1)$, as $F_i\in\mathcal M$, and $\phi_i'(F_i^{-1})=\tilde k_{\boldsymbol e_i}-\gamma\in L^2(0,1)$. Step~4 carries over with $F$ replaced by $F_i$.

For generators defined on an open interval $J$, as in Remark~\ref{rem:two-refs}(iv), the reduction in Step~1 uses the convexity bounds of the proof of Theorem~\ref{thm:WC-RM-BW} with $0$ replaced by a fixed point of $J$, and the `if' part and uniqueness in Step~3 carry over unchanged.
\end{proof}

\begin{proof}[Proof of Proposition~\ref{prop:two-refs-unimodal}]
(i) Let $\Hinv\in\Minv^\eta$ be feasible. Then $\Hinv\in\mathcal S$, and
\begin{align*}
H_g(\Hinv)
&\leq H_g(\Hinv)-\sum_{i=1}^2\lambda_i^*\big(\DB_{\phi_i}(H,F_i)-\ep_i\big)
=\mathcal L_{\boldsymbol\lambda^*}(\Hinv)+\sum_{i=1}^2\lambda_i^*\ep_i\\
&\leq\mathcal L_{\boldsymbol\lambda^*}\big(G^{-1}_{\eta,\boldsymbol\lambda^*}\big)+\sum_{i=1}^2\lambda_i^*\ep_i
=H_g\big(G^{-1}_{\eta,\boldsymbol\lambda^*}\big),
\end{align*}
where the last equality follows from complementary slackness. As $G^{-1}_{\eta,\boldsymbol\lambda^*}$ is feasible, it is a solution. If $H_g(\Hinv)=H_g(G^{-1}_{\eta,\boldsymbol\lambda^*})$, the second inequality is an equality, so that $\Hinv$ maximizes $\mathcal L_{\boldsymbol\lambda^*}$ over $\mathcal S$ and $\Hinv=G^{-1}_{\eta,\boldsymbol\lambda^*}$.

(ii) We first show that $\DB_{\phi_i}(\cdot,F_i)$ is convex and lower semicontinuous on $\Minv$ with respect to the $L^2$-norm. Convexity follows from the convexity of $B_{\phi_i}$ in its first argument. For $\phi_1$, the assumption that $B_{\phi_1}(x,y)$ tends to infinity at finite endpoints of its domain ensures that $x\mapsto B_{\phi_1}(x,y)$, extended by $+\infty$ outside the domain, is lower semicontinuous on $\R$ for every $y$ in the domain. For $\phi_2$, lower semicontinuity follows from continuity on $\R$. If $G_n^{-1}\to\Ginv$ in $L^2(0,1)$, we may pass to a subsequence along which $\DB_{\phi_i}(G_n,F_i)$ converges to its limit inferior and to a further subsequence that converges a.e.; Fatou's lemma then yields $\DB_{\phi_i}(G,F_i)\leq\liminf_n\DB_{\phi_i}(G_n,F_i)$. Being convex and lower semicontinuous, $\DB_{\phi_i}(\cdot,F_i)$ is also weakly lower semicontinuous.

Hence the feasible set $\mathcal K:=\{\Ginv\in\Minv^\eta:\DB_{\phi_i}(G,F_i)\leq\ep_i,\ i=1,2\}$ is convex and weakly closed. It is nonempty, as $H_0^{-1}\in\mathcal K$, and bounded in $L^2(0,1)$, since $B_{\phi_2}(x,y)\geq\frac c2(x-y)^2$ implies $\lVert\Ginv-F_2^{-1}\rVert^2\leq2\ep_2/c$ for $\Ginv\in\mathcal K$. Thus $\mathcal K$ is weakly compact, and the weakly continuous functional $H_g$ attains its maximum on $\mathcal K$ at some $G_*^{-1}$.

On the convex set $\mathcal S$, $H_g$ is real-valued and linear and the constraint functionals $\DB_{\phi_i}(\cdot,F_i)-\ep_i$ are real-valued and convex, and $H_0^{-1}\in\mathcal S$ is a Slater point. By the Lagrange multiplier theorem for convex problems (see, e.g., \citet[Chapter~8]{luenberger1969optimization}), there exists $\boldsymbol\lambda^*\in[0,\infty)^2$ such that $G_*^{-1}$ maximizes $\mathcal L_{\boldsymbol\lambda^*}$ over $\mathcal S$ and $\lambda_i^*\big(\DB_{\phi_i}(G_*,F_i)-\ep_i\big)=0$, $i=1,2$. If $\boldsymbol\lambda^*\neq\boldsymbol0$, $\mathcal L_{\boldsymbol\lambda^*}$ is strictly concave on $\mathcal S$, as $\phi_1$ and $\phi_2$ are strictly convex, so that its maximizer is unique, and (i) applies.
\end{proof}

\end{document}